\documentclass[a4paper]{article}

\usepackage{amsmath,amsthm,amssymb,mathtools}
\usepackage{comment}
\usepackage[T1]{fontenc} 
\usepackage[mathcal]{euscript}
\usepackage{graphicx}
\usepackage[hidelinks]{hyperref}
\usepackage[top=2.7cm,bottom=2.7cm,left=2.05cm,right=2.05cm]{geometry}
\usepackage{tikz}
\usepackage[export]{adjustbox}
\usepackage{appendix}
\usepackage{setspace}
\usepackage{tcolorbox}
\numberwithin{equation}{section}
\theoremstyle{plain}

\theoremstyle{definition}
\newtheorem{remark}{Remark}[section]

\newcommand{\rd}{\mathrm{d}}
\newcommand{\fd}[1]{\frac{\rd #1}{\rd z}}
\newcommand{\sd}[1]{\frac{\rd^2 #1}{\rd z^2}}

\makeatletter
\g@addto@macro\bfseries{\boldmath}
\makeatother

\usetikzlibrary{shapes.geometric,decorations.markings,intersections,calc,positioning}

\tikzset{->-/.style={decoration={
  markings,
  mark=at position .6 with {\arrow{>}}},postaction={decorate}}}

\tikzset{-<-/.style={decoration={
  markings,
  mark=at position .6 with {\arrow{<}}},postaction={decorate}}}

\tikzset{%
    add/.style args={#1 and #2}{
        to path={%
 ($(\tikztostart)!-#1!(\tikztotarget)$)--($(\tikztotarget)!-#2!(\tikztostart)$)%
  \tikztonodes},add/.default={.2 and .2}}
}  

\tikzset{
    extended line/.style={shorten >=-#1,shorten <=-#1},
    extended line/.default=1cm]
}

\tikzset{line through/.style args={#1 parallel to line through #2 and #3 and
length #4}{insert path={%
let \p1=($(#3)-(#2)$),\n1={atan2(\y1,\x1)} in (#1) -- ++ (\n1:#4)}}}

\definecolor{col1}{rgb}{0.4, 0.69, 0.2}
\definecolor{col2}{rgb}{0.96, 0.29, 0.54}

\definecolor{green(ryb)}{rgb}{0.4, 0.69, 0.2}
\definecolor{frenchrose}{rgb}{0.96, 0.29, 0.54}
\definecolor{persianblue}{rgb}{0.11, 0.22, 0.73}
\definecolor{jade}{rgb}{0.0, 0.66, 0.42}
\definecolor{limegreen}{rgb}{0.2, 0.8, 0.2}

\title{Spaces of initial conditions for \\ quartic Hamiltonian systems \\ of Painlev\'e and quasi-Painlev\'e type}
\author{Marta Dell'Atti{\color{jade}$\,^1$} \\[-.5ex] \small{\url{m.dell-atti@uw.edu.pl}} \\[.5ex]
Thomas Kecker{\color{jade}$\,^{2,3}$} \\[-.5ex]  
\small{\url{thomas.kecker@port.ac.uk}} \\[.5ex]
{\color{jade}$\,^1$}{ \small University of Warsaw, Institute of Mathematics, 
}  \\[.1ex] {\small Banacha 2, Warsaw, 02-097, Poland}
\\[1ex] 
{\color{jade}$\,^2$}{ \small School of Mathematics and Physics, University of Portsmouth, Lion Terrace, PO1 3HF, UK}\\[1ex]
{\color{jade}{$\,^{3}$}}{ \footnotesize Corresponding author}
}

\date{December 2025}

\begin{document}

\maketitle

\begin{abstract}
The geometric approach for Painlev\'e and quasi-Painlev\'e differential equations in the complex plane is applied to non-autonomous Hamiltonian systems, quartic in the dependent variables. By computing their defining manifolds (analogue of the Okamoto's space of initial conditions in the quasi-Painlev\'e case), we provide a classification of such systems. We distinguish the various cases by the local behaviour at the movable singularities of the solutions, which are algebraic poles or ordinary poles. The principal cases are categorised by the initial base points of the system in the extended phase space $\mathbb{CP}^2$ and their multiplicities, arising from the coalescence of $4$ simple base points in the generic case. Through the mechanisms of coalescence of base points and degeneration (by setting certain coefficient functions in the Hamiltonian to~$0$), all possible sub-cases of quartic Hamiltonian systems with the quasi-Painlev\'e property are obtained, and are characterised by their corresponding Newton polygons. As particular sub-cases we recover certain systems equivalent to known Painlev\'e equations, or variants thereof. The resulting picture is a multi-faceted description of each case: the local behaviour around singularities, the surface type, and the Newton polygon. 
\end{abstract}

\tableofcontents

\section{Introduction}
\label{sec:intro}

\setlength{\belowdisplayskip}{5pt} \setlength{\belowdisplayshortskip}{5pt}
\setlength{\abovedisplayskip}{5pt} \setlength{\abovedisplayshortskip}{5pt}

The theory of ordinary differential equations and their solutions in the complex plane forms a classical branch of complex analysis that is continuously developing further, often motivated by applications in other sciences, in particular physics. For example, special functions such as Bessel functions were introduced when needed to describe refraction of waves at a circular opening, or Legendre functions to describe quantum wave functions of an electron around a hydrogen atom. These functions all have in common that they are solutions of second-order linear differential equations with non-constant coefficients. While generally not solvable in closed, elementary form, one can write down analytic series solutions, possibly with fixed singularities at points in the complex plane where the coefficient functions of the equation are ill-defined. Such equations are thus considered as being integrable. 
While the special functions named above all describe solutions to linear problems, new phenomena arise when passing to non-linear problems, such as propagation of shallow water waves or light in non-linear me. In this case, the underlying non-linear differential equations can exhibit \textit{movable singularities}, i.e.\ non-analytic points of the solutions whose position is determined by the initial data of the equation. In particular, these points cannot be read off from the equation itself. 
An important class of ordinary differential equations that lies on the boundary of integrability and non-integrability are the Painlev\'e equations, or more generally equations and systems of equations with the Painlev\'e property, which says that all local solutions can be meromorphically continued in the complex plane, possibly punctured at a number of fixed singularities. 

\subsection{Painlev\'e equations}
The six Painlev\'e equations are a by now well-established class of non-linear second-order differential equations, giving rise to meromorphic solutions in the complex plane or a punctured plane. These equations were originally found in the quest to find new transcendental functions defined as solutions of ordinary differential equations, discovered at the beginning of the 20th century by French mathematician P.\ Painlev\'e and his school~\cite{Painleve1900}, while the sixth equation in this list was first found by R. Fuchs in connection with isomonodromic deformations~\cite{Fuchs1907}.

The Painlev\'e equations were obtained by imposing the condition that all solutions allow meromorphic continuation in the (once or twice punctured) plane, on equations in the class 
\begin{equation}
\label{Pclass}
    \sd{y}=R\left(y(z),\fd{y};z\right)\,,
\end{equation}
where the right-hand side is a rational expression in its first two arguments. The classification by Painlev\'e and his school listed $50$ equations from which all equations in the class~\eqref{Pclass} with the Painlev\'e property can be obtained by applying some M\"obius transformation. Only six of these equations could not be solved in terms of formerly known special functions, and are nowadays called the Painlev\'e equations. While originally obtained by this complex analytic motivation, the equations have found abundant applications in more recent times, one important aspect being that are obtained by symmetry reduction of many integrable non-linear partial differential equations such as the KdV, Sine-Gordon, or non-linear Schr\"odinger equation. In this way, the Painlev\'e equations are considered integrable and their solutions, the Painlev\'e transcendents, play a similar role in non-linear physics as do the above mentioned special functions in the linear world. 

In the following we list the six Painlev\'e equations in their usual form:
\begin{align}
\label{Painleve1}
    \text{P}_{\text{I}}\colon &\, \sd{y} =  6\,y^2 + z \,,\\[2ex]
\label{Painleve2}
    \text{P}_{\text{II}}\colon &\, \sd{y} =  2\,y^3 + zy + \alpha \,, \\[2ex]
\label{Painleve3}
    \text{P}_{\text{III}}\colon &\, \sd{y} =  \frac{1}{y} \left( \fd{y} \right)^2 - \frac{1}{z} \fd{y} + \frac{1}{z}(\alpha\, y^2 + \beta) + \gamma\, y^3 + \frac{\delta}{y}\,, \\[1ex]
\label{Painleve4}
    \text{P}_{\text{IV}}\colon &\, \sd{y} =  \frac{1}{2\,y} \left( \fd{y} \right)^2 + \frac{3}{2} \,y^3 + 4\,z\,y^2 +2(z^2-\alpha)y + \frac{\beta}{y}\,, \\[1ex]
\label{Painleve5}
    \text{P}_{\text{V}}\colon &\, \sd{y} =  \frac{3\,y-1}{2\,y(y-1)} \left( \fd{y} \right)^2 - \frac{1}{z} \fd{y} + \frac{(y-1)^2}{z^2} \left( \alpha y + \frac{\beta}{y} \right) + \frac{\gamma\, y}{z} + \frac{\delta\, y(y+1)}{y-1} \,,\\[1ex]
    \begin{split} 
\label{Painleve6}
 \hspace*{-5ex}
   \text{P}_{\text{VI}}\colon &\, \sd{y} =   \frac{1}{2}\left( \frac{1}{y} + \frac{1}{y-1} + \frac{1}{y-z} \right) \left( \fd{y} \right)^2 - \left( \frac{1}{z} + \frac{1}{z-1} + \frac{1}{y-z} \right) \fd{y}  \\[.7ex] & \qquad 
    + \frac{y(y-1)(y-z)}{z^2(z-1)^2} \left( \alpha  + \beta\, \frac{z}{y^2} + \gamma\, \frac{z-1}{(y-1)^2} + \delta\, \frac{z(z-1)}{(y-z)^2} \right), \\[2ex]
    \end{split} 
\end{align}
with $\alpha,\beta,\gamma,\delta \in \mathbb{C}$ constant parameters. The general solutions to these equations are called $\text{P}_{\text{J}}$-transcendents with $\text{J}=\text{I}, \dots, \text{VI}$, however for specific values of the parameters $\alpha,\beta,\gamma,\delta$ solutions exist that can be expressed in terms of known special functions, such as Airy, Bessel, and hypergeometric functions, nowadays listed in the digital library of mathematical functions~\cite{NIST:DLMF}. Painlev\'e's equations I, II and IV have meromorphic solutions in the entire complex plane, while the others are meromorphic only on the universal covering surface of $\mathbb{C} \setminus \{0\}$ (Painlev\'e $\text{III}$ and $\text{V}$) or $\mathbb{C} \setminus \{0,1\}$ (Painlev\'e VI).

An important aspect of the Painlev\'e equations is that they admit a Hamiltonian representation, i.e.\ for each $\text{P}_{\text{J}}$ with $\text{J}=\text{I}, \dots, \text{VI}$ it exists a (non-autonomous and non-unique) Hamiltonian function $H_{\text{J}}(x,y)$ with the phase space variables~$(x(z),y(z))\in \mathbb{C}^2$ expressed as functions of $z\in \mathbb{C}$. 
It is customary to refer to the Hamiltonian systems studied by Okamoto in a series of papers (\cite{okamoto1,okamoto2,okamoto3,okamoto4}) and explicitly given by the following list 
\begin{align} 
H_{\text{I}}\big(x(z),y(z);z\big) &= \frac{1}{2} y^2 - 2 x^3 - zx \,,\\[1ex]
H_{\text{II}}\big(x(z),y(z);z\big) &= \frac{1}{2} y^2 - \left( x^2 - \frac{z}{2} \right) y - \kappa x \,, \\[1ex]
H_{\text{III}}\big(x(z),y(z);z\big) &= \frac{1}{z} \left[ 2 y^2 x^2 - \left( 2 \eta_\infty\, z x^2 + (2\kappa_0 + 1) x - 2 \eta_0\, z \right)y + \eta_\infty (\kappa_0 + \kappa_\infty)zx \right] \,, \\[1ex]
H_{\text{IV}}\big(x(z),y(z);z\big) &= 2xy^2 - \left(x^2 + 2zx + \kappa_0 \right)y + \kappa_\infty x \,,\label{eq:HPIV}\\[1ex]
H_{\text{V}}\big(x(z),y(z);z\big) &= \frac{1}{z} \left[ x(x-1)^2 y^2 - \left( \kappa_0 (x-1)^2 + \kappa_t\, x(x-1) - \eta\, zx \right) y + \kappa (x-1) \right] \,,\\[1ex]
\begin{split} 
H_{\text{VI}}\big(x(z),y(z);z\big) &= \frac{1}{z(z-1)} \left[ x(x-1)(x-z) y^2 - \left[ \kappa_0\, (x-1)(x-z) + \kappa_1\, x(x-z) \right. \right. \\ & \left. \left. ~~+  (\kappa_t-1)\,x(x-1) \right] y + \kappa(x-t) \right],
\end{split} 
\end{align}
even if some of them were already known to Malmquist~\cite{Malmquist}. The Hamiltonians $H_{\text{J}}$, with $\text{J}=\text{I}, \dots, \text{VI}$ are written with the parameters $\kappa, \kappa_0, \kappa_1, \kappa_t, \kappa_{\infty},\eta, \eta_{0}, \eta_{\infty}$, which can be expressed in terms of the parameters $\alpha, \beta, \gamma, \delta$ in~\eqref{Painleve1} - \eqref{Painleve6}. For each of the listed Hamiltonian one then derives the Hamiltonian systems, i.e.\ the system of two first order equations
\begin{equation}
    \fd{y} = \frac{\partial H_{\text{J}}}{\partial x}\,, \qquad \fd{x} = -\frac{\partial H_{\text{J}}}{\partial y}\,.
\end{equation}
Both the Hamiltonian systems $H_\text{J}$ and Painlev\'e equations $\text{P}_{\text{J}}$ are characterised by the fact that their only movable singularities are poles. Roughly speaking, movable singularities are singular points of a solution whose position depend in a parametric way on the initial data for the equation, whereas fixed singularities are those that can be read off from the equation itself (e.g.\ where some coefficient of the equation is itself singular). While movable singularities do not occur in the solutions of linear differential equations, when going to nonlinear equations, movable singularities will occur in general, and generically a solution will have infinitely many movable singularities. 

Going to non-autonomous equations, systems with the Painlev\'e property 
form the class of equations for which the solutions still contain movable poles at worst, apart from fixed singularities, so their solutions are meromorphic functions either on $\mathbb{C}$ or a covering surface thereof. Therefore, locally in a neighbourhood of the singularity $z_*$ the solution has the form of a Laurent series 
\begin{equation}
    x(z)= \sum_{k=k_1}^{\infty} a_k (z-z_*)^k \,, \qquad y(z)= \sum_{k=k_2}^{\infty} b_k (z-z_*)^k \,,  \qquad k_1,k_2 \in \mathbb{Z}, 
\end{equation}
where $\min\{k_1,k_2\} < 0$. One can then ask the question, what types of movable singularities can possibly occur in the solutions of a given differential equation or system of equations. In this work, we are studying a class of Hamiltonian systems endowed with a more general singularity structure: Hamiltonian system with the quasi-Painlev\'e property.

\subsection{Quasi-Painlev\'e equations}\label{sec:quasi_painleve_eqs}
Ordinary differential equations of quasi-Painlev\'e type form an emerging field of study, generalising the Painlev\'e property for classes of complex differential equations in a natural way. While the Painlev\'e equations have been studied extensively in the literature under various aspects, such as their complex analytic properties (see e.g.\ the book~\cite{GromakLaineShimomura+2002}), their role in integrable systems (see e.g.\ chapter 7 in~\cite{Ablowitz_Clarkson_1991}) and their relationship with isomonodromic deformations (see e.g.~\cite{fokas2006painleve}), to name just a few, quasi-Painlev\'e equations are still relatively new and many of their properties remain to be explored. Recent work in the area of quasi-Painlev\'e equations has been published by Filipuk and Stokes~\cite{FilipukStokes,AlexGalinacovering,FILIPUK2024}, as well as the authors of the present article~\cite{MDAKec}. While the Painlev\'e property demands of an equation that all movable singularities of all solutions are poles in the complex plane, the quasi-Painlev\'e property relaxes this by allowing also for movable algebraic singularities to occur in the solutions. When going to equations of quasi-Painlev\'e type, we thus allow for some branching to occur at the movable singularities of the solutions. However, many similarities with the Painlev\'e equations remain, which justify the importance of these equations to be studied in their own right. 

Nonlinear, non-autonomous equations of quasi-Painlev\'e type were introduced starting with two articles by Shimomura~\cite{Shimomura2006,Shimomura2008}, where the name originates, and extended by Filipuk and Halburd~\cite{halburd1,halburd2}, as well as one of the authors of the present article~\cite{Kecker2012,Kecker2016}. This property expresses the idea that any solution can be continued in the plane so that the only movable singularities that can arise (by analytic continuation along finite-length paths) are algebraic poles, i.e.\ they exhibit only finite branching locally. In particular, logarithmic terms $\log(z-z_\ast)$ are prohibited in the expansion of any solution. Globally, however, the branching is very complicated and solutions would in general extend over a Riemann surface with infinitely many sheets. 

Namely, in generalisation of the Painlev\'e property of an equation, Shimomura studies equations
\begin{equation}
\label{shimomura_eqns}
\sd{y} = y^{2k} + z \quad (k>1), \qquad \sd{y} = y^{2k+1} + z y + \alpha \quad (k>2, \quad \alpha \in \mathbb{C}),
\end{equation}
whose solutions have algebraic poles, where the solution is represented by a Puiseux series
\begin{equation}
    \label{puiseux_y}
    y(z) = \sum_{j=j_0}^\infty c_j (z-z_*)^{j/n}, 
\end{equation} 
with $j_0 = -2$, $n=3$ for the first equation in~\eqref{shimomura_eqns} and $j_0=-1$, $n=2$ for the second equation. More specifically, Shimomura shows that all movable singularities that are are obtained by analytic continuation along finite-length curves are algebraic poles of this form, referred to by him as the quasi-Painlev\'e property. The Painlev\'e property is the special case where all movable singularities are indeed ordinary poles, which would be the case for $k=1$ in the equations~\eqref{shimomura_eqns}.
While Shimomura's equations are first examples with the quasi-Painlev\'e property, they do not capture some of the more complex aspects of these equations, such as resonance conditions. Filipuk and Halburd~\cite{halburd1} study the wider class of second-order equations with the quasi-Painlev\'e property of the form
\begin{equation}
\label{polynomial_class}
\sd{y} = P(y;z) = a_N(z) y^N + \sum_{n=0}^{N-2} a_{n}(z) y^n\,,
\end{equation}
where the right-hand side is a polynomial of degree $N$ in $y$ with coefficient functions $a_{n}(z)$ analytic in $z$. For an equation of the form~\eqref{polynomial_class} to be of quasi-Painlev\'e type, certain differential conditions on the coefficient must be satisfied. In the even $N$ case, the only condition is $\sd{a_{N-2}} = 0$, whereas in the odd $N$ case one further relation among the functions $a_n(z)$ is required, corresponding to two different types of possible leading order behaviour at the movable singularities. 
The equations~\eqref{polynomial_class} also admit a Hamiltonian formulation, with $H = \frac{1}{2}(\fd{y})^2 - Q(y;z)$, where $\frac{\partial Q}{\partial y} = P(y;z)$.

One of the authors~\cite{Kecker2016} studies more general polynomial Hamiltonian systems~\eqref{hamiltonian_sys} with the quasi-Painlev\'e property,
\begin{equation}
\label{poly_hamiltonian}
H = H(x(z),y(z);z) = \sum_{(i,j) \in I} \alpha_{ij}(z) x^i y^j\,,
\end{equation}
with analytic coefficient functions $\alpha_{ij}(z)$ in some common domain and certain restricted index sets $I$. Also here, it turns out that the condition of not having logarithmic terms in the series expansions of the solutions at any movable singularity imposes certain conditions on the coefficient functions $\alpha_{ij}(z)$. To find such conditions, one can insert series expansions of the form 
\begin{equation}
\label{puiseux_system}
\begin{split}
    x(z) &= \sum_{j=j_1}^\infty a_j (z-z_*)^{j/n}, \qquad
     y(z) = \sum_{j=j_2}^\infty b_j (z-z_*)^{j/n},
    \end{split}
\end{equation}
into the equation, for integers $j_1$, $j_2$ and $n$ which describe the leading order of the solution at a singularity. There may be various different types of leading order behaviour for singularities, which correspond to different terms in the system of equations balancing. However, it is not always easy to obtain all possible balances, making this process cumbersome. For each leading order behaviour, one can then insert the full series expansions into the equations and find a recurrence relation for the coefficients $a_j$, $b_j$. 

When the recurrence breaks down (called a resonance), we find that a formal series solution exists if and only if the corresponding resonance conditions are satisfied. This method is essentially a generalisation of the Painlev\'e test to equations with solutions involving Puiseux series instead of Laurent series, and we denote it the quasi-Painlev\'e test. Note that for an equation to pass the (quasi-)Painlev\'e test is necessary, but in general not sufficient for an equation to have the (quasi-)Painlev\'e property. However, in~\cite{halburd1} and~\cite{Kecker2016} the authors in fact prove that for the equations in the class~\eqref{polynomial_class} and~\eqref{poly_hamiltonian} the existence of a number of Puiseux series expansions as formal solutions of~\eqref{puiseux_system}, as a consequence of the resonant conditions, is also sufficient for the equation to have the quasi-Painlev\'e property.

\subsection{The geometric approach}
Besides other important properties of the six Painlev\'e equations, Okamoto~\cite{Okamoto1979} introduced the notion of the \textit{space of initial conditions} for each of these equations. This is an extended phase space for these equations, obtained by a $8$-point blow-up of a compact rational surface, for which, after removing certain vertical leaves, the solutions provide a uniform foliation. The configuration of the exceptional curves obtained in this blow-up process give rise to an algebro-geometric theory of these equations. These concepts form the basis of seminal work by Sakai~\cite{Sakai2001}, where differential and discrete Painlev\'e equations are classified according to their surface type and symmetry type.

In recent work by Filipuk and one of the authors~\cite{KeckerFilipuk}, the notion of the space of initial conditions is extended to equations of quasi-Painlev\'e type. Here, it is shown that for certain classes of equations one obtains a regular initial value problem after a finite number of blow-ups, if in addition one changes the role of dependent variables in the system obtained after the last blow-up. However, in the quasi-Painlev\'e case, the space constructed in this way does no longer satisfy the definition of a space of initial conditions in the strict sense, as due to the branching in the solutions the fibration obtained here is not uniform. In the quasi-Painlev\'e case we therefore refer to the notion of {\it defining manifold} for the fibres of the space constructed in the blow-up process.

The Painlev\'e equivalence problem is the question to determine, for a given equation with the Painlev\'e property, to which type of Painlev\'e equation it belongs (i.e.\ to which of the equations it is equivalent under bi-rational transformations). The geometric approach to the Painlev\'e equations initiated by Okamoto and Sakai allows exactly this: by constructing the space of initial conditions for a given equation with the Painlev\'e property, one can compare it to the space of initial conditions for the six Painlev\'e equations in standard form. In particular, by finding a one-to-one correspondence between the irreducible components of the inaccessible divisor arising from the blow-up process, one can obtain an explicit change of coordinates to an suitable Painlev\'e equation. See also the review article~\cite{Kajiwara2017} and the more recent articles~\cite{Dzhamay2021} on this topic.

In~\cite{MDAKec} the authors of the present article show that the geometric approach can be extended, at least to certain extent, to classes of Hamiltonian systems for equations with the quasi-Painlev\'e property. This includes e.g.\ quasi-Painlev\'e equations analogous to $\text{P}_{\text{II}}$ and $\text{P}_{\text{IV}}$. Here, the process of computing the defining manifold for the system (analog to constructing the space of initial conditions in the Painlev\'e case) allows one firstly, to find, within a given class of Hamiltonian systems, the conditions on the coefficients under which these have the quasi-Painlev\'e property, and secondly, to identify the given system with other known systems or equations with the same singularity signature. This last step allows us, like in the Painlev\'e case, to find explicit bi-rational changes of variables to a known equation with the quasi-Painlev\'e property. While here we find explicit correspondences for all cases that occur, it is an open question whether such identification is always possible. 

\subsection{Outline of this article}
The aim of the present article is to give a systematic classification of Hamiltonian systems of quasi-Painlev\'e type, at least for a restricted class of Hamiltonians. By reducing each arising case to a system in a certain standard form we either identify it with a known (quasi-)Painlev\'e system by its singularity signature, or otherwise expose it as a new system. We perform the analysis in this article for non-autonomous, polynomial Hamiltonians in two dependent variables up to degree $4$. Starting from a general quartic, this is done by a process of successive degeneration of the Hamiltonian and subsequent coalescences of base points, as explained further below. 
These two concepts, degeneration and coalescence, occur naturally in our analysis and allow us to cascade down from the most general quartic Hamiltonian with the quasi-Painlev\'e property to a more and more specific one, which in some special cases leads us back to the original Painlev\'e equations. By doing so, we obtain a classification of systems with the quasi-Painlev\'e property, with various new kinds of singularity structures, while also recovering some known as well as finding some new Hamiltonians which lead to  Painlev\'e equations.  

In section~\ref{sec:Hamiltonian_systems},
we review the notion of Okamoto's space of initial conditions for Painlev\'e equations and the extension of this concept to quasi-Painlev\'e equations, the defining manifold. Here, we also place emphasis on the construction of a global symplectic atlas for the space of initial conditions of a given (quasi-)Painlev\'e equation. 

In section~\ref{sec:general_procedure} we outline the general process for our classification of quasi-Painlev\'e Hamiltonian systems. We define the two mechanisms driving the classification: the coalescence and the degeneration. We also introduce Newton polygons and how they can be used to visualise our classification scheme. For each subcase we will apply the regularisation method, by blowing up all base points, that leads to the defining manifold (space of initial conditions in the Painlev\'e cases).

In section~\ref{sec:cubic_systems}, we apply our classification procedure by starting with a review of non-autonomous cubic Hamiltonians with the quasi-Painlev\'e property. All the systems obtained here are in fact related to known Painlev\'e equations (in particular $\text{P}_{\text{I}}$, $\text{P}_{\text{II}}$ and $\text{P}_{\text{IV}}$), so no genuine quasi-Painlev\'e cases arise.

The subsequent section~\ref{sec:quartic_systems} is the main body of this article where we perform a systematic classification of quartic Hamiltonian systems in two dependent variables with the quasi-Painlev\'e property, $H\!=\!H(x,y;z)$, with coefficient functions analytic in $z$ in some common domain $U \in \mathbb{C}$. Starting from a general quartic polynomial, through a process of degeneration of the Hamiltonian and subsequent coalescence of base points, we arrive at various sub-cases of systems with different types of movable poles in their solutions. Under linear equivalence (using all freedom of scaling and shifting the dependent variables) we write down a standardised Hamiltonian in each case, also characterised by their Newton polygon. 

For each of the main cases, we perform the necessary blow-ups of the system to construct the defining manifold 
and find the conditions on the coefficient functions for the system to be of quasi-Painlev\'e type. 
From the intersection diagrams for the system we obtain, we find various relationships between these systems and formerly known Painlev\'e and quasi-Painlev\'e equations. While the general cases presented all have movable algebraic poles as singularities (quasi-Painlev\'e case), in some degenerate cases of the Hamiltonian we find systems that are related in particular to the modified Painlev\'e $\text{III}$ and $\text{V}$, as well as Painlev\'e $\text{I}$, $\text{II}$ and $\text{IV}$. In this way, the Painlev\'e equations are special cases of quasi-Painlev\'e equations where all singularities happen to be ordinary poles. Mixed cases are also possible, where a general solution has some movable singularities that are poles and others that are genuine algebraic poles.

\section{Geometric approach for quasi-Painlev\'e type systems}
\label{sec:Hamiltonian_systems}

In this paper, we will study non-autonomous Hamiltonian systems
\begin{equation} 
\label{hamiltonian_sys}
    \fd{y} = \frac{\partial H}{\partial x}\,, \qquad \fd{x} = -\frac{\partial H}{\partial y}\,, \qquad H = H(x(z),y(z);z)\,,
\end{equation}
where $H$ is a polynomial in $x$ and $y$ with $z$-dependent coefficient functions, as in~\eqref{poly_hamiltonian}.
As already discussed above, the nature of the specific quasi-Painlev\'e system is encoded in the local behaviour of the solutions in the neighbourhood of the movable singularities. Instead of using the quasi-Painlev\'e test described in section~\ref{sec:quasi_painleve_eqs}, 
here we will make use of the second, geometric approach mentioned above, namely by studying the equations via their Okamoto's spaces of initial conditions, or the analogous notion thereof in the case of quasi-Painlev\'e equations, the \textit{defining manifold}. 
Computations to obtain such spaces for certain quasi-Painlev\'e equations were first performed in~\cite{KeckerFilipuk}, e.g.\ for the equation
\begin{equation}
\label{quasi-PI}
    \sd{y} = y^4 + z\, y^2 + a_1(z)\, y + a_0(z)\,,
\end{equation}
a quasi-Painlev\'e analogue of $\text{P}_{\text{I}}$, for which $14$ blow-ups are needed to regularise the equation at points at infinity of the phase space for $(y,\fd{y})$, instead of just $9$ blow-ups in the case of the Painlev\'e equations. Furthermore, a change of dependent and independent variables is required after the last blow-up to obtain a regular system. The solutions of the regular system at points at infinity of the phase space then translate into algebraic poles for the solutions in the original variable, $y$, which in the case of equation~\eqref{quasi-PI} all have the leading order $y(z) \sim (z-z_*)^{-2/3}$ at any movable singularity $z_* \in \mathbb{C}$. 

For more general equations and Hamiltonian systems, several different leading order behaviours exist for algebraic poles in the solutions and one needs to look at the various balances in the equations to obtain a list of all possible leading orders, which can be cumbersome. While the geometric approach studies the equations from an algebro-geometric viewpoint, we note that our aim is somewhat more complex analytic in nature, and we use the geometric approach rather as a tool. Firstly, it allows us to easily find the conditions under which the equations are of quasi-Painlev\'e type, and secondly, it has the additional advantage that all possible leading order behaviours of the solutions at their singularities are easily exposed, as will be clarified in the following sections.

\subsection{Defining manifold}
\label{sec:Okamotos_space}
We briefly recall how to construct the Okamoto's space of initial conditions, but refer to the works~\cite{Okamoto1979,Takano97,Takano99} for a more detailed description. Originally, Okamoto \cite{Okamoto1979} found, for each of the six Painlev\'e equations, a rational surface that is uniformly foliated by the integral curves of the Hamiltonian flow of the system in an extended phase space. Starting from a compact (Hirzebruch) surface, Okamoto obtained the space of initial conditions by removing some inaccessible curves (or vertical leaves) from the surface through a process involving $8$-point blow-ups. Depending on the exact form of the Hamiltonian system and the initial compact surface, the number of blow-ups can deviate from this as the resulting surface after all blow-ups may not be in a minimal form, so that a number of blow-downs may be required. Indeed, the number of points at which the surface is blown-up is not an invariant for the system, whereas the minimal configuration of inaccessible curves to be removed is. In~\cite{Takano97,Takano99} it is shown that the spaces of initial conditions uniquely define the Hamiltonian system for each Painlev\'e equation. As we will see, in the quasi-Painlev\'e case, this uniqueness aspect is no longer true in general, as although the conditions obtained restrict the coefficient functions of the equations, there are generally still arbitrary functions remaining in the Hamiltonian.

From a complex analysis point of view, we are interested in determining the local behaviour of the solutions of the system around the movable singularities. To achieve this, the system, typically exhibiting points of indeterminacy on an augmented phase space, undergoes a process of regularisation, via blow-up transformations. As we already mentioned, the starting point is a polynomial Hamiltonian system with $H\big(x(z),y(z);z\big)$ in the form 
\begin{equation} \label{eq:vector_field}
    \begin{cases}
        \fd{x} = F\big(x(z),y(z);z\big) \\[1ex] 
        \fd{y} = G\big(x(z),y(z);z\big) 
    \end{cases} 
\end{equation}
where $(x,y) \in \mathbb{C}^2$, $z \in \mathbb{C}$ and $F,G$ are rational functions in their first two arguments. To take into account points at infinity, we consider a compactification of the complex plane $\mathbb{C}^2$. For this type of system the compactification can be done in different ways ($\mathbb{CP}^2$, $\mathbb{CP}^1 \times \mathbb{CP}^1$, $\Sigma^{(2)}$), but throughout this paper we will always use $\mathbb{CP}^2$. 

\subsection{Cascades of blow-ups}
We start from the extended phase space $\mathbb{CP}^2$, given by gluing together three affine complex planes identified by the coordinates $(x,y)$, $(u_0,v_0)$ and $(U_0,V_0)$ respectively. As inhomogeneous coordinates these charts are related via the coordinate transformations by:
\begin{equation}\label{eq:CP2}
\includegraphics[width=.2\textwidth,valign=c]{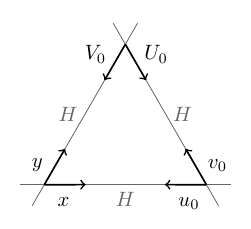} \qquad   
    \begin{aligned}
    & \mathbb{CP}^2 \simeq \mathbb{A}^2_{(x,y)} \cup \mathbb{A}^2_{(u_0,v_0)} \cup \mathbb{A}^2_{(U_0,V_0)}\,, \\[1ex] 
    &[\,1:x:y\,] = [\,u_0:1:v_0\,] = [\,V_0:U_0:1\,]\,, \\[.7ex]
    &u_0 = \frac{1}{x}\,, \qquad V_0 = \frac{1}{y}\,, \qquad v_0 = \frac{y}{x} = \frac{1}{U_0} \,.
\end{aligned}
\end{equation}

In the compactified phase space $\mathbb{CP}^2$, the Hamiltonian system rewritten in all respective charts ($(x,y)$, $(u_0,v_0)$ and $(U_0,V_0)$), may have points of indeterminacy, i.e.\ points at which the vector field (e.g.\ the right hand side of~\eqref{eq:vector_field} in $(x,y)$) has the ill-defined form $\frac{0}{0}$. On further inspection, some of these points turn out to be \emph{base points}, i.e.\ points at which infinitely many flow curves of the vector field coalesce. The system of equations is regularised through the process of blowing-up the phase space at these base points, giving rise to an exceptional curve for each blow-up. In a coordinate chart $(u_i,v_i)$, a blow-up at a point $p\colon (u_i,v_i)=(a,b)$ is performed by
\begin{equation}
  \text{Bl}_p(\mathbb{C}^2) = \{ (u_i,v_i) \times [w_0:w_1] \in \mathbb{C}^2 \times \mathbb{CP}^1: (u_i-a) w_0 = (v_i-b) w_1 \}\,,
\end{equation}
giving rise to two new coordinate charts, namely
$(u_j,v_j)$, $(U_j,V_j)$, according to 
\begin{equation}
\begin{cases}
    u_i = u_j+a = U_j V_j+a  \\[1ex] 
    v_i = u_j\, v_j+b = V_j+b
\end{cases}.
\end{equation}
Here, $U_j = {w_0}/{w_1}$ covers the part of $\mathbb{CP}^1$ modulo where $w_1 \neq 0$, while $v_j = {w_1}/{w_0}$ covers the part where $w_0 \neq 0$. We have the projection onto the first component,
\begin{equation}
    \pi_p : \text{Bl}_p(\mathbb{C}^2) \to \mathbb{C}^2, \quad (u_i,v_i) \times [w_0,w_1] \mapsto (u_i,v_i).
\end{equation}
Away from the point $p$, the blow-up is a one-to-one map,
\begin{equation}
    \text{Bl}_p(\mathbb{C}^2) \setminus \pi_p^{-1}(p) \quad \longleftrightarrow \quad \mathbb{C}^2 \setminus \{ p\}.
\end{equation}
The set $E = \pi_p^{-1}(p)$ is the exceptional curve introduced by the blow-up, which as a point set in coordinates is given by 
\begin{equation}
    E = \{ u_j = 0 \} \cup \{ V_j = 0 \}.
\end{equation}
By definition, exceptional curves have self-intersection number $-1$, while $H$, the hyperplane divisor in~\eqref{eq:CP2}, has self-intersection number $+1$. The intersection between $H$ and any exceptional curve, as well as the intersection between two different exceptional curves, is $0$. Furthermore, the linear extension of the intersection of these two types of divisors gives rise to the intersection form on the Picard lattice of the $n$-times blown up space thus constructed,
\begin{equation}
    \text{Pic}(X_n) = \mathbb{Z} H \oplus \mathbb{Z} E_1 \oplus \cdots \oplus \mathbb{Z} E_n\,,
\end{equation}
\begin{equation}
H \cdot H = 1, \qquad E_i \cdot H = 0 \quad\forall i, \qquad E_i \cdot E_j = -\delta_{ij}\,. \end{equation}
The appearance of the exceptional curve $E$ after a blow-up affects the self-intersection of the lines involved. For instance, blowing-up the point $p$ with coordinates $(u_i,v_i)=(0,0)$ at the intersection of the lines $L_1$ and $L_2$ -- in general, linear combinations of $E_i$ and $H$ -- with self-intersection $L_1 \cdot L_1$ and $L_2 \cdot L_2$ respectively:
\begin{equation*}
    \includegraphics[width=.6\textwidth]{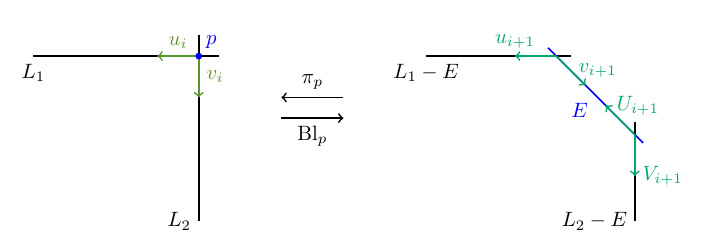}
\end{equation*}
After the blow-up, the self-intersection of the proper transform of the two lines is $(L_1-E)\cdot(L_1-E)$ and $(L_2-E)\cdot(L_2-E)$ respectively. After a blow-up, new base points may arise on the exceptional curve and usually, to completely remove the indeterminacy, a composition of finitely many successive changes of coordinates is required.
We will refer to the composition of blow-ups starting from a base point in $\mathbb{CP}^2$ as a \emph{cascade}. Sometimes, the blow-up of a base point can give rise to several new base points on the exceptional curve. In this case, a cascade of blow-ups can split into various branches. We keep track of all blow-ups by drawing the intersection diagrams, depicting the irreducible components of the exceptional divisor for all the cases analysed. Furthermore, for the principal cases we will provide the cascades of points at which we successively blow-up the face, explicitly giving the coordinates of the points. 
In the Painlev\'e case the resulting (minimal) configuration of the curves with self-intersection $-2$ determines the \emph{surface type}, represented by a Dynkin diagram. In the cases we analyse not falling in the Painlev\'e class, we will obtain a configuration of $-2$ curves and one $-3$ curve, giving rise to a sort of analogue of the Dynkin diagram. E.g.\ for the case that we will denote with Q1 we have the following representations:
\begin{equation*}
\includegraphics[width=.25\textwidth,valign=c]{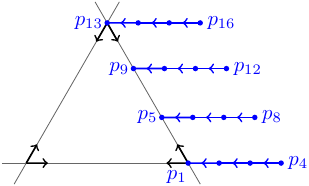} \qquad \includegraphics[width=.65\textwidth,valign=c]{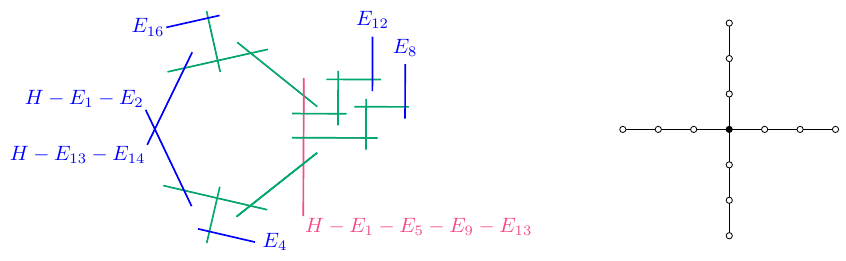}
\end{equation*}
where on the left we display the cascades of blow-ups originating at the base points $p_1$, $p_5$, $p_9$ and $p_{13}$ in $\mathbb{CP}^2$; in the middle we represent the corresponding intersection diagram, where the blue lines are curves with self-intersection $-1$, the green lines self-intersection $-2$ and the red line self-intersection $-3$; on the right the analogue of the Dynkin diagram is shown, with the black node for the only $-3$ curve, and the white nodes for the usual $-2$ curves.

\subsection{Intermediate changes of coordinates}\label{sec:intermediate_changes_coordinates}

In work by Shioda and Takano~\cite{Takano97} and Matano, Matuyima and Takano~\cite{Takano99} it is shown that to each of the Painlev\'e equations is associated a global symplectic atlas on the space of initial conditions, obtained by gluing the coordinate charts obtained after the final blow-up in each cascade, together with the original chart. This means that, in each chart $(u_i,v_i)$ of the space, the system derives from a polynomial Hamiltonian with respect to the canonical symplectic $2$-form
\begin{equation}
\label{ramified_2-form}
\omega = \rd y \wedge \rd x = k\, \rd u_i \wedge \rd v_i \quad \text{ or } \quad \omega = \rd y \wedge \rd x = k\, \rd U_i \wedge \rd V_i \,,
\end{equation}
where $k \in \mathbb{Z}$, and $|k|$ denotes the ramification number of the space of initial conditions (for $\text{P}_{\text{II}}$--$\text{P}_{\text{VI}}$ in fact $k=1$, only for $\text{P}_{\text{I}}$ we have $k=2$). Note, however, that blow-up transformations are in general not canonical.
So, in order to arrive at suitable coordinate charts, it may be necessary to make certain intermediate changes of variables in the blow-up process in order for the denominator of the $2$-form $\omega$ to remain a monomial. In particular, for certain equation in the quasi-Painlev\'e class, it is shown by Filipuk and Stokes~\cite{FilipukStokes} that one can achieve, using certain intermediate coordinate changes in the blow-up process, that the system remains polynomial Hamiltonian, with $2$-form more generally given by
\begin{equation}
\begin{aligned}
\omega = \rd y \wedge \rd x & = k\, u_i^m \,\rd u_i \wedge \rd v_i \quad 
\text{ or } \quad \omega = \rd y \wedge \rd x = k\, V_i^m\, \rd U_i \wedge \rd V_i\,.
\end{aligned}
\end{equation}
Here, $k \in \mathbb{Z}$ and the power $m \in \mathbb{N}$ is determined by the nature of the movable algebraic poles, in particular we have $m=n-1$ for algebraic poles around which the solution is given by a power series 
\begin{equation}
\label{frac_power_sol}
y(z) = \sum_{j=j_0}^\infty c_j (z-z_\ast)^{j/n}, \quad j_0 \in \mathbb{Z}\,,
\end{equation} convergent in a cut neighbourhood of $z_\ast$. Furthermore, we need to assume here that $n$ is the smallest possible integer of a solution~\eqref{frac_power_sol}, i.e.\ if $\gcd(j,n)>1$ for all $j$ with non-zero coefficients $c_j$, $n$ would have to be reduced. 
We need two possible types of intermediate changes of variables. The first one, also called twist in~\cite{FilipukStokes}, is needed when base points other than the one to be blown up are visible in a coordinate chart resulting in the coefficient of the $2$-form $\omega$ to become non-monomial. This is given in the respective charts by
\begin{equation}\label{eq:generic_twist}
    u = \widetilde{u} \,, \quad 
    v = \dfrac{1}{\widetilde{v}} 
\qquad \text{ or } \qquad 
    U = \dfrac{1}{\widetilde{U}}  \,, \quad 
    V = \widetilde{V} \,,
\end{equation}
effectively sending a point at $v=0$ (respectively $U=0$) on the exceptional curve to infinity.
The other type of intermediate coordinate change was introduced in~\cite{iwasaki} for Painlev\'e \text{I} and applied to some quasi-Painlev\'e type systems in~\cite{AlexGalinacovering} and is needed when the space of initial conditions becomes ramified. In this case we apply a $k$-fold covering transformation,
\begin{equation}\label{eq:generic_covering}
    u = \dfrac{\widehat{u}}{\widehat{v}} \,, \quad 
    v = \widehat{v}^k  
  \qquad \text{ or } \qquad 
    U = \widehat{U}^k \,, \quad 
    V = \dfrac{\widehat{V}}{\widehat{U}} \,. 
\end{equation}
Once all base points are completely resolved, with a polynomial Hamiltonian in each final coordinate chart and $\omega$ in the form~\eqref{ramified_2-form}, one can read off the behaviour of the a solution at a movable singularity in a straightforward way. For a solution of the form~\eqref{frac_power_sol}, to denote the leading-order behaviour as well as the fractional power of~$z-z_*$ in which it is expanded, we use the following notation in the remainder of this article,
\begin{equation}\label{eq:power_series_form}
    y(z) = c(z-z_*)^{p} + \mathcal{P}_h\big((z-z_*)^{1/n} \big), \qquad c=c_{j_0}, \qquad p = \frac{j_0}{n}\,,
\end{equation}
where $h \in \mathbb{C}$ denotes a free parameter in the series expansion: Usually, the coefficients $c_j$, $j>j_0$, are computable recursively, with one exception due to a resonance in the recursion, giving rise to a $1$-parameter family of series solutions. Here and in the following, we will write
\begin{equation}
    \mathcal{P}_h( \zeta ) = \sum_{j=j_0+1}^\infty c_j(h)\, \zeta^j
\end{equation}
for any $1$-parameter family of power series in $\zeta = (z-z_*)^{1/n}$ with coefficients $c_j$ that in general can depend on $h$, a free parameter. The parameter $h$, together with the position $z_*$ of a movable singularity, are determined by values~$y(z_0)$ and $\fd{y}(z_0)$ at some nearby point $z_0$, so prescribing the position $z_*$ of a movable singularity and the value of $h$ together are equivalent to giving initial conditions for $y$ and $\fd{y}$ at an ordinary point $z_0$.

We will see in the following that for the quartic Hamiltonian systems obtained in the classification in section~\ref{sec:quartic_systems}, we find coordinate charts after the final blow-up in each blow-up cascade where $n$ can take on different values, here in particular $n\!\in\!\{1,2,3,4,5\}$, so that in the quasi-Painlev\'e case we obtain solutions having a mix of different types of algebraic poles in general. The Painlev\'e cases are distinguished in that we have $n=1$ for all exponents. We will see below that some of the Painlev\'e equations, or variants thereof, emerge as special cases in this classification.

\section{General outline of the procedure}
\label{sec:general_procedure}
We start from a general non-autonomous quartic Hamiltonian, where we assume the coefficients of the quartic terms to be constant, to avoid fixed singularities. The generic quartic Hamiltonian initially has four base points on the line at infinity for the affine chart $(x,y)$, i.e.\ in $\mathbb{CP}^2$ on the line  represented as a point set by $\{u_0 = 0 \} \cup \{ V_0 = 0 \}$. We generate sub-cases by two different mechanisms, namely coalescence of two base points on the one hand, and a degeneration in the Hamiltonian on the other. We describe these processes by making use of two different tools: via the Newton polygons associated with the form of the polynomial Hamiltonian in the original system, and via the power series in which the solutions are expressed at the level of the regularised system. 
\subsection{Coalescence of base points and degeneration }
By changing a certain parameter in the system of equations, two previously separated base points can coalesce into a single point, with the effect that the two blow-up cascades merge into one cascade. This is a limiting process with the result that the behaviour of the solution around the movable singularity $z_*$, obtained on the exceptional curves after the final blow-up in these cascades, changes in the merged cascade. 

Such a merge can happen, for example, if two base points emerge on an exceptional curve $E_i = \{u_i=0\}$ after a blow-up, with coordinates $(u_i,v_i) = (0, a \pm b)$, say, where $a$ and $b$ are certain parameters (or functions) in the Hamiltonian (which in principle can be $z$ dependent). Letting $b \to 0$, these two base points merge and their cascades of blow-ups coalesce into a single cascade with a number of blow-ups less or equal to the sum of blow-ups in the original cascades, e.g.\ 
\begin{equation}\label{eq:coalescence}
\vcenter{\hbox{\begin{tikzpicture}
    \path (0,0) node (a1) {\includegraphics[width=.24\textwidth]{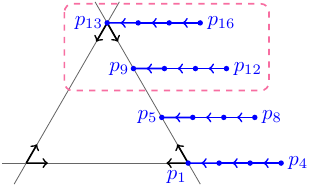}} -- ++(6,0) node (a2) {\includegraphics[width=.24\textwidth]{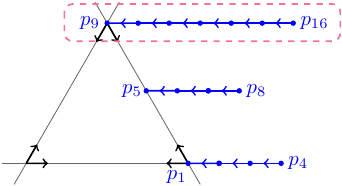}};
    \draw[thick,->,cyan!80!gray] (a1) -- (a2);
\end{tikzpicture}}}
\end{equation}
The types of branched movable singularities obtained by analysing the system on the final exceptional curve of this cascade may also be different to the singularities before the coalescence. In~\eqref{eq:coalescence}, on the left the solution in the two highlighted cascades is expressed as the power series in $(z-z_*)^{1/2}$, while the solution in the merged case on the right is represented by a power series in $(z-z_*)^{1/3}$. With the nomenclature explained below the coalescence showed in~\eqref{eq:coalescence} is from the case $\text{Q1}=(\tfrac{1}{2},\tfrac{1}{2},\tfrac{1}{2},\tfrac{1}{2})$ to the case $\text{Q2}=(\tfrac{1}{2},\tfrac{1}{2},\tfrac{1}{3})$. 

We observe that, while a coalescence can change the shape of this polygon, it does not change the (generic) genus of the algebraic curve the polynomial defines. In general we observe that, the cascade resulting from the coalescence of two cascades which have solutions in power series in $(z-z_*)^{1/n}$ and $(z-z_*)^{1/m}$, respectively, yields solutions whose behaviour is described as a power series in $(z-z_*)^{1/\ell}$ where: 
\begin{equation} 
\includegraphics[width=.33\textwidth,valign=c]{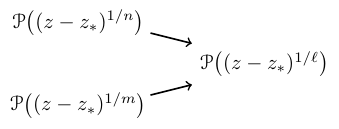}\,, \hspace{10ex}
    \ell = n + m - 1\,. 
\end{equation}

A degeneration, instead, is obtained if the leading order behaviour of a system in coordinates $(u_i,v_i)$ (or $(U_i,V_i)$) of a blow-up changes when a certain parameter or coefficient function in the Hamiltonian is set to $0$. Usually, an additional cancellation then happens in the rational expression of the corresponding system, and the next-to-leading order terms exhibit one or more different base points. For example, suppose that the system $(u_i,v_i)$ is of the form
\begin{equation}
    \fd{u_i} = \frac{a v_i + u_i (b - v_i^2) + \mathcal{O}(u_i^2)}{u_i^3}\,, \qquad \fd{v_i} = \frac{c + v_i + \rd u_i v_i + \mathcal{O}(u_i^2)}{u_i^2}\,,
\end{equation}
where $a,b,c,d$ are certain parameters or functions in the Hamiltonian. The system has a single base point at $(u_i,v_i)=(0,0)$. Letting $a \to 0$, we can cancel a factor of $u_i$ in the equation for $\fd{u_i}$ and the new system now exhibits two base points at $(u_i,v_i) = (0,\sqrt{b})$ and $(u_i,v_i) = (0,-\sqrt{b})$. This happens whenever the parameter in question (here $a$) is multiplying a term of leading order in the vector field that determines the position of a base point in that coordinate chart. In this case, the cascade of blow-ups from the original base point splits into shorter cascades for the two new base points, e.g.  
\begin{equation}\label{eq:degeneration}
\vcenter{\hbox{\begin{tikzpicture}
    \path (0,0) node (a1) {\includegraphics[width=.24\textwidth]{Q2_casc.pdf}} -- ++(6,0) node (a2) {\includegraphics[width=.24\textwidth]{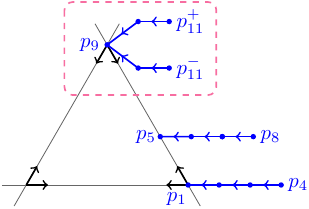}};
    \draw[thick,->,orange!80!gray] (a1) -- (a2);
\end{tikzpicture}}}
\end{equation}
Also here, the branching of movable singularities after the degeneration is different to that of the original system. In~\eqref{eq:degeneration}, a cascade of blow-ups from which a solution is obtained with algebraic poles of the type $(z-z_*)^{-1/3}$, splits into two cascades, each of which gives rise to ordinary poles $(z-z_*)^{-1}$ in the solution. With the nomenclature below, this transformation goes from the case $\text{Q2}=(\tfrac{1}{2},\tfrac{1}{2},\tfrac{1}{3})$ to the case $\text{Q2.a1}=(\tfrac{1}{2},\tfrac{1}{2},1,1)$.

In general, we observe that, if two cascades originate from the degeneration of the cascade corresponding to power series solutions in $(z-z_*)^{1/\ell}$, this gives rise to solutions in power series of $(z-z_*)^{1/n}$ and $(z-z_*)^{1/m}$ respectively, where: 
\begin{equation}\label{eq:power_deg}  
\includegraphics[width=.33\textwidth,valign=c]{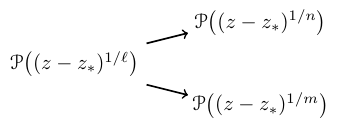}\,, \hspace{10ex} \ell=n+m+1\,. 
\end{equation} 
While this seems to be a somewhat opposite effect as seen in the coalescence of base points above, it is important to note that the two processes are not inverses of one another.

Using the mechanisms of coalescence and degeneration we obtain a whole map of quasi-Painlev\'e Hamiltonian systems with different singularity structures:
\begin{equation*}
    \begin{tikzpicture}[scale=0.47]
\path (6.4,6) node (Q5) {\hyperref[sec:1_quadrupole]{Q5}} -- ++(10,0) node (Q3bi) {\hyperref[sec:Q3b]{Q3.b1}} -- ++(4,0) node (Q3bii) {\hyperref[sec:Q3b]{Q3.b2}}-- ++(4,0) node (Q3biii) {\hyperref[sec:Q3b]{Q3.b3}}; 
\path (Q5) -- ++(-4,0) node (Q5a) {\hyperref[sec:Q5a]{Q5.a1}};
\path (Q5a) -- ++(3,-3) node (Q5bi) {\hyperref[sec:Q5b]{Q5.b2}} -- ++(4,0) node (Q5bii) {\hyperref[sec:Q5b]{Q5.b3}};
\path (Q5) -- ++(3,3) node (Q3) {\hyperref[sec:2_double_points]{Q3}} -- ++(4,0) node (Q3ai) {\hyperref[sec:Q3a]{Q3.a1}} -- ++(4,0) node (Q3aii) {\hyperref[sec:Q3a]{Q3.a2}};
\path (Q5) -- ++(-3,3) node (Q4) {\hyperref[sec:1_triple_1_simple_point]{Q4}} -- ++(-4,0) node (Q4ai) {\hyperref[sec:case_Q4_subcase_a]{Q4.a1}} -- ++(-4,0) node (Q4aii) {\hyperref[sec:case_Q4_subcase_a]{Q4.a2}};
\path (Q5) -- ++(0,6) node (Q2) {\hyperref[sec:1_double_2_simple]{Q2}} -- ++(4,0) node (Q2ai) {\hyperref[sec:Q2a]{Q2.a1}} -- ++(4,0) node (Q2aii) {\hyperref[sec:Q2a]{Q2.a2}};
\path (Q2) -- ++(-4,1) node (Q1) {\hyperref[sec:4_simple_points]{Q1}};
\node[below=-.1 of Q1] (a) {\footnotesize $\left(\frac{1}{2},\frac{1}{2},\frac{1}{2},\frac{1}{2}\right)$};
\node[below=-.1 of Q5] (a) {\footnotesize $\left(\frac{1}{5}\right)$};
\node[below=-.1 of Q2] (a) {\footnotesize $\left(\frac{1}{2},\frac{1}{2},\frac{1}{3}\right)$};
\node[below=-.1 of Q2ai] (a) {\footnotesize $\left(\frac{1}{2},\frac{1}{2},1,1\right)$};
\node[below=-.1 of Q2aii] (a) {\footnotesize $\left(\frac{1}{2},\frac{1}{2},1\right)$};
\node[below=-.1 of Q4] (a) {\footnotesize $\left(\frac{1}{2},\frac{1}{4}\right)$};
\node[below=-.1 of Q4ai] (a) {\footnotesize $\left(\frac{1}{2},\frac{1}{2},1\right)$};
\node[below=-.1 of Q4aii] (a) {\footnotesize $\left(\frac{1}{2},\frac{1}{2}\right)$};
\node[below=-.1 of Q3] (a) {\footnotesize $\left(\frac{1}{3},\frac{1}{3}\right)$};
\node[below=-.1 of Q3ai] (a) {\footnotesize $\left(\frac{1}{3},1,1\right)$};
\node[below=-.1 of Q3aii] (a) {\footnotesize $\left(\frac{1}{3},1\right)$};
\node[below=-.1 of Q3bi] (a) {\footnotesize $\left(1,1,1,1\right)$};
\node[below=-.1 of Q3bii] (a) {\footnotesize $\left(1,1,1\right)$};
\node[below=-.1 of Q3biii] (a) {\footnotesize $\left(1,1\right)$};
\node[below=-.1 of Q5a] (a) {\footnotesize $\left(\frac{1}{3},1\right)$};
\node[below=-.1 of Q5bi] (a) {\footnotesize $\left(1,1\right)$};\node[below=-.1 of Q5bii] (a) {\footnotesize $\left(1\right)$};
\path (Q1) -- ++(0.8,0) coordinate (q1ar); 
\path (Q2) -- ++(-.25,-1.5) coordinate (q2arl); 
\path (Q2) -- ++(.25,-1.5) coordinate (q2arr);
\path (Q3) -- ++(-0.3,-1.5) coordinate (q3ar); 
\path (Q4) -- ++(.3,-1.5) coordinate (q4ar); 
\path (Q5) -- ++(-.8,0) coordinate (q5ar); 
\path (Q5a) -- ++(0,-1.5) coordinate (q5aB); 
\path (Q5a) -- ++(0,-3) coordinate (q5aBB); 
\path (Q3ai) -- ++(0,-1.5) coordinate (q3aiar); 
\path (Q4ai) -- ++(0.15,.8) coordinate (q4aiP); 
\path (Q4aii) -- ++(0.15,.8) coordinate (q4aiiP); 
\path (Q5bi) -- ++(0,.8) coordinate (q5biP); 
\path (Q5bii) -- ++(0,.8) coordinate (q5biiP); 
\path (Q3bi) -- ++(.6,.8) coordinate (q3biP); 
\path (Q3bii) -- ++(0,.8) coordinate (q3biiP); 
\path (Q3biii) -- ++(0,.8) coordinate (q3biiiP); 
\draw[->,thick,cyan!80!gray] (q1ar) -- (Q2);
\draw[->,thick,cyan!80!gray] (q2arr) -- (Q3);
\draw[->,thick,cyan!80!gray] (q2arl) -- (Q4);
\draw[->,thick,cyan!80!gray] (q3ar) -- (Q5);
\draw[->,thick,cyan!80!gray] (q4ar) -- (Q5);
\draw[->,thick,cyan!80!gray] (Q2ai) -- (Q2aii);
\draw[->,thick,cyan!80!gray] (Q3ai) -- (Q3aii);
\draw[->,thick,cyan!80!gray] (Q4ai) -- (Q4aii);
\draw[->,thick,cyan!80!gray] (Q3bi) -- (Q3bii);
\draw[->,thick,cyan!80!gray] (Q3bii) -- (Q3biii);
\draw[->,thick,cyan!80!gray] (Q5bi) -- (Q5bii);
\draw[->,thick,orange!80!gray] (Q2) -- (Q2ai);
\draw[->,thick,orange!60!gray] (Q3) -- (Q3ai);
\draw[->,thick,orange!60!gray] (Q4) -- (Q4ai);
\draw[->,thick,orange!60!gray] (q5aB) -- (q5aBB);
\draw[->,thick,cyan!80!gray] (q5aBB) -- (Q5bi);
\draw[->,thick,orange!60!gray] (q3aiar) -- (Q3bi);
\draw[->,thick,orange!60!gray] (q5ar) -- (Q5a);
\node[blue] (a) at (q5biP) {\footnotesize $\text{P}_{\text{II}}$};
\node[blue] (a) at (q5biiP) {\footnotesize $\text{P}_{\text{I}}$};
\node[blue] (a) at (q4aiP) {\footnotesize $\text{quasi-P}_{\text{IV}}$};
\node[blue] (a) at (q4aiiP) {\footnotesize $\text{quasi-P}_{\text{II}}$};
\node[blue] (a) at (q3biP) {\footnotesize $\text{P}_{\text{V}}$};
\node[blue] (a) at (q3biiP) {\footnotesize $\text{P}_{\text{III}}$};
\node[blue] (a) at (q3biiiP) {\footnotesize $\text{P}_{\text{III}}$};
\path (Q3bii) -- ++(0,7) coordinate (leg) -- ++(2.8,0.05) coordinate (coal);
\path (leg) -- ++(0,-.75) coordinate (leg1) -- ++(2.94,0.05) coordinate (deg);
\draw[->,thick,cyan!80!gray] (leg) -- ++(1,0); 
\draw[->,thick,orange!60!gray] (leg1) -- ++(1,0); 
\node[black] (a) at (coal) {\footnotesize coalescence};
\node[black] (a) at (deg) {\footnotesize degeneration};
\end{tikzpicture}
\end{equation*}
Each case is labelled by $\text{Q}$ referring to the form of the Hamiltonian (quartic), an increasing numeric index is added to distinguish between cases obtained by a coalescence, and an alphabetic index is introduced to identify cases resulting from a degeneration. Furthermore, each case is characterised by an array of rational exponents, following a nomenclature that is a simpler version of the one proposed in~\cite{FILIPUK2024}. In our case, the length of the array represents the number of different possible behaviours around the movable singularities in the regularised system without taking into account the possible ramifications, and each element in the array corresponds to the power $k$ of the elementary term in the power series $\mathcal{P}\big((z-z_*)^k\big)$.

Since we revise as well the systems associated with cubic Hamiltonians, here we depict the corresponding diagram in terms of behaviour of power series: 
\vspace*{-1ex}
\begin{equation*}
    \begin{tikzpicture}
        \path (0,0) node (C1) {\hyperref[sec:C1]{C1}} -- ++(2,0) node (C2) {\hyperref[sec:C2]{C2}}-- ++(2,0) node (C3) {\hyperref[sec:C3]{C3}}; 
        \path (C1) -- ++(0,.4) coordinate (C1P); 
\path (C2) -- ++(0,.4) coordinate (C2P); 
\path (C3) -- ++(0,.4) coordinate (C3P); 
\node[blue] (a) at (C1P) {\footnotesize $\text{P}_{\text{IV}}$};
\node[blue] (a) at (C2P) {\footnotesize $\text{P}_{\text{II}}$};
\node[blue] (a) at (C3P) {\footnotesize $\text{P}_{\text{I}}$};
\node[below=-.1 of C1] (a) {\footnotesize $\left(1,1,1\right)$};
\node[below=-.1 of C2] (a) {\footnotesize $\left(1,1\right)$};
\node[below=-.1 of C3] (a) {\footnotesize $\left(1\right)$};
\draw[->,thick,cyan!80!gray] (C1) -- (C2);
\draw[->,thick,cyan!80!gray] (C2) -- (C3);
    \end{tikzpicture}
\end{equation*}

The description provided by the power series is not sufficient to justify the selection of the admissible cases and sub-cases generated by the above mentioned mechanisms. This is why we introduce an additional representation, provided by the Newton polygons.

\subsection{Newton polygons}
The Hamiltonian $H(x,y;z)$ defines a ($z$-dependent) algebraic curve in $\mathbb{CP}^2$. The generic genus of this curve can be determined from the Newton polygon of the polynomial in $x$ and $y$ given by the Hamiltonian. Namely, if we write
\begin{equation}
    H(x,y;z) = \sum_{j,k} \alpha_{jk}(z) \,x^j y^k,
\end{equation}
the Newton polygon is the convex hull of the set of points $(j,k)$ for which $\alpha_{jk} \neq 0$. The genus is then given by the number of lattice points in the interior of the polygon.  Here it is an example of the Newton polygon associated with the polynomial Hamiltonian in $x$ and $y$ for a genus $3$ curve: 
\begin{equation*}
H(x,y)=x^3y+\alpha_{30}\,x^3+\alpha_{20}\,x^2+\alpha_{11}\,xy+\alpha_{02}\,y^2+\alpha_{10}\,x+\alpha_{01}\,y \qquad 
     \includegraphics[width=.14\textwidth,valign=c]{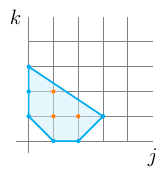} 
\end{equation*}
\vspace*{-3ex}

\noindent
The blue dots at coordinates $(j,k)$ correspond to the non-zero element $x^jy^k$ in the polynomial, while the orange dots are points of the lattice in the interior of the polygon.
In particular, we are interested in studying systems described by quartic Hamiltonians. Therefore, there are several constraints on the admissible Newton polygons for the cases here studied. For instance, we represent three admissible cases of Newton polygon, alongside with the geometrical constraints to which they are subject:
\vspace*{-1ex}
\begin{equation*}
    \includegraphics[width=.59\textwidth]{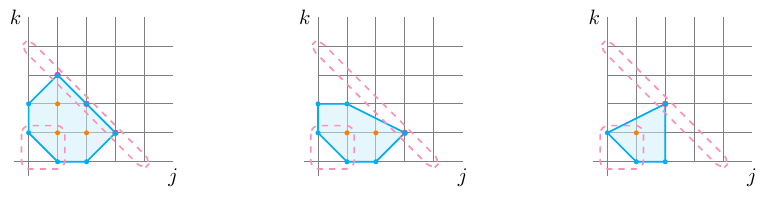}
\end{equation*}
\vspace*{-4ex}

\noindent
for the cases $\text{Q}1$ (eq.~\eqref{quartic1}) (left), $\text{Q}4.\text{a}1$~(eq.~\eqref{Ham_Q4a}) (middle), $\text{Q}3.\text{b}2$~(eq.~\eqref{22Ham_subcase3b_3}) (right). These constraints are shown in the the figure via highlighted curves surrounding some points of the lattice $(j,k)$, $j$ and $k$ representing the element $x^jy^k$ in the expression of $H(x,y)$. The curve around the fourth diagonal fixes the highest admissible power in the polynomial, quartic in the case shown. The region within the curve drawn around the bottom left part of the lattice will be fixed independently of the specific case under study. These constraints limit the number and the shapes of the polygons associated with the analysed cases. The exploitation of such constraints allows us to exclude from our study the potential case $\text{Q5.b1}$ (see the scheme on page~\pageref{eq:power_deg}), that we have not observed in our analysis and that is in principle admissible with the description provided by the power series.

The two mechanisms previously described to navigate the possible cases are the coalescence and degeneration of branches. The framework offered by Newton polygons is particularly effective in this context to describe both transformations, since the two possible transformations are represented as their action on the shape of the Newton polygons: 
\begin{equation*}
    \includegraphics[width=.58\textwidth]{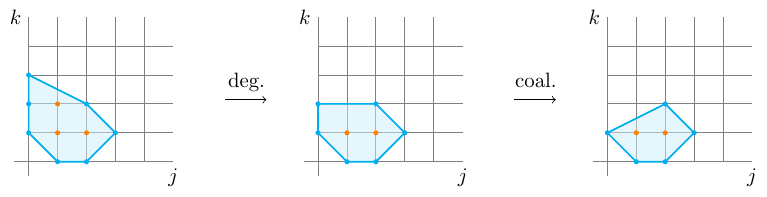}
\end{equation*}
\vspace*{-4ex}

\noindent
The degeneration (left to middle) transforms the polygon such that the genus of the curve associated with the polynomial decreases (from $3$ to $2$ in the figure). The coalescence (middle to right) is a transformation of the polygon leaving the genus of the curve invariant. 
The scheme given on page~\pageref{eq:power_deg} is then reproduced with the Newton polygons: 
\begin{equation*}
    \begin{tikzpicture}[scale=0.5]
\path (6.4,6) node (Q5) {\hyperref[sec:1_quadrupole]{Q5}} -- ++(10,0) node (Q3bi) {\hyperref[sec:Q3b]{Q3.b1}} -- ++(4,0) node (Q3bii) {\hyperref[sec:Q3b]{Q3.b2}}-- ++(4,0) node (Q3biii) {\hyperref[sec:Q3b]{Q3.b3}}; 
\path (Q5) -- ++(-4,0) node (Q5a) {\hyperref[sec:Q5a]{Q5.a1}};
\path (Q5a) -- ++(3,-5) node (Q5bi) {\hyperref[sec:Q5b]{Q5.b2}} -- ++(4,0) node (Q5bii) {\hyperref[sec:Q5b]{Q5.b3}};
\path (Q5) -- ++(3,5) node (Q3) {\hyperref[sec:2_double_points]{Q3}} -- ++(4,0) node (Q3ai) {\hyperref[sec:Q3a]{Q3.a1}} -- ++(4,0) node (Q3aii) {\hyperref[sec:Q3a]{Q3.a2}};
\path (Q5) -- ++(-3,5) node (Q4) {\hyperref[sec:1_triple_1_simple_point]{Q4}} -- ++(-4,0) node (Q4ai) {\hyperref[sec:case_Q4_subcase_a]{Q4.a1}} -- ++(-4,0) node (Q4aii) {\hyperref[sec:case_Q4_subcase_a]{Q4.a2}};
\path (Q5) -- ++(0,10) node (Q2) {\hyperref[sec:1_double_2_simple]{Q2}} -- ++(4,0) node (Q2ai) {Q2.a1} -- ++(4,0) node (Q2aii) {Q2.a2};
\path (Q2) -- ++(-4,1) node (Q1) {\hyperref[sec:4_simple_points]{Q1}};
\node[below=-.1 of Q1] (a) {\includegraphics[scale=.4]{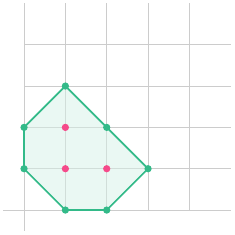}};
\node[below=-.1 of Q5] (a) {\includegraphics[scale=.4]{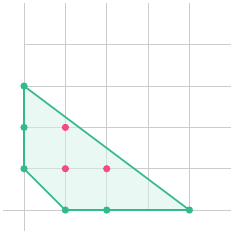}};
\node[below=-.1 of Q2] (a) {\includegraphics[scale=.4]{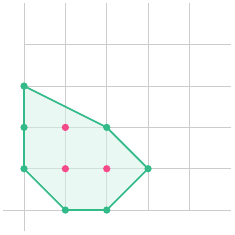}};
\node[below=-.1 of Q2ai] (a) {\includegraphics[scale=.4]{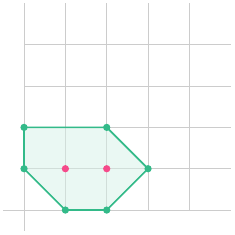}};
\node[below=-.1 of Q2aii] (a) {\includegraphics[scale=.4]{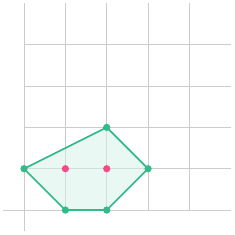}};
\node[below=-.1 of Q4] (a) {\includegraphics[scale=.4]{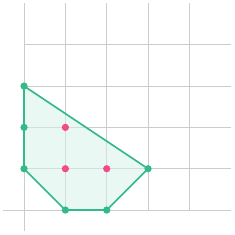}};
\node[below=-.1 of Q4ai] (a) {\includegraphics[scale=.4]{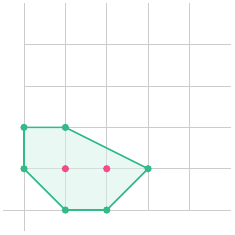}};
\node[below=-.1 of Q4aii] (a) {\includegraphics[scale=.4]{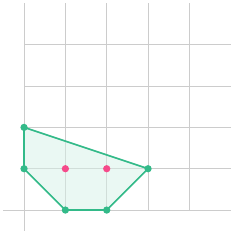}};
\node[below=-.1 of Q3] (a) {\includegraphics[scale=.4]{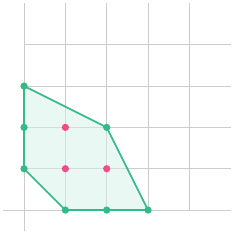}};
\node[below=-.1 of Q3ai] (a) {\includegraphics[scale=.4]{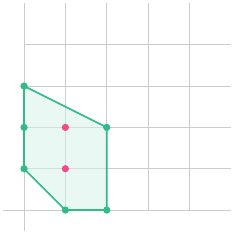}};
\node[below=-.1 of Q3aii] (a) {\includegraphics[scale=.4]{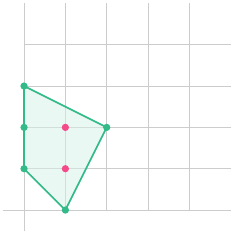}};
\node[below=-.1 of Q3bi] (a) {\includegraphics[scale=.4]{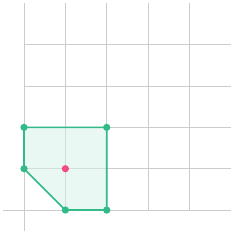}};
\node[below=-.1 of Q3bii] (a) {\includegraphics[scale=.4]{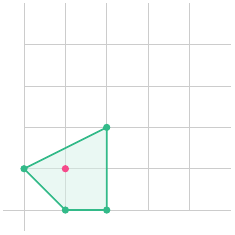}};
\node[below=-.1 of Q3biii] (a) {\includegraphics[scale=.4]{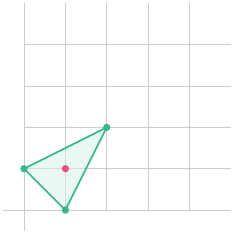}};
\node[below=-.1 of Q5a] (a) {\includegraphics[scale=.4]{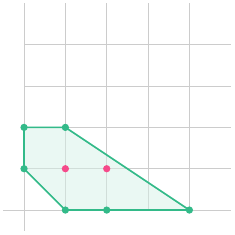}};
\node[below=-.1 of Q5bi] (a) {\includegraphics[scale=.4]{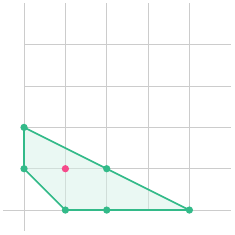}};
\node[below=-.1 of Q5bii] (a) {\includegraphics[scale=.4]{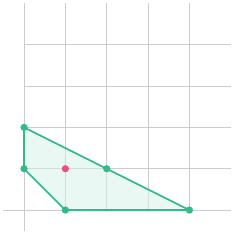}};
\path (Q1) -- ++(0.8,0) coordinate (q1ar); 
\path (Q2) -- ++(-.25,-3.9) coordinate (q2arl); 
\path (Q2) -- ++(.25,-3.9) coordinate (q2arr);
\path (Q3) -- ++(-0.3,-3.9) coordinate (q3ar); 
\path (Q4) -- ++(.3,-3.9) coordinate (q4ar); 
\path (Q5) -- ++(-.8,0) coordinate (q5ar); 
\path (Q5a) -- ++(0,-3.9) coordinate (q5aB); 
\path (Q5a) -- ++(0,-5) coordinate (q5aBB); 
\path (Q3ai) -- ++(0,-3.9) coordinate (q3aiar); 
\path (Q4ai) -- ++(0.15,.8) coordinate (q4aiP); 
\path (Q4aii) -- ++(0.15,.8) coordinate (q4aiiP); 
\path (Q5bi) -- ++(0,.8) coordinate (q5biP); 
\path (Q5bii) -- ++(0,.8) coordinate (q5biiP); 
\path (Q3bi) -- ++(.6,.8) coordinate (q3biP); 
\path (Q3bii) -- ++(0,.8) coordinate (q3biiP); 
\path (Q3biii) -- ++(0,.8) coordinate (q3biiiP); 
\draw[->,thick,cyan!80!gray] (q1ar) -- (Q2);
\draw[->,thick,cyan!80!gray] (q2arr) -- (Q3);
\draw[->,thick,cyan!80!gray] (q2arl) -- (Q4);
\draw[->,thick,cyan!80!gray] (q3ar) -- (Q5);
\draw[->,thick,cyan!80!gray] (q4ar) -- (Q5);
\draw[->,thick,cyan!80!gray] (Q2ai) -- (Q2aii);
\draw[->,thick,cyan!80!gray] (Q3ai) -- (Q3aii);
\draw[->,thick,cyan!80!gray] (Q4ai) -- (Q4aii);
\draw[->,thick,cyan!80!gray] (Q3bi) -- (Q3bii);
\draw[->,thick,cyan!80!gray] (Q3bii) -- (Q3biii);
\draw[->,thick,cyan!80!gray] (Q5bi) -- (Q5bii);
\draw[->,thick,orange!80!gray] (Q2) -- (Q2ai);
\draw[->,thick,orange!60!gray] (Q3) -- (Q3ai);
\draw[->,thick,orange!60!gray] (Q4) -- (Q4ai);
\draw[->,thick,orange!60!gray] (q5aB) -- (q5aBB);
\draw[->,thick,cyan!80!gray] (q5aBB) -- (Q5bi);
\draw[->,thick,orange!60!gray] (q3aiar) -- (Q3bi);
\draw[->,thick,orange!60!gray] (q5ar) -- (Q5a);
\node[blue] (a) at (q5biP) {\footnotesize $\text{P}_{\text{II}}$};
\node[blue] (a) at (q5biiP) {\footnotesize $\text{P}_{\text{I}}$};
\node[blue] (a) at (q4aiP) {\footnotesize $\text{quasi-P}_{\text{IV}}$};
\node[blue] (a) at (q4aiiP) {\footnotesize $\text{quasi-P}_{\text{II}}$};
\node[blue] (a) at (q3biP) {\footnotesize $\text{P}_{\text{V}}$};
\node[blue] (a) at (q3biiP) {\footnotesize $\text{P}_{\text{III}}$};
\node[blue] (a) at (q3biiiP) {\footnotesize $\text{P}_{\text{III}}$};
\path (Q3bii) -- ++(0,11) coordinate (leg) -- ++(2.8,0.05) coordinate (coal);
\path (leg) -- ++(0,-.75) coordinate (leg1) -- ++(2.94,0.05) coordinate (deg);
\draw[->,thick,cyan!80!gray] (leg) -- ++(1,0); 
\draw[->,thick,orange!60!gray] (leg1) -- ++(1,0); 
\node[black] (a) at (coal) {\footnotesize coalescence};
\node[black] (a) at (deg) {\footnotesize degeneration};
\end{tikzpicture}
\end{equation*}
For the cubic case we have the following representation with Newton polygons: 
\begin{equation*}
    \begin{tikzpicture}
        \path (0,0) node (C1) {\hyperref[sec:C1]{C1}} -- ++(2,0) node (C2) {\hyperref[sec:C2]{C2}}-- ++(2,0) node (C3) {\hyperref[sec:C3]{C3}}; 
        \path (C1) -- ++(0,.4) coordinate (C1P); 
\path (C2) -- ++(0,.4) coordinate (C2P); 
\path (C3) -- ++(0,.4) coordinate (C3P); 
\node[blue] (a) at (C1P) {\footnotesize $\text{P}_{\text{IV}}$};
\node[blue] (a) at (C2P) {\footnotesize $\text{P}_{\text{II}}$};
\node[blue] (a) at (C3P) {\footnotesize $\text{P}_{\text{I}}$};
\node[below=-.1 of C1] (a) {\includegraphics[scale=.4]{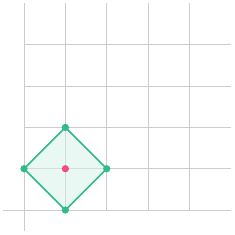}};
\node[below=-.1 of C2] (a) {\includegraphics[scale=.4]{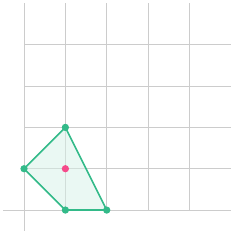}};
\node[below=-.1 of C3] (a) {\includegraphics[scale=.4]{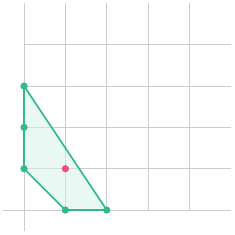}};
\draw[->,thick,cyan!80!gray] (C1) -- (C2);
\draw[->,thick,cyan!80!gray] (C2) -- (C3);
    \end{tikzpicture}
\end{equation*}

\subsection{Regularisation through cascades of blow-ups }\label{sec:outline_general_scheme}

In the analysis below, starting from general cubic and quartic Hamiltonian, we consider each possible coalescence of base points and degeneration of branches as sub-cases. For each sub-case, we determine the intersection diagram and the conditions on the coefficient functions to be satisfied in order for the system to be of quasi-Painlev\'e type. Note that we are not actually proving here that the systems have the quasi-Painlev\'e property. This was done e.g.\ for some systems of quasi-Painlev\'e type in the authors' recent article~\cite{MDAKec}, where certain auxiliary functions are constructed to prove that the intermediate exceptional divisors from the blow-up process are inaccessible for the flow of the vector field. While we note that this can be done in principle also for all the cases considered in this article, the focus here is to get a comprehensive overview of the movable singularities for the individual systems. 

This proceeds along the following lines:
\begin{enumerate}
    \item Blow-up all base points to resolve the indeterminacies of the vector field in each affine chart of $\mathbb{CP}^2$ (e.g.\ the right hand side of~\eqref{eq:vector_field} in the chart $(x,y)$). For each base point, a whole cascade of blow-ups may be necessary. Furthermore, a cascade can sometimes split into several sub-cascades (or branches).
    \item The obtained configuration of exceptional divisors obtained in the blow-up process is gathered in the intersection diagram. The essential information, i.e. the (self-)intersection numbers between divisors is extracted in term of a Dynkin-style diagram. The defining manifold for the system is the $n$-times blown up space with the inaccessible exceptional curves removed. 
    \item For each cascade of blow-ups, compute the Hamiltonian and canonical $2$-form $\omega =dy \wedge \rd x$ for the system after the final blow-up. Here, one can achieve that the Hamiltonian remains polynomial in the dependent variables of all final coordinate charts $(u_{\!f},v_{\!f})$, while the $2$-form takes on the form $k\,u_{\!f}^m \,\rd u_{\!f} \wedge \rd v_{\!f}$. In order for the $2$-form to stay in this simple form, certain intermediate transformations may be required after some blow-ups (the above mentioned twists and $k$-fold coverings), which we will point out in the text where they occur. However, this is done solely to keep the equations simple, all the blow-ups could also be performed without this additional step, but would be computationally more involved in some cases.
    \item To avoid logarithmic singularities in the solutions of the system one needs an additional condition satisfied that ensures that a certain cancellation takes place in the final system of each blow-up cascade. Under this condition the system can be converted into a regular initial value problem (after interchange of the dependent and independent variable). This ensures that the solutions can be expressed as Laurent series in some fractional power of $z-z_\ast$ (Puiseux series), where $z_\ast$ is the position of a movable singularity. These conditions are equivalent to passing the quasi-Painlev\'e test, as outlined in section~\ref{sec:quasi_painleve_eqs}. 
    \item To obtain the singular behaviour of the solution in the original affine coordinate chart $(x(z),y(z))$, one needs to unravel all the changes of coordinates introduced by the blow-ups. We do this for each case, indicating the leading order and series expansion of the solutions in the form of~\eqref{eq:power_series_form} for each cascade of blow-ups. While the full series solutions can also be obtained in this way, one can also get this by inserting a formal series into the system and solving a recurrence relation for the coefficients. In this way this process is equivalent to performing a (quasi-)Painlev\'e test on the system.
\end{enumerate}

In certain sub-cases, where the singularities obtained from all cascades of blow-ups are poles, we can relate the system to one of the known Painlev\'e equations. While in case for cubic Hamiltonians we find systems equivalent to Painlev\'e's equations $\text{I}$, $\text{II}$ and $\text{IV}$, in the quartic case we also recover systems equivalent to the (modified) Painlev\'e~$\text{III}$ and $\text{V}$ equations. In appendix~\ref{app:tables}, we list all the cubic and quartic cases with their representations in terms of power series, Newton polygons and surface type.

\section{Brief review of cubic Hamiltonian systems}
\label{sec:cubic_systems}
To introduce our method used for the classification of quartic quasi-Painlev\'e type systems in section~\ref{sec:quartic_systems} and for reasons of completeness we outline our analysis by reviewing cubic Hamiltonian systems. Here, it turns out that all systems of quasi-Painlev\'e type actually have the Painlev\'e property, as all singularities of the solutions are poles (rather than algebraic poles).  We start from a cubic (C) Hamiltonian of the form
\begin{equation}
\begin{aligned} 
H_{\text{C}}(x(z),y(z);z) &=\alpha_{30}\, x^3 + \alpha_{21}\, x^2y + \alpha_{12}\, xy^2 + \alpha_{03}\, y^3 + \alpha_{20}(z)\, x\,^2 + \alpha_{11}(z)\, xy  +\alpha_{02}(z)\, y^2 \\
&~~ + \alpha_{10}(z)\, x + \alpha_{01}(z)\, y,
\end{aligned}
\end{equation}
where the coefficients $\alpha_{ij}(z)$, $i+j \leq 2$ are analytic functions, while the coefficients of the cubic terms are constants, $\alpha_{30}$, $\alpha_{21}$, $\alpha_{12}$, $\alpha_{03} \in \mathbb{C}$. In the generic case, the Hamiltonian system derived from $H_{\text{C}}$ will initially have three base points when compactified on $\mathbb{CP}^2$. We then consider the cases where the base points coalesce one by one. If we allowed the coefficients of the cubic terms to be functions as well, this would in general introduce fixed singularities into the systems of equations, which we want to exclude here. 

\begin{figure}[t]
    \centering
    \includegraphics[width=.87\textwidth]{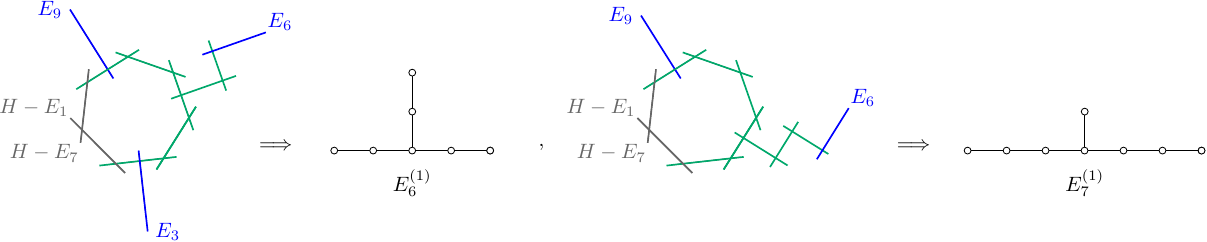}
    \caption{Rational surface constructed for $H_{\text{C}1}$ and its corresponding intersection diagram (left), coinciding with the Dynkin diagram $E_6^{(1)}$ for $\text{P}_{\text{IV}}$. On the right the same for $H_{\text{C}2}$, whose intersection diagram coincides with $E_7^{(1)}$ for $\text{P}_{\text{II}}$. }
    \label{fig:cubic_1_2}
\end{figure}

\begin{figure}[t]
    \centering
    \includegraphics[width=.6\textwidth]{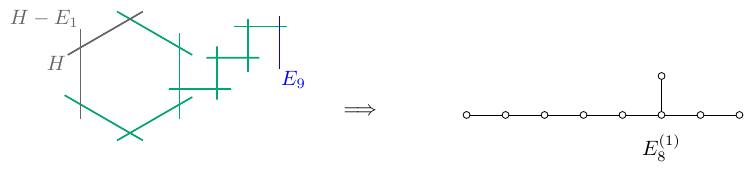}
    \caption{Rational surface constructed for $H_{\text{C}3}$ and its corresponding intersection diagram (left), coinciding with the Dynkin diagram $E_8^{(1)}$ for $\text{P}_{\text{I}}$.  }
    \label{fig:cubic_3}
\end{figure}

For each sub-case, we recover a Painlev\'e equation, in particular $\text{P}_{\text{I}}$ (figure \ref{fig:cubic_3}), $\text{P}_{\text{II}}$ (right of figure \ref{fig:cubic_1_2}) and $\text{P}_{\text{IV}}$ (left of figure \ref{fig:cubic_1_2}). The coefficients $\alpha_{jk}(z)$ are in general analytic functions in $z$, but these are fixed by the resonance conditions for the system to have the Painlev\'e property. Note also that for quadratic Hamiltonian systems the equations are linear and do not have movable singularities, therefore have the Painlev\'e property by default. 

\begin{figure}[t]
    \centering
    \includegraphics[width=.7\textwidth]{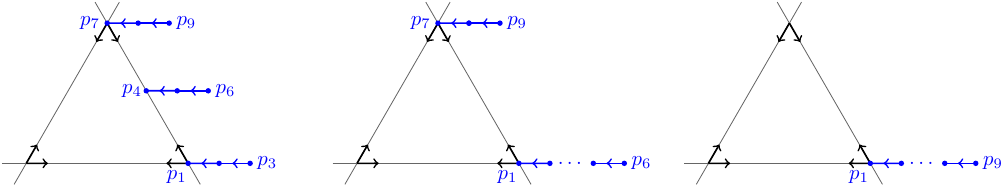}
    \caption{Cascades of blow-up transformations for the Hamiltonian systems related to (from left to right) $H_{\text{C}1}$, $H_{\text{C}2}$, $H_{\text{C}3}$.}
    \label{fig:cubic_cascades}
\end{figure}

Taking only the homogeneous cubic part of the Hamiltonian, we can factorise this as
\begin{equation}
\begin{aligned} 
H_{\text{C}}^{\text{hom}}(x(z),y(z);z) &= \alpha_{30}\, x^3 + \alpha_{21}\, x^2y + \alpha_{12}\, xy^2 + \alpha_{03}\, y^3 = \prod_{i=1}^3 \left( \alpha_i \,x + \beta_i \,y \right), \quad \alpha_i, \beta_i \in \mathbb{C}\,, 
\end{aligned} 
\end{equation}
we identify linear factors under the equivalence relation
\begin{equation}
    (\alpha_i,\beta_i) \sim (\lambda \,\alpha_i, \lambda \,\beta_i)\,, \quad  \lambda \in \mathbb{C}\,, 
\end{equation}
that is we consider the coefficients of these factors as points $[\alpha_i:\beta_i] \in \mathbb{CP}^2$. Under this equivalence the factorisation is unique and we distinguish the following three cases:

\begin{enumerate}
    \item All three points $[\alpha_i:\beta_i] \in \mathbb{CP}^2$, $i=1,2,3$, are pairwise distinct.
    \item There are two distinct points, one of which is a double point
    \item There is only one triple point
\end{enumerate}

In the following we consider all cases in turn. In each case we bring the Hamiltonian into some standard form, by shifting and / or re-scaling the dependent variables $x,y$.

\subsection{\texorpdfstring{Case C1: 3 simple points ($\text{P}_{\text{IV}}$)}{C1}}\label{sec:C1}

By a M\"obius transformation, we can arrange the three points $[\alpha_i\colon\! \beta_i] \in \mathbb{CP}^1$ to be $[1\colon\! 0]$, $[1\colon\! 1]$ and $[0\colon\! 1]$. The Hamiltonian thus takes the form
\begin{equation}
\label{C1_ham}
H_{\text{C}1}(x(z),y(z);z) = x(x-y)y + \alpha_{11}(z) \,xy + \alpha_{10}(z) \,x + \alpha_{01}(z) \,y \,,
\end{equation}
where the coefficients $a_{20}$ and $a_{02}$ are set to $0$ by an additional shift in $x$ and $y$.
In the extended phase space of the system $\mathbb{CP}^2$ there are three base points all on the line $\{u_0=0 \} \cup \{ V_0 =0 \}$, with coordinates 
\begin{equation}
    p_1 \colon (u_0,v_0) = (0,0)\,, \quad p_4 \colon (u_0, v_0)=(0,1)\,, \quad p_7\colon (U_0,V_0) =(0,0)\,,  
\end{equation}
and from each base point a cascade made up of two more points to blow-up arises. The cascades are as follows: 
\begin{align}
    p_1 &\,\,\leftarrow\,\, p_2\colon (u_1,v_1)=(0,0)\,\, \leftarrow\,\, p_3\colon (u_2, v_2) = (0, -\alpha_{10}) \,, \\
    p_4 &\,\,\leftarrow\,\, p_5\colon (u_5,v_5)=(0,\alpha_{11})\,\, \leftarrow\,\, p_6\colon (u_6, v_6) = \left(\alpha_{01}+\alpha_{10}-\fd{\alpha_{11}}, 0\right) \,, \\
    p_7 &\,\,\leftarrow\,\, p_8\colon (U_8,V_8)=(0,0)\,\, \leftarrow\,\, p_9\colon (U_9, V_9) = (\alpha_{01}, 0)  \,, 
\end{align}
and the conditions for the system to have the Painlev\'e property are
\begin{equation}
    \fd{\alpha_{01}}= 0 \,, \qquad \fd{\alpha_{10}} = 0 \,, \qquad \sd{\alpha_{11}} =0\,. 
\end{equation}
With these conditions implemented, \eqref{C1_ham} is similar to the Okamoto Hamiltonian $H_{\text{IV}}$ in~\eqref{eq:HPIV}.  
\begin{remark}
With a different normalisation for the cubic term in~\eqref{C1_ham}, such that the base point are symmetrically distributed on the line at infinity, one could alternatively consider the Hamiltonian
\begin{equation}
\begin{aligned} 
H_{\text{C}1}(x(z),y(z);z) &= \frac{1}{3} \left( x^3 + y^3 \right) + \widetilde{\alpha}_{11}(z) \,xy + \widetilde{\alpha}_{10}(z) \,x + \widetilde{\alpha}_{01}(z) \,y\,. 
\end{aligned}
\end{equation}
Under the conditions $\sd{\widetilde{\alpha}_{11}}=0$, $\fd{\widetilde{\alpha}_{01}}=0$ and $\fd{\widetilde{\alpha}_{10}}=0$ this becomes the cubic Hamiltonian studied in~\cite{Kecker2015}. 
\end{remark}

\subsection{\texorpdfstring{Case C2: 1 double point, 1 simple point ($\text{P}_{\text{II}}$)}{C2}} \label{sec:C2}
The Hamiltonian representing this case is
\begin{equation}
\begin{aligned} 
H_{\text{C}2}(x(z),y(z);z) &= xy^2 + \beta_{20}(z) \,x^2 + \beta_{10}(z) \,x + \beta_{01}(z) \,y \,,
\end{aligned}
\end{equation}
which can be seen as the structure obtained by the coalescence of two of the three branches for $H_{\text{C}1}$ in figure \ref{fig:cubic_cascades}, namely $\{p_1,p_2,p_3\}$ and $\{p_4,p_5,p_6\}$, into a single longer branch $\{p_1, \dots, p_6\}$, as represented in the same figure for $H_{\text{C}2}$.  

By re-scaling $x\mapsto \beta_{20}^{-1/2}x$ and $y\mapsto \beta_{20}^{1/4}y$ one can set the function $\beta_{20} = 1$. Then this system is regularised by two cascades, arising from the points in $\mathbb{CP}^2$
\begin{equation}
    p_1 \colon (u_0,v_0) = (0,0)\,, \quad p_7\colon (U_0,V_0) =(0,0)\,,  
\end{equation}
with $6$ and $3$ blow-ups respectively. The shortest cascade is:
\begin{align*}
    p_7 \,\,&\leftarrow\,\, p_8\colon (U_8,V_8)=(0,0)\,\, \leftarrow\,\, p_9\colon (U_9, V_9) = (-\beta_{01}, 0)  \,,
\end{align*}
whereas the longest cascade is given by
\begin{align}
\begin{split} 
    p_1 \,\,&\leftarrow\,\, p_2\colon (U_1,V_1)=(0,0)\,\, \leftarrow\,\, p_3\colon (U_2, V_2) = (-1, 0)  \,\,\leftarrow\,\,p_4\colon (U_3, V_3) = (0,0)\,\,\leftarrow \\ &\leftarrow\,\, p_5\colon (U_4,V_4)=(-\beta_{10},0)\,\, \leftarrow\,\, p_6\colon (U_5, V_5) = \left(-\beta_{01}+\fd{\beta_{10}}, 0\right) \,. 
    \end{split}
\end{align}
The conditions emanating from these two cascades for the system to be of Painlev\'e type are
\begin{equation}
    \fd{\beta_{01}} =0 \,, \qquad \sd{\beta_{10}}=0\,. 
\end{equation}
With these conditions implemented this turns out to be the standard Hamiltonian for $\text{P}_{\text{II}}$.

\subsection{\texorpdfstring{Case C3: 1 triple point ($\text{P}_{\text{I}}$)}{C3}} \label{sec:C3}
The last Hamiltonian we consider in the cubic case is 
\begin{equation}
H_{\text{C}3}(x(z),y(z);z) = y^3 + \gamma_{20}(z)\, x^2 + \gamma_{11}(z) xy + \gamma_{01}(z) \,y \,, 
\end{equation}
result of the coalescence of the three branches $\{p_1,p_2,p_3\}$, $\{p_4,p_5,p_6\}$, $\{p_7,p_8,p_9\}$ of the case $H_{\text{C}1}$ in figure \ref{fig:cubic_cascades} into the single long branch $\{p_1,\dots,p_9\}$ of $H_{\text{C}3}$ in the same figure. 

First re-scaling $x$ we can let $\gamma_{20}=1$. Then shifting $x \to x + {\gamma_{11}}/{2}$ one can absorb the $xy$ term, giving rise to a term $(\gamma_{11}^2 \, y^2)/4$, which in turn can be absorbed by a shift in $y$, leaving us with the Hamiltonian
\begin{equation}
H_{\text{C}3}(x(z),y(z);z) = y^3 + x^2  + \gamma(z) \,y\,.
\end{equation}
The sole cascade in this case is given by the following highly degenerate sequence of $9$ blow-ups 
\begin{align}
\begin{split} 
    p_1&\colon (u_0,v_0)=(0,0) \,\,\leftarrow\,\, p_2\colon (U_1,V_1)=(0,0)\,\, \leftarrow\,\, p_3\colon (U_2, V_2) = (0, 0)  \,\,\leftarrow \\ &\leftarrow\,\,p_4\colon (U_3, V_3) = (-1,0)\leftarrow\,\, p_5\colon (U_4,V_4)=(0,0)\,\, \leftarrow\,\, p_6\colon (U_5, V_5) = (0, 0) \,\, \leftarrow \,\,  \\
   &\leftarrow  \,\, p_7\colon (U_6,V_6)=(0,0) \,\,\leftarrow \,\, p_8\colon (U_7,V_7)=(-\gamma,0)\,\, \leftarrow\,\, p_9\colon (U_8, V_8) = \left( \fd{\gamma}, 0\right)  \,. 
\end{split}
\end{align}
With the resonance condition $\sd{\gamma}=0$ satisfied, in the case $\fd{\gamma} \neq 0$, under a suitable re-scaling of $z$ this becomes the Hamiltonian for the first Painlev\'e equation,
\begin{equation}
    \sd{y} = 6y^2 + z\,,
\end{equation}
while in the case where also $\fd{\gamma}=0$, the equation can be integrated in terms of elliptic functions and we would have needed only $8$ blow-ups.

\section{Quartic Hamiltonian systems}
\label{sec:quartic_systems}
We now explore polynomial quartic (Q) Hamiltonian systems
\begin{equation}
\begin{aligned}
H_{\text{Q}}(x(z),y(z);z) &= a_{40}\, x^4 + a_{31}\, x^3\, y + a_{22}\, x^2 y^2 + a_{13}\, xy^3 + a_{04}\, y^4 \\ &~~
+ a_{30}(z)\, x^3 + a_{21}(z)\, x^2y  + a_{12}(z)\, xy^2 + a_{03}(z)\, y^3 \\ & ~~+ a_{20}(z)\, x\,^2 + a_{11}(z)\, xy  +a_{02}(z)\, y^2  + a_{10}(z)\, x + a_{01}(z)\, y,
\end{aligned}
\end{equation}
quartic in $x$ and $y$, where the coefficients $\alpha_{ij}(z)$, $i+j \leq 3$ are analytic functions, while the coefficients of the quartic terms are assumed constant. Again, this assumption is made to avoid the occurrence of fixed singularities in the equations of motion.

As before, we factorise the quartic homogeneous part of the Hamiltonian into linear forms, 
\begin{equation}
\label{H_quartic_hom}
\begin{aligned} 
H_{\text{Q}}^{\text{hom}}(x(z),y(z);z) &= a_{40}\, x^4 + a_{31}\, x^3\, y + a_{22}\, x^2 y^2 + a_{13}\, xy^3 + a_{04}\, y^4 = \prod_{i=1}^4 \left( \alpha_i \,x + \beta_i \,y \right),
\end{aligned}
\end{equation}
with $\alpha_i, \beta_i \in \mathbb{C}$. Again we identify linear factors under the equivalence relation
\begin{equation}
    (\alpha_i,\beta_i) \sim (\lambda \,\alpha_i, \lambda \,\beta_i)\,, \quad  \lambda \in \mathbb{C}\,, 
\end{equation}
that is, we consider $[\,\alpha_i:\beta_i\,]$, $i \in \{1,2,3,4\}$, as points in $\mathbb{CP}^1$. Therefore, we thus distinguish the following five cases, where for each case we bring the quartic part into some standard form:

\begin{enumerate}
    \item All four points $[\,\alpha_i:\beta_i\,] \in \mathbb{CP}^1$, $i=1,2,3,4$, are pairwise distinct.
    \item There are three distinct points, one of which is a double point.
    \item There are two distinct points, both of which are double points.
    \item There are two distinct points, one of which is a triple point, the other a simple point.
    \item All the $[\,\alpha_i:\beta_i\,]$, $i=1,2,3,4$, define the same point in $\mathbb{CP}^1$, a quadruple point.
\end{enumerate}
We now look at all the cases in detail and produce their defining manifolds, as well as the conditions on the remaining coefficient functions in the Hamiltonian for the system to have the quasi-Painlev\'e property. Sub-cases arise where certain coefficient functions multiplying the leading order terms in the system of equations after a blow-up, defining any new base points on the exceptional curve, vanish. This is turn gives rise to a degeneration of the Hamiltonian. For each case we give the cascade of blow-ups, the configuration of the exceptional divisors arising from these, and a (Dynkin-like) diagram that reduces the irreducible components of the inaccessible divisor class to blobs, filled in according to the self-intersection number, and connected to other blobs according to the intersection between these divisor components. Here, a hollow blob denotes a curve of self-intersection number $-2$, while for a filled-in blob the corresponding curve has self-intersection number $-3$.
\begin{figure}[t]
    \centering
    \includegraphics[width=.95\textwidth]{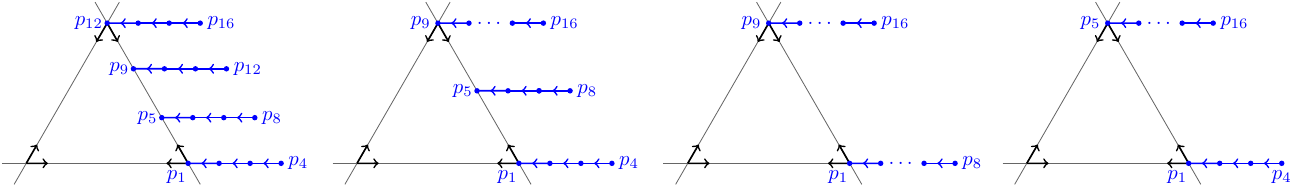}
    \caption{Cascades of blow-up transformations for the Hamiltonian systems related to (from left to right) $H_{\text{Q}1}$ in section~\ref{sec:4_simple_points}, $H_{\text{Q}2}$ in section~\ref{sec:1_double_2_simple}, $H_{\text{Q}3}$ in section~\ref{sec:2_double_points} and $H_{\text{Q}4}$ in section~\ref{sec:1_triple_1_simple_point}.}
    \label{fig:quartic_cascades}
\end{figure}

As we will see below, when the conditions on the coefficients are implemented for systems to be of quasi-Painlev\'e type, the solutions of the equations of motions in the plane in general have singularities that are movable algebraic poles expressed as Laurent series in $(z-z_*)^{1/n}$, with $n\in \{1,2,3,4,5\}$, and solutions with mixed types of algebraic poles can occur. 
While in the general case the equations give rise to branched solutions, we will see below that in some special cases we also obtain equations of Painlev\'e type, related to the modified third and fifth Painlev\'e equations. 

\subsection{\texorpdfstring{Case Q1: 4 simple points}{Q1}}\label{sec:4_simple_points}
By a M\"obius transformation $z \mapsto \frac{a z + b}{c z + d}$  we can achieve that three of the four points $[\,\alpha_i\colon \beta_i\,]$, $i=1,2,3,4$ are $[\,1 : 0\,]$, $[\,0 : 1\,]$ $[\,1 : -1\,]$, while the fourth point is  $[\,t : 1\,]$, where $t \in \mathbb{C} \setminus \{0,1\}$ is a parameter that cannot be fixed. Therefore, we consider the Hamiltonian
\begin{equation}\label{quartic1}
\begin{aligned}
H_{\text{Q}1}(x(z),y(z);z) =&\, xy(x-y)(x-t\,y) + a_{21}(z)\, x^2y + a_{12}(z)\, xy^2 + a_{20}(z)\, x^2 + a_{11}(z)\, xy \\ & + a_{02}(z)\, y^2 + a_{10}(z)\, x + a_{01}(z)\, y.
\end{aligned}
\end{equation}
Note that through an additional shift in $x$ and $y$, we are able to set the coefficient functions $a_{30}(z)$ and $a_{03}(z)$ to $0$.
The base points of this Hamiltonian system extended to $\mathbb{CP}^2$ are given by
\begin{equation}\label{quartic1_basepoints_CP2}
    p_1 \colon (u_0,v_0) = (0,0)\,, \quad p_5 \colon (u_0,v_0) = (0,1)\,, \quad p_9 \colon (U_0,V_0) = (t\,,0)\,, \quad p_{13} \colon (U_0,V_0) = (0,0)\,,
\end{equation}
where the points $p_5$ and $p_9$ are visible in both coordinate charts $(u_0,v_0)$, $(U_0,V_0)$. We will see below that each of these points can be regularised by a cascade of $4$ blow-ups if certain conditions on the coefficient functions are satisfied, leading to an additional cancellation in the system after the last blow-up in each cascade. 
The first two cascades of blow-ups are 
\begin{align}
\begin{split} 
        &\hspace*{-1.5ex} p_1 \,\, \leftarrow \,\, p_2 \colon \left(u_1,v_1\right) = \left(0\,,0\right)   \,\, \leftarrow \,\, p_3 \colon \left(u_2,v_2\right) = (0\,, -a_{20} ) \,\, \leftarrow  \,\, p_4 \colon \left(u_3,v_3\right) = \left(0\,,-a_{10}+a_{20}\,a_{21} \right), 
    \end{split} \\[1ex]
    \begin{split} 
       &\hspace*{-1.5ex} p_5 \,\, \leftarrow \,\, p_6 \colon \left(u_6,v_6\right) = \left(0\,,\frac{a_{12}+a_{21}}{t-1}\right) \,\, \leftarrow \,\, p_7 \colon \left(u_7,v_7\right) = ( 0\,,h_7 ) \,\, \leftarrow  \,\, p_8 \colon \left(u_8,v_8\right) = ( 0\,,h_8 ),
       \end{split} 
\end{align}
with
\begin{align*}
    h_7 &= \frac{a_{02}+a_{12}\big(a_{12}+(t+1)a_{21}\big)}{t-1}+(t-1)(a_{11}+a_{20})
    \,, \\[1ex]
    h_8 &= \frac{1}{1-t}\left(a_{01}+a_{10}\right)-\frac{1}{(1-t)^2}\left(\fd{a_{12}}+\fd{a_{21}}\right)+\frac{a_{21}\big( 2t\,a_{21}^2 + (t^2+4t+1)a_{12}^2 +3t(t+1)a_{12}\,a_{21} \big) }{(1-t)^5}\\[1ex]
    &\hspace{2ex}
    +\frac{(3t-1)a_{20}\,a_{21} +2t(a_{11}\,a_{21}+a_{20}\,a_{12}+(1+t)(a_{02}\,a_{21}+a_{11}\,a_{12}))}{(1-t)^3}- \frac{a_{12}^3}{(1-t)^4}
    \,.
\end{align*}
The third cascade is 
\begin{align}
    \begin{split}
         \hspace*{-1ex} p_9 \,\, &\leftarrow \,\, p_{10} \colon \left(U_{10},V_{10}\right) = \left( \frac{a_{12}+t\,a_{21}}{1-t}\,, 0\right) \,\,  \leftarrow \,\, p_{11} \colon \left(U_{11},V_{11}\right) = \left( h_{11}\,, 0\right) \,\, \leftarrow  \,\, p_{12} \colon \left(U_{12},V_{12}\right) =\left( h_{12}\,, 0\right)\!,
    \end{split} 
    \end{align}
with 
\vspace*{-1ex}

\begin{equation*}
    \begin{aligned}
        h_{11} &= \frac{a_{12}^2+a_{21}^2}{(1-t)^3}-\frac{a_{12}\,a_{21}}{(1-t)^2}+\frac{a_{02}+t\,a_{11}+t^2 a_{20}}{t(1-t)}
        \,,    \\[1ex]
        h_{12} &= \frac{a_{01}+t\,a_{10}}{t(1-t)}+\frac{1}{t(1-t)^2}\left(\fd{a_{12}}+t\,\fd{a_{21}}\right)
        +\frac{a_{21}\left( t(1+t)a_{21}^2+(1+4t+t^2)a_{12}\,a_{21}+3(t+1)a_{12} \right)}{(1-t)^5}\\[1ex]
        &\hspace{2ex} + \frac{(3t-1)a_{02}\,a_{12}+2\,t^2(a_{21}(a_{02}+a_{20})+a_{11}\,a_{12})+t^2(1+t)(a_{12}\,a_{20}+a_{11}\,a_{21})}{t^2(1-t)^3} 
        \,. 
    \end{aligned}
\end{equation*}
Lastly, the fourth cascade is
    \begin{align}
    \begin{split}
          \hspace*{-2ex} p_{13} \, &\leftarrow \, p_{14} \colon\! \left(U_{14},V_{14}\right) = \left( 0 \,, 0 \right) \, \leftarrow \, p_{15} \colon\! \left(U_{15},V_{15}\right) = \left( -\frac{a_{02}}{t}\,, 0 \right) \,  \leftarrow \, p_{16} \colon\! \left(U_{16},V_{16}\right) = \left( \frac{a_{02}\,a_{12}-t\,a_{01}}{t^2}\,, 0 \right)\!,
    \end{split}
\end{align}
\begin{figure}[t]
    \centering
    \includegraphics[width=.65\textwidth]{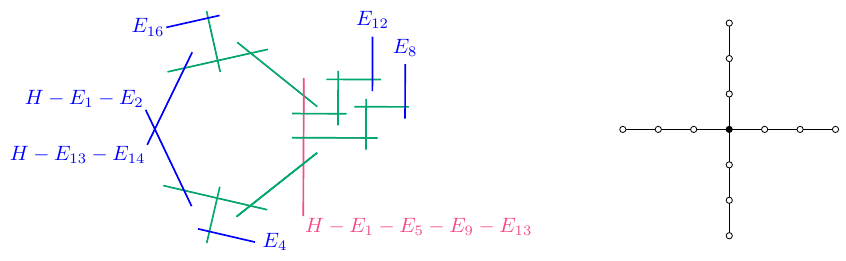}
    \caption{Rational surface constructed for $H_{\text{Q}1}$ (left) and its corresponding intersection diagram (right). }
    \label{fig:quartic_1}
\end{figure}
The conditions for the system \eqref{quartic1} to be of quasi-Painlev\'e type are 
\begin{equation}
\label{quartic1_conds}
    \fd{a_{20}}= 0\,,  \quad  \fd{a_{02}}=0 \,, \quad  \fd{}\!\left((t-1)^2 a_{11}+ a_{12}^2+(t+1)a_{12}\,a_{21}+ t\,a_{21}^2 \right) =0 \,. 
\end{equation}
Hence, we let 
\begin{equation*}
a_{20}(z)=a_{20}\,,\quad  a_{02}(z) = a_{02}\,, \quad  a_{11}(z) = c - \frac{1}{(t-1)^2}\left( a_{12}(z)^2+(t+1)a_{12}(z)\,a_{21}(z)+ t\,a_{21}(z)^2 \right)   \,, 
\end{equation*}
where $a_{20},a_{02},c \in \mathbb{C}$ are constants. This means that of the $8$ coefficient functions in the Hamiltonian~\eqref{quartic1}, the quasi-Painlev\'e conditions determine $3$ of these functions, with $5$ arbitrary (analytic) functions remaining. Unlike the Painlev\'e case, it is a general feature of quasi-Painlev\'e equations that not all coefficient functions are determined (up to re-scaling and shift of $z$). In principle, by eliminating $x(z)$ (or $y(z)$) one could deduce a second-order equation from the Hamiltonian system, but due to the higher degrees of $x$ and $y$ this would involve solving an algebraic equation. In this case, the equation obtained would take the general form
\begin{equation}
    \left( \sd{y} + f\!\left(y,\fd{y}\,;z\right) \right)^{\!\!2} + g\!\left(y,\fd{y}\,;z\right) = 0,
\end{equation}
for certain functions $f$ and $g$, polynomial in their first two arguments. However, the expressions for $f$ and $g$ are somewhat long so are omitted here.

The coordinate charts $(x,y)$, $(u_4,v_4)$, $(u_8,v_8)$, $(u_{12},v_{12})$, $(u_{16},v_{16})$  can be glued to form a symplectic atlas, forming the space of initial conditions for the system~\eqref{quartic1}, with symplectic $2$-form
\begin{equation}
\omega = \rd y \wedge \rd x = u_4\, \rd u_4 \wedge v_4 = u_8\, \rd u_8 \wedge v_8 = u_{12}\, \rd u_{12} \wedge \rd v_{12} = u_{16}\, \rd u_{16} \wedge \rd v_{16}\,,
\end{equation}
such that in each chart the system is given by a polynomial Hamiltonian.

To obtain the behaviour of the solution at a movable singularity $z_* \in \mathbb{C}$, one needs to inspect the system of equations after the last blow-up in each cascade, with the conditions~\eqref{quartic1_conds} implemented and an additional change of independent variable. Namely, we let $u_4$, the variable defining the exceptional curve $E_4 = \{u_4 =0\}$, become the independent variable, so that $z=z(u_4)$ and $v_4=v_4(u_4)$. In these variables, we have a regular initial value problem
\vspace{1ex}
\begin{equation}
\label{reg_init_val_prob}
\begin{cases}
    \dfrac{\rd z}{\rd u_4} = u_4 \left( 1 + u_4 \,F_1(z,v_4;u_4) \right) \\[2ex]
    \dfrac{\rd v_4}{\rd u_4} = \dfrac{F_2(z,v_4;u_4)}{1 + u_4 \,F_1(z,v_4;u_4)}
\end{cases}, 
\hspace{8ex} \begin{cases}
    z(0) = z_*  \\[1ex]
    v_4(0) = h
\end{cases} ,
\end{equation}
\vspace{1ex}

\noindent 
where $F_1$ and $F_2$ are analytic functions in their arguments (in fact polynomial in $u_4,v_4$) and $h \in \mathbb{C}$ is the parameter where the solution intersects the exceptional curve $E_4$.

Therefore, the leading order of solutions passing through the exceptional curve $E_4$ is
\begin{equation}
    z(u_4) = z_* + \frac{1}{2} u_4^2 + \mathcal{O}(u_4^3) \,, \qquad v_4(u_4) = h + \mathcal{O}(u_4) \,,
\end{equation}
where the right-hand sides are convergent power series in $u_4$ since the system~\eqref{reg_init_val_prob} has local analytic solutions. In the next step, one needs to take the square root and invert the series to obtain the solution of $u_4$, $v_4$ as a series expansion in $(z-z_*)^{1/2}$. With the notation introduced in~\eqref{eq:power_series_form}, we can write
\begin{equation}
    u_4 = \sqrt{2} (z-z_*)^{1/2} + \mathcal{P}_h\big( (z-z_*)^{1/2} \big), \qquad v_4 = h + \mathcal{P}_h \big( (z-z_*)^{1/2} \big).
\end{equation}
Lastly, to obtain the solution in the original variables $(x,y)$ one needs to undo all the changes of variable introduced by the blow-ups, 
\begin{equation}
   x(z)=\frac{1}{u_4}\,, \qquad y(z)=-a_{20}\,u_4+\left( a_{20}\,a_{21}(z)-a_{10}(z)\right) u_4^2 + u_4^3\,v_4 \,,
\end{equation}
yielding 
\begin{equation}
    E_4\colon ~~  x(z)=\frac{1}{\sqrt{2}}(z-z_*)^{-1/2}+\mathcal{P}_h\big( (z-z_*)^{1/2} \big) \,, \qquad y(z)=- a_{20}\,\sqrt{2}(z-z_*)^{1/2}+\mathcal{P}_h\big( (z-z_*)^{1/2} \big)\,, 
\end{equation}
with the parameter $h$ appearing in the higher orders of $z-z_*$.

For singularities where the solution passes through the final exceptional curve $E_8$, the behaviour of the solutions is  \begin{equation} \begin{split} E_8\colon ~ & x(z)= \frac{1}{\sqrt{2(t-1)}} (z-z_*)^{-1/2} + \mathcal{P}_h\big( (z-z_*)^{1/2} \big), \\[1ex]  & y(x)= \frac{1}{\sqrt{2(t-1)}} (z-z_*)^{-1/2} + \mathcal{P}_h\big( (z-z_*)^{1/2} \big),   \end{split}  \end{equation} whereas for the singularities lying on the exceptional curve $E_{12}$ the behaviour of the solutions is given by \begin{equation}  \begin{split}  E_{12}\colon ~~ & x(z)= \frac{t}{\sqrt{2t\,(1-t)}}(z-z_*)^{-1/2}+\mathcal{P}_h\big( (z-z_*)^{1/2} \big) \,, \\[1ex] & y(x)= \frac{t}{\sqrt{2t\,(1-t)}}(z-z_*)^{-1/2}+\mathcal{P}_h\big( (z-z_*)^{1/2} \big) \,,  \end{split}  \end{equation} and lastly, along the last exceptional curve $E_{16}$ we have   
\begin{equation} 
\begin{split} 
E_{16}\colon ~~ & x(z) =   \frac{i\,a_{02}\sqrt{2}}{\sqrt{t}}(z-z_*)^{1/2} + \mathcal{P}_h\big( (z-z_*)^{1/2} \big) \,, \quad y(z) = -\frac{i}{\sqrt{2t}}(z-z_*)^{-1/2}+\mathcal{P}_h\big( (z-z_*)^{1/2} \big) \,. 
\end{split} 
\end{equation}
Thus, the system derived from the Hamiltonian~\eqref{quartic1}, under the conditions~\eqref{quartic1_conds}, is of quasi-Painlev\'e type with all movable singularities being square-root type algebraic poles. Here, for different values of $t$ we find Hamiltonian systems with isomorphic defining manifolds, which however are not related by birational transformations. 

\begin{remark}
The defining manifold for the Hamiltonian system~\eqref{quartic1}, with the homogeneous quartic part replaced by $H_4 = x^4 - y^4$, was already presented in \cite{KeckerFilipuk}, while the system itself was introduced in \cite{Kecker2016} as an example of a Hamiltonian system with the quasi-Painlev\'e property whose solutions are of the leading order $x(z),y(z) \sim c(z-z_\ast)^{-1/2}$ at any singularity $z_\ast \in \mathbb{C}$. The defining manifold is obtained by $4$ cascades of blow-ups of the $4$ initial base points arising on $\mathbb{CP}^2$, 
\begin{equation}
    (u_0,v_0) \in \{ (0,1), (0,-1), (0,i), (0,-i) \}.
\end{equation}
The resonance conditions in this case are
\begin{equation}
    \fd{} \left( 2\,a_{20} - a_{12}^2 \right) = 0\,, \qquad  \fd{a_{11}} = 0 \,, \qquad  \fd{} \left( 2\,a_{02} + a_{21}^2 \right) = 0.
\end{equation} 
This system is equivalent to~\eqref{quartic1} only for specific values of the parameter $t$, namely for those $t$ for which one can find a M\"obius transformation mapping the four points $z \in \{0,1,t,\infty\}$ into $\{ 1,-1,i,-i \}$,
\begin{equation}
    z \mapsto \frac{a\,z + b}{c\,z + d}\,, \qquad a,b,c,d \in \mathbb{C}\,, \qquad a\,d - b\,c = 1 \,.
\end{equation}
The parameters $a,b,c,d$ and $t$ are fixed by imposing one particular choice of the mapping, say 
\begin{equation}
    0 \mapsto i \,, \quad 1 \mapsto 1 \,, \quad t \mapsto -i \,, \quad \infty \mapsto -1\,. 
\end{equation}
With this choice the M\"obius transformation becomes 
\begin{equation}
    z \mapsto \frac{\frac{1-i}{2}z +i}{-\frac{1-i}{2}z+1}\,, \quad \text{with } t=2\,. 
\end{equation}
Under a permutation of the points $\{0,1,t,\infty\}$ one obtains $4! = 24$ different values for $t$ in total. This also shows that for other values of $t$ one obtains inequivalent systems but with isomorphic defining manifolds. 
\end{remark}
{\begin{figure}[t]
    \centering
    \includegraphics[width=.61\textwidth]{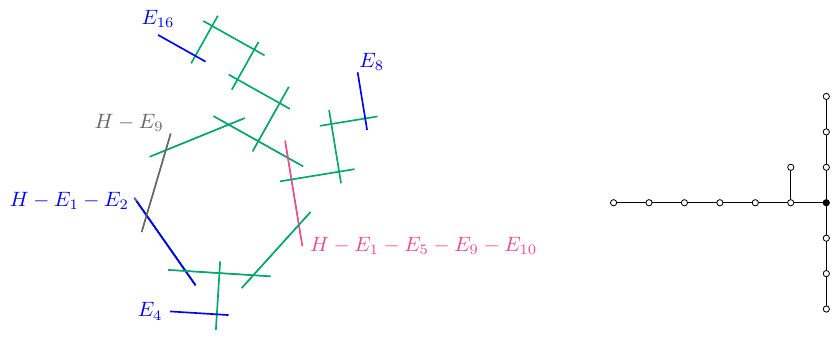}
    \caption{Rational surface constructed for $H_{\text{Q}2}$ (left) and its corresponding intersection diagram (right).}
    \label{fig:quartic_2}
\end{figure}
}

\subsection{Case Q2: 1 double point, 2 simple points}
\label{sec:1_double_2_simple}
Here we can achieve that the points $[\,\alpha_i:\beta_i\,]$, $i=1,2,3,4$, corresponding to the linear factors of the homogeneous quartic part of the Hamiltonian, become $[\,0:1\,]$, $[\,1:-1\,]$ and $[\,1:0\,]$ (twice). This leads us to a standard Hamiltonian of the form 
\begin{equation}
\label{H_quartic_case2}
\begin{aligned}
H_{\text{Q}2}(x(z),y(z);z) =&\, x^2 (x - y) \,y + b_{21}(z)\, x^2 \,y + b_{03}(z)\, y^3 + b_{20}(z)\, x^2 + b_{11}(z)\, xy + b_{02}(z)\, y^2 \\ 
&  + b_{10}(z)\, x + b_{01}(z)\, y\,, 
\end{aligned} 
\end{equation}
where the coefficient functions $b_{30}(z)$ and $b_{12}(z)$ have already been eliminated by appropriate shifts in $x$ and $y$. In the following, we will assume that $b_{03} \neq 0$, the case where $b_{03} = 0$ needs to be treated as a sub-case below (degeneration). The Hamiltonian~\eqref{H_quartic_case2} is obtained from the previous case, with four single points, by a coalescence of two cascades of blow-ups into one cascade, specifically the cascades originated in $p_9$ and $p_{12}$ in section \ref{sec:4_simple_points} by letting $t \to 0$. 

The three base points initially found after extending the phase space to $\mathbb{CP}^2$ are 
\begin{equation}
     p_{1}\colon (u_0,v_0) = (0,0)\,, \qquad p_5\colon (u_0,v_0) = (0,1) \,, \qquad p_9\colon(U_0,V_0) = (0,0)\,.
\end{equation}
The first two base points can be resolved in a cascade of $4$ blow-ups each, while the third point needs $8$ blow-ups, as follows. The cascades originating from $p_1$ and $p_{5}$, respectively, are 
\begin{align} 
    \begin{split}
        \hspace*{-1ex}p_1 \,\, &\leftarrow \,\, p_{2}\colon\!\left(u_{1},v_{1} \right) = \left(0, 0  \right) \,\, \leftarrow \,\, p_{3}\colon\!\left(u_{2},v_{2} \right) = \left(0\,,-b_{20}\right) \,\, \leftarrow \,\,p_{4}\colon\!\left(u_{3},v_{3} \right) = \big( 0\,,b_{20}\,b_{21}-b_{10} \big), 
    \end{split} \\[.5ex]
    \begin{split}
        \hspace*{-1ex}p_{5} \,\, &\leftarrow \,\, p_{6}\colon\!\left(u_{5},v_{5} \right) = \left(0, b_{03}+b_{21} \right) \,\, \leftarrow \,\, p_{7}\colon\!\left(u_{6},v_{6} \right) = \left(0\,, f_{7} \right) \,\,  \leftarrow  \,\,p_{8}\colon\!\left(u_{7},v_{7} \right) = \left(0\,, f_{8} \right), 
    \end{split}
\end{align}
where 
\begin{align*}
    f_{7}&=2 \,b_{03} \big(b_{21}+2 b_{03}\big)+b_{02}+b_{11}+b_{20} \,, \\[1ex] 
    f_{8}&= b_{01}+b_{10}-\fd{b_{03}}- \fd{b_{21}}+ b_{21}\big( b_{02}+6 \,b_{03}^2-b_{20} \big) +b_{03} \big(b_{21}^2 + 3 \,b_{02}+5 \,b_{03}^2+2 b_{11}+b_{20}\big)\,.
\end{align*}
We give the blow-ups for the cascade originating from the base point $p_9$ below. Here, as an additional requirement to keep the Hamiltonian function polynomial with respect to the symplectic $2$-form in the system after the final blow-up, we need to implement a twist of the form~\eqref{eq:generic_twist}, as discussed in section~\ref{sec:intermediate_changes_coordinates}. In this case, the twist happens after the $2$nd blow-up in the cascade originated at $p_9$, which introduces an additional intermediate change of variables $(\widetilde{U}_{10},\widetilde{V}_{10})$, implemented as
\begin{equation}
   {U}_{10}  = \frac{1}{\widetilde{U}_{10}}\,, \qquad {V}_{10} = \widetilde{V}_{10}\,.
\end{equation}
With this, the cascade of blow-ups is
\begin{equation} \label{eq:case2_casc_long}
    \begin{aligned}
        \!\!\!\!& p_9 \,\, \leftarrow \,\, p_{10} \colon (u_{9},v_{9})= (0,0) \,\, \leftarrow \,\, p_{11} \colon (\widetilde{U}_{10},\widetilde{V}_{10})= \Big(\frac{1}{b_{03}},0\Big)  \,\,\leftarrow \,\,  p_{12} \colon (U_{11} ,V_{11} )= \left(-1,0\right) \,\,\leftarrow \\
        & \leftarrow \,\, p_{13} \colon (U_{12},V_{12})= \Big(-\frac{b_{02}}{b_{03}}-b_{03}-\,b_{21},0\Big) \,\, \leftarrow \,\, p_{14} \colon (U_{13},V_{13})= \big( f_{14}  ,0\big) \,\,\leftarrow \\[.5ex]
        & \leftarrow \,\, p_{15} \colon (U_{14},V_{14})= \big( f_{15} ,0\big) \,\, \leftarrow \,\, p_{16} \colon (U_{15},V_{15})= \big( f_{16} ,0\big)\,, 
    \end{aligned}
\end{equation}
where 
\begin{align*}
    f_{14} &= -\big(b_{02} + b_{11}+2\, b_{03}\, b_{21}+2\, b_{03}^2 \big) -\frac{1}{3\,b_{03}} \fd{b_{03}} \,, \\[1ex] 
    f_{15} &=-\left( \fd{b_{03}} +b_{03} \big(5\,b_{03}^2+b_{21}^2+3\, b_{02}+2\, b_{11}+b_{20}\big) 
        +b_{01} +b_{21} \left(b_{02}+6\, b_{03}^2\right) \right) \,, \\[1ex]
    f_{16} &= \fd{b_{02}} -\fd{}(b_{21}\,b_{03})+\frac{1}{3}\, \fd{b_{03}} \left( \frac{b_{02}}{b_{03}} -13\, b_{03} \right) -2\,b_{03}^3(10\,b_{21} + 7\,b_{03})\\[1ex]
        &\qquad -b_{03}\left( 2\,b_{01}+ b_{10}+ 2\,b_{21}(3\,b_{02}+b_{11})+\fd{b_{21}} \right) -b_{03}^2 \big( 10\,b_{02}+3\,b_{20} + 6\,b_{11} + 6\,b_{21}^2 \big)
        \,. 
\end{align*}
 \begin{figure}
        \centering
        \includegraphics[width=.6\textwidth]{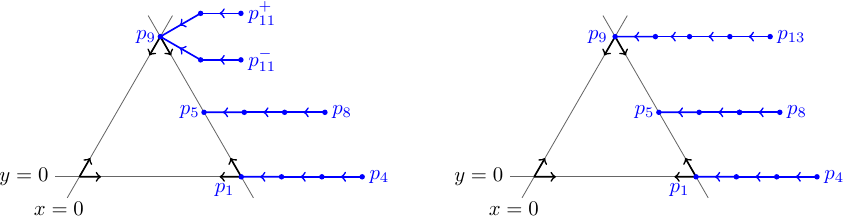}
        \caption{Cascades of blow-ups transformation for $H_{\text{Q}2}$ with $b_{03}=0$, giving rise to two different branches with $b_{02} \neq 0$ (left) or just one branch under the coalescence $b_{02} \to 0$ (right).}
        \label{fig:quartic_cascades_subcases2}
    \end{figure} 
The surface diagram of the defining manifold is depicted in figure \ref{fig:quartic_2} and the conditions obtained after the cascades for the system to be of quasi-Painlev\'e type are 
\vspace*{-1ex}
\begin{equation}
\begin{aligned}
    \fd{b_{20}}  &= 0\,, \qquad 
    \frac{\mathrm{d}}{\mathrm{d}z}\Big( b_{11} +b_{02} + 2b_{03}(b_{03}+b_{21}) \Big) = 0 \,, \qquad 
    b_{03}\, \sd{b_{03}} - \left(\fd{b_{03}}\right)^{\!\!2}  = 0\,.
\end{aligned}
\end{equation}
The first condition says that $b_{20}(z) = b_{20} \in \mathbb{C}$ is constant, while the second condition gives $b_{11}(z) = c - b_{02}(z) - 2b_{03}(b_{03}+b_{21})$, with $c \in \mathbb{C}$. The last condition implies that 
$b_{03}(z) = a\exp(b\, z) $,  
where $a,b \in \mathbb{C}$ are constants. 
 
The $2$-form $\omega$ in the final charts of each cascade of blow-ups is given by
\begin{equation}
    \omega = \rd y \wedge \rd x = u_4 \,\rd u_4 \wedge \rd v_4 = u_8 \,\rd u_8 \wedge \rd v_8 = u_{16}^2 \,\rd u_{16} \wedge \rd v_{16}.
\end{equation}
In this case, the behaviour of the solutions about the movable singularities is not the same for the three branches. Indeed, for the solutions going through $E_4$ and $E_8$ in figure~\ref{fig:quartic_2} the singularities are of square-root type, while through $E_{16}$ in figure~\ref{fig:quartic_2} they are of cubic-root type. In particular, along $E_4$ and $E_8$ we have 
\begin{align}
\begin{split}\label{eq:lead_ord_case2_1}
\hspace*{-1.8ex} E_4\colon  & x(z) = \frac{1}{\sqrt{2}}(z-z_*)^{-1/2} + \mathcal{P}_h\big( (z-z_*)^{1/2} \big)      \,,\quad 
y(x) =  -b_{20}\sqrt{2}(z-z_*)^{1/2}  + \mathcal{P}_h\big( (z-z_*)^{1/2} \big)     \,, 
\end{split} \\[1ex]
\begin{split} \label{eq:lead_ord_case2_2}
\hspace*{-1.8ex}E_8\colon  & x(z) = -\frac{i}{\sqrt{2}}(z-z_*)^{-1/2}+\mathcal{P}_h\big( (z-z_*)^{1/2} \big)      \,,\quad 
y(x) = -\frac{i}{\sqrt{2}}(z-z_*)^{-1/2}+\mathcal{P}_h\big( (z-z_*)^{1/2} \big)         \,, 
\end{split} 
\end{align}
and along $E_{16}$
\begin{equation} 
\begin{split} 
\hspace*{-1.8ex}E_{16}\colon  & x(z) =  \Big( \frac{b_{03}(z_*)}{3}\Big)^{\!\!1/3}(z-z_*)^{-1/3}+\mathcal{P}_h\big( (z-z_*)^{1/3} \big)     \,, \\ 
&y(x) = \Big( \frac{1}{9\,b_{03}(z_*)}\Big)^{\!\!1/3} (z-z_*)^{-2/3} + \mathcal{P}_h((z-z_*)^{1/3}) \,. \\
\end{split}
\end{equation}

\subsubsection{\texorpdfstring{Degeneration $\text{Q2} \to \text{Q2.a1}$ and coalescence $\text{Q2.a1} \to \text{Q2.a2}$}{Q2.a1 Q2.a2}}\label{sec:Q2a}

\begin{figure}[t]
        \centering
        \includegraphics[width=.5\textwidth]{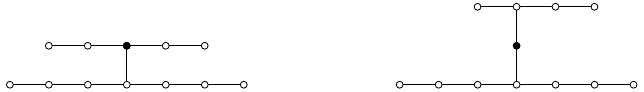}
        \caption{Intersection diagrams associated with the cascades shown in figure \ref{fig:quartic_cascades_subcases2} (left and right respectively), where we observe the coalescence of the short cascades for $H_{\text{Q}2}^{\text{a}1}$ in~\eqref{H_quartic_case2a1}, from the points $\{p_9, \dots, p^{\pm}_{11}\}$ (left) to $\{p_9, \dots, p_{13}\}$ with $H_{\text{Q}2}^{\text{a}2}$ in~\eqref{H_quartic_case2a12} (right). }
        \label{fig:inter_diag_subcases_2a}
    \end{figure}

When the coefficient $b_{03}$ in the Hamiltonian \eqref{H_quartic_case2} vanishes, we observe a structural change in the third cascade of blow-ups of the point $p_9$ in the generic case Q2. This is due to fact that if the function $b_{03} = 0$ in~\eqref{eq:case2_casc_long} an additional cancellation occurs in the system involving $(u_9,v_9)$. The cascades of blow-ups to be performed are schematised in figure~\ref{fig:quartic_cascades_subcases2}. 
For convenience the Hamiltonian will be rewritten by letting $b_{02} = (\widetilde{b}_{02})^2$.
\begin{equation}
\label{H_quartic_case2a1}
H_{\text{Q}2}^{\text{a}1}(x(z),y(z);z) =\, x^2 (x - y) \,y + b_{21}(z)\, x^2 \,y +  b_{20}(z)\, x^2 + b_{11}(z)\, xy + \widetilde{b}_{02}^2(z)\, y^2 + b_{10}(z)\, x + b_{01}(z)\, y\,. 
\end{equation}
    When $\widetilde{b}_{02}(z) \neq 0$, the branch starting from $p_9$ splits in two branches (left of figure~\ref{fig:quartic_cascades_subcases2}), each with two more points at which to blow up the surface. In this case, we obtain the following conditions for the quasi-Painlev\'e property of the system:
    \begin{equation}
        \fd{b_{20}}=0\,, \quad  \fd{}\left(b_{11}+\widetilde{b}_{02}^2\right)=0\,, \quad  \fd{}\left( \frac{b_{01}}{\widetilde{b}_{02}} + \widetilde{b}_{02}\,b_{21} \right)=0\,, \quad  \widetilde{b}_{02}\,\sd{\widetilde{b}_{02}}-\left(\fd{\widetilde{b}_{02}}\right)^{\!\!2}=0\,,
    \end{equation}
    from which we find that $b_{20}(z)=b_{20}$ is a constant, $\widetilde{b}_{02}(z)= \exp(az+b) $, $b_{11}(z)= c- \widetilde{b}_{02}(z)^2 $ and $b_{01}(z)=\widetilde{b}_{02}(z) (d - \widetilde{b}_{02}(z)\,b_{21}(z))$, with $a,b,c,d \in \mathbb{C}$.
Here, the leading order behaviour of the solution on the exceptional curves $E_4$ and $E_8$ is the same as in~\eqref{eq:lead_ord_case2_1} and~\eqref{eq:lead_ord_case2_2}, while we find the following for the two new cascades:
\begin{equation}
\begin{aligned}
E_{11}^{\pm}\colon ~ x(z)=\pm \,\widetilde{b}_{02}(z_*)+\mathcal{P}_h\big( (z-z_*)^{1} \big)\,, &\qquad  & y(z)= \pm \frac{1}{2\,\widetilde{b}_{02}(z_*)}(z-z_*)^{-1}+\mathcal{P}_h\big( (z-z_*)^{1} \big)\,.  
\end{aligned}
\end{equation}
If instead $\widetilde{b}_{02}(z)=0$, while assuming $b_{01}(z) \neq 0$, we observe the coalescence of the above-mentioned branches, giving rise to a single branch originating at $p_9$ and composed of a cascade of $5$ blow-ups (right of figure~\ref{fig:quartic_cascades_subcases2}). The Hamiltonian is 
\begin{equation}
\label{H_quartic_case2a12}
H_{\text{Q}2}^{\text{a}2}(x(z),y(z);z) =\, x^2 (x - y) \,y + b_{21}(z)\, x^2 \,y +  b_{20}(z)\, x^2 + b_{11}(z)\, xy + b_{10}(z)\, x + b_{01}(z)\, y\,, 
\end{equation}
and in this case the conditions for the quasi-Painlev\'e property are 
\begin{equation}
    \fd{b_{20}}=0\,, \qquad \fd{b_{11}}=0\,, \qquad b_{01}\,\sd{b_{01}}-\left(\fd{b_{01}}\right)^{\!\!2}=0\,,
\end{equation}
yielding $b_{20}(z)=b_{20}$, $b_{11}(z)=b_{11}$ and $b_{01}(z)=\exp (az+b)$, with $b_{20},b_{11},a,b \in \mathbb{C}$ constants. 
Again the behaviour of the solution along the exceptional curves $E_4$ and $E_8$ is reported in~\eqref{eq:lead_ord_case2_1} and~~\eqref{eq:lead_ord_case2_2}, while along the exceptional curve $E_{13}$ the leading order is 
\begin{equation}
E_{13}\colon ~ x(z)=b_{01}(z_*)(z-z_*) + \mathcal{P}_h((z-z_*)^1)\,,  \qquad y(z)=\frac{1}{b_{01}(z_*)}(z-z_*)^{-2}+ \mathcal{P}_h((z-z_*)^{1})\,. 
\end{equation}
The corresponding intersection diagrams are reported in figure \ref{fig:inter_diag_subcases_2a}.
  
\subsection{Case Q3: 2 double points}\label{sec:2_double_points}
By a M\"obius transformation, the base point $p_5$ from the previous case can be slid along the line at infinity in figure~\ref{fig:quartic_cascades_subcases2} to take any coordinates $(u_0,v_0) = (0,c)$, (respectively $(U_0,V_0)=(c^{-1},0)$), $c \in \mathbb{C} \setminus \{0\}$. 
\begin{figure}[t]
    \centering
    \includegraphics[width=.7\textwidth]{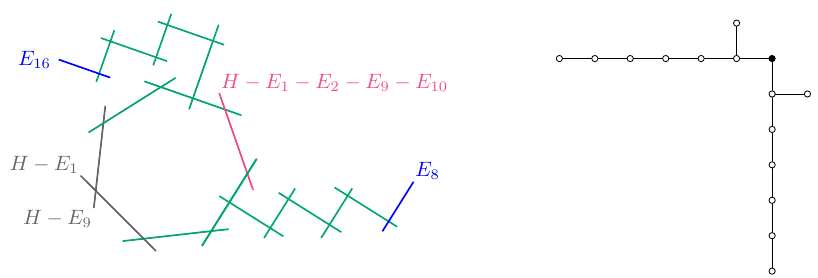}
    \caption{Rational surface constructed for $H_{\text{Q}3}$ (left) and its corresponding intersection diagram (right). }
    \label{fig:quartic_3}
\end{figure}
If we then let $c \to 0$ the point coalesces with the other single point, $p_1$, to form another double base point. In this case the homogeneous quartic part of the Hamiltonian becomes $x^2 y^2$. The general Hamiltonian in this case is
\begin{equation}
\label{22Ham}
H_{\text{Q}3}(x(z),y(z);z) = x^2y^2 + c_{30}(z)\,x^3 + c_{03}(z)\,y^3 + c_{20}(z)\,x^2 + c_{11}(z) \,xy + c_{02}(z)\, y^2 + c_{10}(z)\, x + c_{01}(z)\, y \,.
\end{equation}
Here, by a shift in $x$ and $y$ we have set the coefficients $c_{21}(z)$ and $c_{12}(z)$ to $0$. By an additional scaling $x \mapsto c(z)x$, $y \mapsto c(z)^{-1} y$ we can achieve that $c_{30}(z) = c_{03}(z) =: c_3(z)$, unless either $c_{30}$ or $c_{03}$, or both are identically $0$, which will be treated as separate sub-cases, by subsequent degeneration of the Hamiltonian, below.
The system~\eqref{22Ham} presents two base points in $\mathbb{CP}^2$ with coordinates  
\begin{equation}
    p_1 \colon (u_0,v_0) = (0,0)\,, \qquad p_9 \colon (U_0,V_0) = (0,0)\,, 
\end{equation}
giving rise to the following two cascades of blow-ups. The cascade originating at the point $p_1$ is 
\begin{equation} \label{eq:case3_casc_1}
    \begin{aligned}
        \!\!&\!\! p_1\,\, \leftarrow \,\, p_2 \colon \left(U_1,V_1\right) = \left(0\,,0\right) \,\, \leftarrow \,\, p_3 \colon \left(\widetilde{u}_2,\widetilde{v}_2\right) = \left(0\,,-\frac{1}{c_{3}}\right)\,\,  \leftarrow \,\, p_4 \colon \left(u_3,v_3\right) = \left(0\,,0\right) \,\, \leftarrow \\
        & \leftarrow \,\, p_5 \colon \left(u_4,v_4\right) =\left(0\,,-\frac{c_{20}}{c_{3}}\right) \,\, \leftarrow \,\, p_6 \colon \left(u_5,v_5\right) = \left(0\,, c_{11}-c_{3}^2 + \frac{1}{3 c_3} \fd{c_3} \right) \,\, \leftarrow \\
        &\leftarrow \,\, p_7 \colon \left(u_6,v_6\right) = \big(0\,, c_{10}-c_{3}\,c_{02}   \big) \,\, \leftarrow \,\, p_8 \colon \left(u_7,v_7\right) = \left(0\,,f_{8}\right) \,, 
    \end{aligned}
\end{equation}
where 
\begin{equation}
    f_{8} = 2\,c_{3}^2\,c_{20}-c_{3}\,c_{01} -c_{11}\,c_{20} + \fd{c_{20}}  - \frac{1}{3}\,\frac{c_{20}}{c_3} \fd{c_3} \,. 
\end{equation}
In order to obtain a polynomial Hamiltonian after the last blow-up transformation, we implemented the twist~\eqref{eq:generic_twist} for $(u_2,v_2)\mapsto (\widetilde{u}_2,\widetilde{v}_2)$.

Due to the symmetry in the Hamiltonian, the second cascade is similar:
\begin{equation}
    \begin{aligned} \label{eq:case3_casc2}
        &p_9 \,\, \leftarrow \,\, p_{10} \colon \left(u_{10},v_{10}\right) = \left(0\,,0\right) \,\, \leftarrow \,\, p_{11} \colon \left(\widetilde{U}_{11},\widetilde{V}_{11}\right) = \left(-\frac{1}{c_{3}},0\right) \, \, \leftarrow \,\, p_{12} \colon \left(U_{12},V_{12}\right) = \left(0\,,0\right) \,\, \leftarrow \\
        &\hspace*{-2ex}  \leftarrow \,\, p_{13} \colon \left(U_{13},V_{13}\right) =\left( -\frac{c_{02}}{c_{3}},0\right) \,\, \leftarrow \,\, p_{14} \colon \left(U_{14},V_{14}\right) = \left(  c_{3}^2 - c_{11} - \frac{1}{3 c_3} \fd{c_3},0\right)  \,\, \leftarrow\\[.5ex]
        &\hspace*{-2ex} \leftarrow \,\, p_{15} \colon \left(U_{15},V_{15}\right) = \big( c_{01}- \,c_{3} ,0\big) 
         \leftarrow \,\, p_{16} \colon \left(U_{16},V_{16}\right) = \left(f_{16} ,0\right) \,, 
    \end{aligned}
\end{equation}
where 
\begin{equation}
    f_{16} = 2\,c_{3}^2\,c_{02}-c_{3}\,c_{10}-c_{11}\,c_{02} - \fd{c_{02}}  + \frac{1}{3}\,\frac{c_{02}}{c_3} \fd{c_3} \,,
\end{equation}
and involving the twist~\eqref{eq:generic_twist} for $(U_{11},V_{11})\mapsto (\widetilde{U}_{11},\widetilde{V}_{11})$.

The intersection diagram associated with the system is shown in figure~\ref{fig:quartic_3}.
The conditions obtained after the cascades for the system to be of quasi-Painlev\'e type are
\begin{equation}
    \begin{aligned}
        \fd{}\left( c_3^2 -c_{11} \right)&=0 \,, \qquad 
        c_3\,\sd{c_{3}} - \left(\fd{c_3}\right)^{\!\!2}= 0\,. 
    \end{aligned}
\end{equation}
Since $c_3(z) \neq 0$, the latter condition gives $
c_3(z) = \exp(a z + b), $
with $a, b \in \mathbb{C}$ integration constants. Furthermore, the first condition becomes
$
c_{11}(z) = c + \exp(2 (a z + b))$, with $c \in \mathbb{C}$ constant.
Under these conditions, the solutions at the movable singularities for the branch starting from $p_1$ (third diagram in figure~\ref{fig:quartic_cascades}) has leading order behaviour
\begin{equation}
\begin{split} 
E_8\colon ~ &x(z)= \big(-9\, c_3(z_*)  \big)^{-1/3} (z-z_*)^{-2/3} + \mathcal{P}_h((z-z_*)^{1/3})\,, \\ &y(z)=\left(-\frac{c_3(z_*)}{3}\right)^{\!1/3} (z-z_*)^{-1/3}+\mathcal{P}_h\big( (z-z_*)^{1/3} \big)\,. 
\end{split} 
\end{equation}
The behaviour of the solutions through the exceptional curve after the last blow-up of the cascade originating from $p_9$ (third diagram in figure~\ref{fig:quartic_cascades}) is
\begin{equation}
\begin{split} 
   E_{16}\colon ~ &x(z)=\left( \frac{c_3(z_*)}{3}\right)^{\!1/3} (z-z_*)^{-1/3} +\mathcal{P}_h\big( (z-z_*)^{1/3} \big) \,, \\ 
   &y(x)=-\big( 9 \, c_3(z_*) \big)^{-1/3} (z-z_*)^{-2/3}+ \mathcal{P}_h((z-z_*)^{1/3})\,. 
   \end{split}
\end{equation}
 The symplectic $2$-form, in the coordinate charts after the final blow-up for each cascade is
\begin{equation}
\omega = \rd y \wedge \rd x = u_8^2 \, \rd u_8 \wedge \rd v_8 = V_{16}^2 \, \rd U_{16} \wedge \rd V_{16}\,.
\end{equation}

\subsubsection{\texorpdfstring{Degeneration $\text{Q3} \to \text{Q3.a1}$ and coalescence $\text{Q3.a1} \to \text{Q3.a2}$}{Q3.a1 Q3.a2}}\label{sec:Q3a}
This case refers to the occurrence where the coefficient functions $c_{30} \neq c_{03}$ in the Hamiltonian \eqref{22Ham}, and one of them vanishes. By a further scaling $x \mapsto c(z) \,x$ and $y \mapsto c(z)^{-1} y$, which leaves the quartic term invariant, we can achieve that $c_{03}(z) = c_{20}(z)$, assuming that $c_{20}(z) \neq 0$. Here, for convenience in the blow-up calculation we let $c_{03}(z) = c_{20}(z) = -\widetilde{c}(z)^2$, so that the Hamiltonian becomes 
\begin{equation}
\label{22Ham_subcase_c30=0}
H_{\text{Q}3}^{\text{a1}}(x,y;z) = x^2y^2 -\widetilde{c}(z)^2\left( y^3 + \,x^2 \right) + c_{11}(z) \,xy + c_{02}(z)\, y^2 + c_{10}(z)\, x + c_{01}(z)\, y \,. 
\end{equation}
\begin{figure}
        \centering
        \includegraphics[width=.65\textwidth]{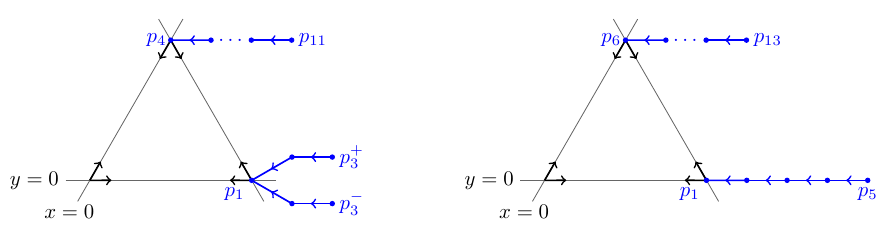}
        \caption{Cascades of blow-up transformation for $H_{\text{Q}3}^{\text{a}1}$ in \eqref{22Ham_subcase_c30=0} (left) and $H_{\text{Q}3}^{\text{a}2}$ in~\eqref{22Ham_subcase_c30=0_ii} (right).} 
        \label{fig:quartic_cascades_subcases3}
    \end{figure} 
The degeneration $c_{30}=0$ gives rise to a splitting of the cascade of blow-ups emanating from the base point $p_9: (u_0,v_0)=(0,0)$. The corresponding blow-up diagram is shown in figure \ref{fig:quartic_cascades_subcases3} (left). This splitting leads to a change of leading-order behaviour for solutions passing through the final exceptional curves in the two sub-cascades. The conditions from the three cascades, for the system to be of quasi-Painlev\'e type, are
\begin{equation}
    \sd{\widetilde{c}}\,\widetilde{c}-\left(\fd{\widetilde{c}}\right)^{\!\!2} =0\,, \qquad \fd{c_{11}}=0\,, \qquad \fd{}\!\left(\frac{c_{10}}{\widetilde{c}}\right)=0  \,.
\end{equation}
These can be simplified and integrated to give
\begin{equation}
\widetilde{c}(z) = \exp(a z +b)\,, \qquad  c_{11}(z) = c_{11},  \qquad c_{10}(z) = c \cdot \widetilde{c}(z) \,, 
\end{equation}
where $a,b,c,c_{11} \in \mathbb{C}$ are constants.

The $2$-form in the coordinates after final blow-ups in each cascade is of the form
\begin{equation}
    \omega = \rd y \wedge \rd x = -V_{11}^2 \, \rd U_{11} \wedge \rd V_{11} = \rd u_{3}^+ \wedge \rd v_{3}^+ = \rd u_{3}^- \wedge \rd v_{3}^-.
\end{equation}
The singularities arising from the exceptional curve after the cascade of blow-ups from $p_4$ (left of figure~\ref{fig:quartic_cascades_subcases3}) is of the form 
\vspace*{-2ex}
\begin{equation} 
\begin{split} 
E_{11}\colon ~ & x(z) = \left( -\frac{\widetilde{c}(z_*)^2}{3} \right)^{\!1/3} \!(z-z_\ast)^{-1/3} + \mathcal{P}_h\big( (z-z_*)^{1/3} \big)\,, \\ 
& y(z) = \big(-3\, \widetilde{c}(z_*)\big)^{-2/3} (z-z_\ast)^{-2/3} + \mathcal{P}_h((z-z_*)^{1/3})\,,
\end{split} 
\end{equation}
while the singularities from the split cascades emanating from $p_1$ (left of figure~\ref{fig:quartic_cascades_subcases3}) are now ordinary poles, 
\begin{equation} 
E_{3}^{\pm}\colon ~ x(z) = \frac{\pm 1}{2\,\widetilde{c}(z_\ast)} (z-z_\ast)^{-1}+\mathcal{P}_h\big( (z-z_*)^{1} \big)\,, \qquad y(z) = -\widetilde{c}(z_\ast) + \mathcal{P}_h\big((z-z_\ast)^1\big)\,,
\end{equation}
with two different types of residues for $x$ at the movable singularity $z_*$.

\begin{figure}[t]
    \centering
     \includegraphics[width=.52\textwidth]{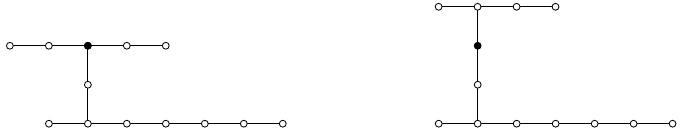}
    \caption{The mechanism of coalescence of the two small branches identified with the points $\{p_1, \dots, p_{3}^{\pm}\}$ in figure \ref{fig:quartic_cascades_subcases3} (left) for $H_{\text{Q}3}^{\text{a}1}$ to the single branch $\{p_1, \dots, p_{5}\}$ in figure \ref{fig:quartic_cascades_subcases3} (right) for $H_{\text{Q}3}^{\text{a}2}$.}
    \label{fig:inter_diag_subcase_3a}
\end{figure}

In the case where the coefficients $c_{03}(z) \neq c_{20}(z)=0$ we observe the coalescence of the two branches of the diagram on the left of figure~\ref{fig:quartic_cascades_subcases3}. Here, if $c_{10}(z) \neq 0$, we can re-scale $x$ and $y$ so that $c_{10}(z) = c_{03}(z)$ and the Hamiltonian takes the form
\begin{equation}
\label{22Ham_subcase_c30=0_ii}
H_{\text{Q}3}^{\text{a}2}(x(z),y(z);z) = x^2y^2 -c_{03}(z) \left(y^3 + x\right)  + c_{11}(z) \,xy + c_{02}(z)\, y^2 + c_{01}(z)\, y \,. 
\end{equation}
Performing the blow-ups, we get the intersection diagrams in figure~\ref{fig:inter_diag_subcase_3a}  for the cascades schematised in figure \ref{fig:quartic_cascades_subcases3} respectively to the left and to the right. 
The resonance conditions are
\begin{equation}
    \fd{c_{11}} = 0, \qquad \sd{c_{03}}\,c_{03} - \left(\fd{c_{03}}\right)^{\!\!2} = 0,
\end{equation}
resulting in $c_{03}(z) = \exp(a z + b)$ and $c_{11} (z)= c_{11}$, with the constants $a,b,c_{11} \in \mathbb{C}$.
The leading order behaviours for the solutions at the two types of movable singularity are given by
\begin{equation}
\begin{aligned}
E_{13}\colon~ x(z) &= \left(\frac{c_{03}(z_\ast)}{3}\right)^{\!1/3}(z-z_\ast)^{-1/3}+\mathcal{P}_h((z-z_*)^{1/3}) \,, \\ 
y(z) &= -\left(9 \,c_{03}(z_\ast)\right)^{\!-1/3}(z-z_\ast)^{-2/3} +\mathcal{P}_h\big( (z-z_\ast)^{1/3}  \big) , 
\end{aligned} 
\end{equation}
for the long branch terminating at the point $p_{13}$ on the right of figure~\ref{fig:quartic_cascades_subcases3} and 
\begin{equation}
E_{5}\colon ~x(z) = -\frac{1}{c_{03}(z_\ast)}(z-z_\ast)^{-2}+\mathcal{P}_h((z-z_\ast)^{1}), \qquad y(z) = -c_{03}(z_\ast)(z-z_\ast)+\mathcal{P}_h((z-z_\ast)^{1}) \,,
\end{equation}
for the cascade ending at the point $p_5$ in the same diagram. 

\subsubsection{\texorpdfstring{Degeneration $\text{Q3.a1} \to \text{Q3.b1}$ ($\text{P}_{\text{V}}$) and coalescences $\text{Q3.b1} \to \text{Q3.b2 } (\text{P}_{\text{III}}) \to \text{Q3.b3} $ ($\text{P}_{\text{III}}$)}{Q3.b1, Q3.b2, Q3.b3}}\label{sec:Q3b}
In this case, both cascades of blow-ups~\eqref{eq:case3_casc_1} and~\eqref{eq:case3_casc2} split into two, each of which give rise to ordinary poles as singularities for the system of equations, as represented on the left of figure~\ref{fig:quartic_3b_3c_3d}. The Hamiltonian, after re-scaling of $x$ and $y$, is
\begin{equation}
\label{22Ham_subcase3b_1}
    H_{\text{Q}3}^{\text{b}1}(x(z),y(z);z) = x^2y^2 -\widetilde{c}(z)^2\left( x^2 + \,y^2 \right) + c_{11}(z) \,xy + c_{10}(z)\, x + c_{01}(z)\, y,
\end{equation}
Performing the blow-ups, we obtain the following conditions, 
\begin{equation}
    \widetilde{c}\,\,\sd{\widetilde{c}}  - \left(\fd{\widetilde{c}}\right)^{\!\!2} = 0\,, \qquad \fd{c_{11}} = 0\,, \qquad \fd{}\!\left( \frac{c_{10}}{\widetilde{c}} \right) = 0\,, \qquad \fd{}\!\left( \frac{c_{01}}{\widetilde{c}} \right) = 0\,,
\end{equation}
which, when implemented, result in a system with meromorphic solutions, i.e.\ it has the Painlev\'e property. Namely, the leading order of the solutions are 
\begin{equation}
  E_{3}^{\pm}\colon ~  x(z) = \pm \frac{1}{2\,\widetilde{c}(z_*)} (z-z_\ast)^{-1} + \mathcal{P}_h\big( (z-z_*)^{1} \big)\,, \qquad y(z) = \pm \,\widetilde{c}(z_\ast) + \mathcal{P}_h((z-z_\ast)^{1}) \,,
\end{equation}
for the branches emerged by the base point $p_1$ in the left of figure~\ref{fig:quartic_3b_3c_3d}\,, and 
\begin{equation}
 E_{6}^{\pm}\colon ~  x(z) = \mp \, \widetilde{c}(z_*) + \mathcal{P}_h((z-z_\ast)^{1})\,, \qquad  y(z) = \pm \frac{1}{2\,\widetilde{c}(z_*)} (z-z_\ast)^{-1} + \mathcal{P}_h\big( (z-z_*)^{1} \big)\,, 
\end{equation}
for the branches with origin in $p_4$ (left of figure~\ref{fig:quartic_3b_3c_3d}).

\begin{remark}
The intersection diagram obtained for the Hamiltonian system~\eqref{22Ham_subcase3b_1} is that of the extended affine Weyl group of type $D_5^{(1)}$, which it shares with the Painlev\'e V equation (left of figure~\ref{fig:quartic_3b_3c_3d}). In fact, our Hamiltonian~\eqref{22Ham_subcase3b_1}, with the normalisation $\widetilde{c}(z) = \exp(z)$, $c_{10}(z) = a \exp(z)$, $c_{01}(z) = b \exp(z)$, $c_{11}(z) = c$ with constants $a,b,c \in \mathbb{C}$,
\begin{equation}
\label{22Ham_subcase3b_2}
H_{\text{Q}3}^{\text{b}1}(x,y;z) = x^2 y^2 - \exp(2z) \left( x^2 + y^2 \right) + c\, xy + \exp(z) (a \,x + b\, y),
\end{equation}
is related to the modified Painlev\'e V equation (see~\cite[p.29]{GromakLaineShimomura+2002}),
\begin{equation} 
\label{mPV}
\frac{\rd^2 w}{\rd t^2} = \left( \frac{1}{2w} + \frac{1}{w-1} \right) \left( \frac{\rd w}{\rd t}\right)^{\!\!2} +(w-1)^2 \left( \alpha\,w+\frac{\beta}{w} \right) + \gamma \, \exp(t)\, w+ \frac{\delta \, \exp(2t) \,w (w+1)}{w-1} \,.
\end{equation}
In particular, letting  
\begin{equation}
x(z) = \text{exp}\!\left( \frac{t}{2} \right)\!\!\left(\frac{w(t)+1}{w(t)-1}\right)\,, \qquad z = \frac{t}{2} \,,
\end{equation}
in~\eqref{22Ham_subcase3b_2} and eliminating $y(z)$ from the Hamiltonian system obtained, we can deduce the same second-order equation for $w(t)$ with the parameters $\alpha,\beta,\gamma,\delta$ being related to $a,b,c$ in~\eqref{22Ham_subcase3b_2} in the following way,
\begin{equation} 
\begin{split}
    \alpha &= \frac{1}{32} \left(1+b+c\right)^2 \,,\qquad  \beta = -\frac{1}{32} \left(1-b+c\right)^2 \,, \qquad  \gamma = a \,,   \qquad  \delta = -2 \,. 
\end{split} 
\end{equation}
Thus, with~\eqref{22Ham_subcase3b_2} we have found a very symmetric Hamiltonian system for the modified Painlev\'e V equation.

\begin{figure}[t]
    \centering
    \includegraphics[width=0.8\textwidth]{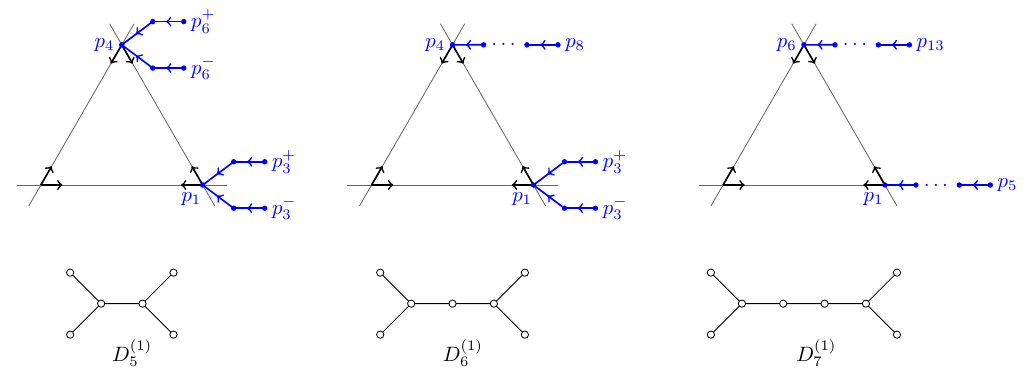}
    \caption{Cascades of blow-ups transformation for $H_{\text{Q}3}^{\text{b}1}$ (left), $H_{\text{Q}3}^{\text{b}2}$ (middle) and $H_{\text{Q}3}^{\text{b}3}$ (right) giving rise to three different intersection diagrams that are associated with the Dynkin diagrams of $D_5^{(1)}$ for $\text{P}_{\text{V}}$, $D_6^{(1)}$ and $D_7^{(1)}$ both for $\text{P}_{\text{III}}$ respectively.}
    \label{fig:quartic_3b_3c_3d}
\end{figure}

We observe a coalescence in~\eqref{22Ham_subcase3b_2} if we restore the coefficients $c_{20}$ multiplying $x^2$ and $c_{02}$ multiplying $y^2$ and let $c_{02} = 0$ (or $c_{20}=0$ which is similar). This corresponds to the coalescence of the two branches of blow-ups originating in~$p_4$ (left of figure~\ref{fig:quartic_3b_3c_3d}) producing the blow-up structure shown in the middle of figure~\ref{fig:quartic_3b_3c_3d}. In this case, we can re-scale $x \mapsto \widetilde{c}(z) \,x$, $y \mapsto \widetilde{c}(z)^{-1}y$ in order to set the coefficient $\widetilde{c}(z)=-1$, which is assumed in the following, yielding the Hamiltonian
\begin{equation}
\label{22Ham_subcase3b_3}
    H_{\text{Q}3}^{\text{b}2}(x(z),y(z);z) = x^2 y^2 - x^2 + c_{11}(z) \,xy + c_{10}(z)\, x + c_{01}(z)\, y\,. 
\end{equation}
Here, the conditions for the equations of motion to be of Painlev\'e type are 
\begin{equation}
\fd{c_{11}} = 0\,, \qquad \fd{c_{10}} = 0\,, \qquad c_{01} \, \sd{c_{01}} - \left(\fd{c_{01}}\right)^{\!\!2} = 0\,,
\end{equation}
which amount to $c_{01}(z) = \exp(az+b)$, $c_{10}(z) = c$ and $c_{11}(z) = d$ with constants $a,b,c,d \in \mathbb{C}$. In the following, we assume $a\neq0$ and shift and re-scale $z$ so that $c_{01}(z) = \exp(z)$. In this form, the resulting system of equations is related to the modified Painlev\'e III equation~\cite[p.22]{GromakLaineShimomura+2002} of type $D_6^{(1)}$. Namely, $x(z)$ satisfies the following second-order equation,
\begin{equation}
    \sd{x}= \frac{1}{x}\!\left( \fd{x}\right)^{\!\!2} - 2c\, x^2 + 4\, x^3 - \exp(z) \left(1+d\right) - \frac{\exp(2z)}{x} \,. 
\end{equation}
The leading order for the solutions at the movable singularities then are 
\begin{equation}
E_{8}\colon ~    x(z) = \exp(z_\ast) (z-z_\ast)+ \mathcal{P}_h\big((z-z_\ast)^1\big) \,, \qquad y(z) = - \exp(-z_\ast)(z-z_\ast)^{-2} + \mathcal{P}_h\big((z-z_\ast)^{1}\big)
\end{equation}
for the branch originating at $p_4$ (centre of figure~\ref{fig:quartic_3b_3c_3d}) and 
\begin{equation}
 E_{3}^{\pm}\colon ~   x(z) = \frac{\pm1}{2} (z-z_\ast)^{-1}+ \mathcal{P}_h\big( (z-z_*)^{1} \big) \,, \qquad y(z) = \pm 1 + \mathcal{P}_h\big((z-z_\ast)^1\big)\,,
\end{equation}
for the branches emerging from $p_1$ (centre of figure~\ref{fig:quartic_3b_3c_3d}).

With the choice $c_{20}=0$ in~\eqref{22Ham_subcase3b_3} we obtain a system related to the 
$D_7^{(1)}$ form of the Painlev\'e III equation. This corresponds to the other two cascades of blow-ups coalescing (right of figure~\ref{fig:quartic_3b_3c_3d}). If $c_{10}, c_{01} \neq 0$ we can re-scale $x$ and $y$ so that $c_{10}(z)=c_{01}(z)=: \widetilde{c}(z)$, yielding
\begin{equation}
\label{22Ham_subcase3_d5}
    H_{\text{Q}3}^{\text{b}3}(x(z),y(z);z) = x^2y^2 + c_{11}(z) \,xy + \widetilde{c}(z)\left( x + y \right) \,.
\end{equation}
Here, the conditions for the system to be of Painlev\'e type are 
\begin{equation}
\fd{c_{11}}=0 \,, \qquad \widetilde{c}\,\sd{\widetilde{c}}  - \left(\fd{\widetilde{c}}\right)^{\!\!2} = 0\,,
\end{equation}
giving $\widetilde{c}(z)=\exp(az+b)$ and $c_{11}(z)=c$, with constants $a,b,c \in \mathbb{C}$. Again, assuming $a\neq0$, we re-scale $z$ so that $\widetilde{c}(z)=\exp(z)$. The second-order equation satisfied by $x(z)$ is 
\begin{equation}
    \sd{x}=\frac{1}{x}\left( \fd{x} \right)^{\!\!2} - 2\exp(z)\,x^2-\exp(z)(1+c)-\frac{\exp(2z)}{x} \,.
\end{equation}
The leading order behaviour of the solutions at the movable singularities for the cascade starting at $p_1$ (right of figure~\ref{fig:quartic_3b_3c_3d}) is
\begin{equation}
   E_{5}\colon ~  x(z)=-\exp(-z_*)(z-z_*)^{-2} + \mathcal{P}_h\big((z-z_*)^{1}\big)\,, ~~~ y(z)=-\exp(z_*)(z-z_*)+ \mathcal{P}_h\big((z-z_*)^1\big)\,, 
\end{equation}
and for the cascade emerging from $p_6$ (right of figure~\ref{fig:quartic_3b_3c_3d}),
\begin{equation}
  E_{13}\colon ~   x(z)=\exp(z_*)(z-z_*)+ \mathcal{P}_h\big((z-z_*)^1\big)\,, ~~~ y(z) = - \exp(-z_*)(z-z_*)^{-2} + \mathcal{P}_h\big((z-z_*)^{1}\big) \,. 
\end{equation}
\end{remark}
\subsection{Case Q4: 1 triple point, 1 simple point}\label{sec:1_triple_1_simple_point}
If, starting from the Hamiltonian in section~\ref{sec:1_double_2_simple}, we slide the base point along the line at infinity and let it coalesce with the point $p_9$ (rather than $p_1$ in the previous section), we end up with the case of one triple and one single base point. The Hamiltonian in this instance can be taken to be
\begin{equation}
\label{Ham_Q4_general}
    H_{\text{Q}4}(x(z),y(z);z) = x^3 y + d_{12}(z) x y^2 + d_{03}(z) y^3 + d_{20}(z) x^2 + d_{11}(z) xy + d_{02}(z) y^2 + d_{10}(z) x + d_{01}(z) y \,.
\end{equation}
Here, by appropriate shifts in $x$ and $y$ we have eliminated the coefficients of the terms $x^3$ and $x^2y$, and by an additional re-scaling of the variables $x$ and $y$ we can set the coefficient $d_{03}(z)=1$. The case where $d_{03}=0$ will be treated separately below as a degeneration. 
\begin{figure}[t]
    \centering
    \includegraphics[width=.7\textwidth]{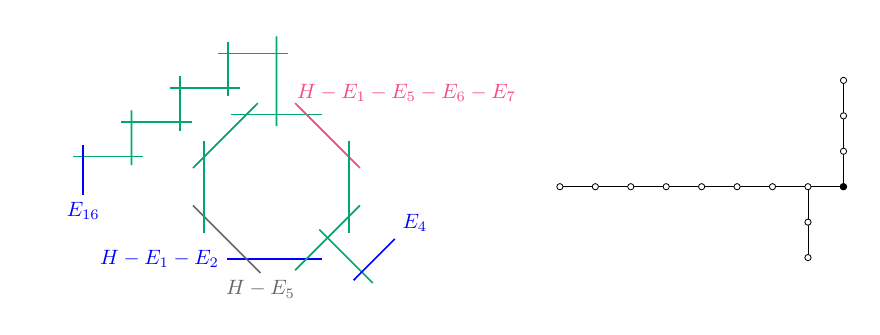}
    \caption{Rational surface constructed for $H_{\text{Q}4}$ (left) and its corresponding intersection diagram (right). }
    \label{fig:quartic_4}
\end{figure}

The cascades of blow-ups are given by two branches, originating at the base points 
\begin{equation}
    p_1\colon (u_0,v_0) = (0,0) \,, \qquad p_{5}\colon (U_0,V_0) = (0,0) \,,
\end{equation}
as depicted on the right of figure \ref{fig:quartic_cascades}. The first cascade is given by 
\begin{equation}
    \begin{aligned}
         \hspace*{-1ex} p_{1} &\,\, \leftarrow \,\, p_{2} \colon \left(u_{1},v_{1}\right) = \left(0\,,0\right) \,\, \leftarrow \,\, p_{3} \colon \left(u_{2},v_{2}\right) = \left(0\,,-d_{20}\right) \,\, \leftarrow \,\, p_{4} \colon \left(u_{3} ,v_{3} \right) = \left(0\,,-d_{10}\right) \,. 
    \end{aligned}
\end{equation}
The cascade originating from $p_5$ is given in terms of the coordinates as
\begin{equation}
    \begin{aligned}
         p_5 &\,\, \leftarrow \,\, p_6 \colon \left(u_6,v_6\right) = \left(0\,,0\right) \,\, \leftarrow \,\, p_7 \colon \left(u_7,v_7\right) = \big(0\,,0\big) \,\, \leftarrow \,\, p_8 \colon \left(\widehat{u}_8,\widehat{v}_8\right) = \left(0\,,1\right) \,\, \leftarrow \\[.5ex] 
        &\hspace*{-2ex}\leftarrow \,\, p_9 \colon \left(u_9,v_9\right) =\left(0\,,-\frac{d_{12}}{2}\right) \,\, \leftarrow \,\, p_{10} \colon \left(u_{10},v_{10}\right) = \left(0\,,\frac{d_{12}^2}{8}\right) \,\,\leftarrow
        \\[.5ex] 
        &\hspace*{-2ex}\leftarrow \,\, p_{11} \colon \left(u_{11},v_{11}\right) = \left(0\,,\frac{d_{02}}{2}\right) \,\,\leftarrow \,\, 
        p_{12} \colon \left(u_{12},v_{12}\right) = \left(0\,,\frac{d_{11}}{2}- \frac{d_{02}\,d_{12}}{4}- \frac{d_{12}^4}{128} \right) \,\, \leftarrow \\[.5ex] 
        &\hspace*{-2ex}\leftarrow \,\, p_{13} \colon \left(u_{13},v_{13}\right) = \left(0\,, \frac{d_{20}}{2}-\frac{1}{8}\fd{d_{12}} \right) \,\, \leftarrow \,\, p_{14} \colon \left(u_{14},v_{14}\right) = \big(0\,, f_{14} \big) \,\, \leftarrow \\[.5ex] 
        &\hspace*{-2ex}\leftarrow\,\, p_{15} \colon \left(u_{15},v_{15}\right) = \big(0\,, f_{15} \big) \,\,\leftarrow \,\, p_{16} \colon \left(u_{16},v_{16}\right) = \big(0\,, f_{16} \big) \,\, \,, 
    \end{aligned}
\end{equation}
where 
\begin{equation*}
    \begin{aligned}
        f_{14} &= \frac{1}{16}\left( \frac{d_{12}}{2}\right)^{\!\!6}-\frac{d_{12}^2}{16} \left( d_{11} - \frac{d_{02}\,d_{12}}{2} \right)+ \frac{d_{12}}{4} \left( d_{20} - \frac{1}{12}\fd{d_{12}} \right) 
        +\frac{d_{02}^2}{8}-\frac{d_{01}}{2}\,,     \\[1ex]
        f_{15} &= -\frac{d_{10}}{2}-\frac{1}{4}\fd{d_{02}}-\frac{d_{12}^2}{24}\fd{d_{12}}\,, \\[1ex]
        f_{16} &= \frac{d_{12}^2}{16} \left({d_{01}}-\frac{3}{4} \,d_{02}^2\right)
        +\frac{d_{12}}{8} \left(\fd{d_{02}}+ d_{02}\, d_{11} - 2d_{10}\right) -\frac{d_{11}^2}{8}-\frac{1}{2}\fd{d_{11}}-\frac{5}{128} \left(\frac{d_{12}}{2}\right)^{\!\!8}  \\
        &\hspace{2ex}
        + \left(\frac{d_{12}^3}{32}+ \frac{d_{02}}{4} \right)\left(\frac{3}{4}\, \fd{d_{12}}-d_{20}\right)
        -\frac{3}{16}\left(\frac{d_{12}}{2}\right)^{\!\!4} \left(d_{11} - \frac{d_{12}\,d_{02}}{2}  \right).
    \end{aligned}
\end{equation*}
In the long branch we have implemented the intermediate change of variables $(u_8,v_8) \mapsto (\widehat{u}_8,\widehat{v}_8)$ as a $2$-fold covering~\eqref{eq:generic_covering},
\begin{equation}
    {u}_8 = \frac{\widehat{u}_8}{\widehat{v}_8}\,, \qquad {v}_8 = -\widehat{v}_8^2\,,
\end{equation}
The conditions on the coefficients for the Hamiltonian system to be of quasi-Painlev\'e type are 
\begin{equation}
    \fd{d_{20}} = 0 \,, \qquad \sd{d_{12}} = 0\,,
\end{equation}
therefore $d_{20}(z)=d_{20}$ and $d_{12}(z)=az+b$, with $d_{20},a,b \in \mathbb{C}$ constants. 
The $2$-form in the coordinates after final blow-ups in each cascade is of the form
\begin{equation}
    \omega = \rd y \wedge \rd x = u_{4} \, \rd u_{4} \wedge \rd v_{4} = 2\,u_{16}^3 \, \rd u_{16} \wedge \rd v_{16} \,.
\end{equation}
The solutions at the movable singularities have the following behaviour
\begin{equation}\label{eq:lead_ord_quartic_4_general_E4}
    E_4 \colon ~ x(z)= \frac{1}{\sqrt{2}}(z-z_*)^{-1/2}+ \mathcal{P}_h\big( (z-z_*)^{1/2} \big) \,, \qquad y(z) = -d_{20}\sqrt{2}(z-z_*)^{1/2}+\mathcal{P}_h\big( (z-z_*)^{1/2} \big) \,, 
\end{equation}
along the exceptional curve $E_4$ in figure~\ref{fig:quartic_4}, and 
\begin{equation}
    E_{16} \colon ~ x(z)= \frac{1}{\sqrt{2}}(z-z_*)^{-1/2}+\mathcal{P}_h((z-z_*)^{1/4}) \,, \qquad y(z) = \frac{1}{2\sqrt{2}}(z-z_*)^{-3/4} + \mathcal{P}_h((z-z_*)^{1/4}) \,, 
\end{equation}
after the last blow-up in the long cascade in figure~\ref{fig:quartic_4}. 

\begin{figure}[t]
    \centering
    \includegraphics[width=0.55\textwidth]{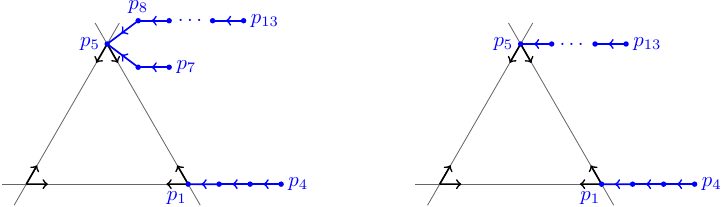}
    \caption{Cascades of blow-ups for $H_{\text{Q}4}^{\text{a}1}$ in~\eqref{Ham_Q4a} (left) and for $H_{\text{Q}4}^{\text{a}2}$ in~\eqref{Ham_Q4b}, both reproducing the cases studied in~\cite{MDAKec}.}
    \label{fig:quartic_4_subcases_cascades}
\end{figure}

\subsubsection{\texorpdfstring{Degeneration $\text{Q4} \to \text{Q4.a1}$ ($\text{quasi-P}_{\text{IV}}$) and coalescence $\text{Q4.a1} \to \text{Q4.a2}$  ($\text{quasi-P}_{\text{II}}$)}{Q4.a1 Q4.a2}}
\label{sec:case_Q4_subcase_a}
In this sub-case, where $d_{03}=0$, we can use the scaling freedom $x \to c x$, $y \to c^{-3} y$ to let $d_{12}=1$. With this, our Hamiltonian becomes
\begin{equation}
\label{Ham_Q4a}
    H_{\text{Q}4}^{\text{a}1}(x(z),y(z);z) = x^3 y + x y^2 + d_{20}(z) x^2 + d_{11}(z) xy + d_{02}(z) y^2 + d_{10}(z) x + d_{01}(z) y \,,
\end{equation}
whose cascades of blow-ups to obtain a regular system are depicted in figure~\ref{fig:quartic_4_subcases_cascades} (left). This is related to a Hamiltonian $H^{\text{qsi-P}_{\text{IV}}}_2$ for the equation of quasi-Painlev\'e $\text{IV}$ type introduced by the authors in~\cite[sec.~4.3]{MDAKec}, repeated here in variables $(u,v)$,
\begin{equation}\label{eq:H_2_qsi-P_IV}
    H^{\,{\text{qsi-P}}_{\text{IV}}}_2(u(z),v(z);z) = \frac{1}{2}\,u^2\, v -u \left(v^3 +\frac{\beta_4(z)}{4}\, v^2 +\frac{\beta_3(z)}{3}\, v +\beta_0(z) \right)-\frac{\beta_2(z)}{2}\, v^2 - \beta_1 (z) \,v +\beta_5(z)\,.
\end{equation}
The two Hamiltonians are related via the change of variables 
\begin{equation}
    x = -d_{02}(z)+ \frac{v}{2 a^2}\,, \qquad y = a\,u\,,
\end{equation}
with $a \in \mathbb{C}$ a constant solving the equation 
\begin{equation}
   a^5=-\frac{1}{8}\,,
\end{equation}
and the relations connecting the parameters $d_{01}(z), \dots, d_{20}(z)$ and $\beta_0(z),\dots, \beta_5(z)$ are 
\begin{equation}
\begin{split}
    \beta_0 & = a(d_{02}^3+d_{02}d_{11} - d_{01}) 
    \,,\quad \beta_1 = \frac{2\,d_{02}\,d_{20}-d_{10}}{2a^2}\,, \quad \beta_2=-\frac{d_{20}}{2\,a^4}\,, \\[1ex] 
    \beta_3 &= -\frac{3(3\,d_{02}^2+d_{11})}{2\,a}\,, \quad \beta_4 = \frac{3\,d_{02}}{a^3}\,,\quad  \beta_5 = d_{02}^2\,d_{20} - d_{02}\,d_{10}\,,
\end{split}
\end{equation}
where we have added an overall function $\beta_5(z)$, depending on $z$ only, in the Hamiltonian~\eqref{eq:H_2_qsi-P_IV}, which is not important for the dynamics of the system.

\begin{figure}[t]
    \centering
    \includegraphics[width=0.6\textwidth]{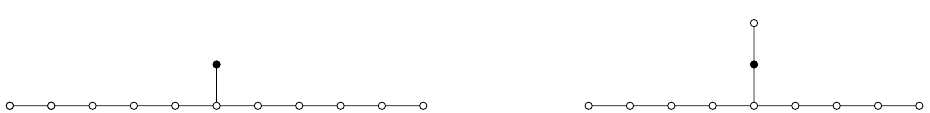}
    \caption{Intersection diagrams associated with $H_{\text{Q}4}^{\text{a}1}$ in \eqref{Ham_Q4b} (left) and $H_{\text{Q}4}^{\text{a}2}$ in \eqref{Ham_Q4a} (right).}
    \label{fig:quartic_4_subcases_inter_diag}
\end{figure}

The conditions for the Hamiltonian system derived from~\eqref{Ham_Q4a} to be of quasi-Painlev\'e type are 
\begin{equation}
    \fd{d_{20}}= 0 \,, \qquad \sd{d_{02}}=0\,, \qquad \fd{}\!\left(  d_{02}^3+d_{02}\, d_{11}  - d_{01} \right)=0\,, 
\end{equation}
from which $d_{20}(z)=d_{20}$, $d_{02}(z)=az+b$, $d_{01}(z)= d_{02}(z)\,d_{11}(z)+d_{02}(z)^3 + c $, where $d_{02}, a, b, c \in \mathbb{C}$ are constants. 
The $2$-form after the final blow-ups in each cascade is
\begin{equation}
    \omega = \rd y \wedge \rd x = u_4 \, \rd u_4 \wedge \rd v_4 = \rd U_7 \wedge \rd V_7 = -V_{13}\,\rd U_{13} \wedge \rd V_{13} \,,
\end{equation}
The behaviour of the solutions at movable singularities along $E_4$ is analogous to that already treated in~\eqref{eq:lead_ord_quartic_4_general_E4}. However, the behaviour of the solutions along $E_7$ is 
\begin{equation}
    E_7 \colon ~ x(z) = -d_{02}(z_*)+\mathcal{P}_h\big( (z-z_*)^{1} \big)     \,, \qquad y(z)= -(z-z_*)^{-1}+\mathcal{P}_h\big( (z-z_*)^1 \big)      \,,
\end{equation}
whereas along the last exceptional curve we have 
\begin{equation}
    E_{13} \colon ~ x(z) =  -\frac{i}{\sqrt{2}}(z-z_*)^{-1/2}+\mathcal{P}_h\big( (z-z_*)^{1/2}\big)   \,, \qquad y(z)=  \frac{1}{2}(z-z_*)^{-1}+\mathcal{P}_h\big( (z-z_*)^{1}\big)     \,. 
\end{equation}
The intersection diagram associated with this case is depicted at the left of figure~\ref{fig:quartic_4_subcases_inter_diag}. 

By letting $d_{12}=0$, we re-scale $x$ and $y$ so that $d_{02}=1$, the Hamiltonian now being
\begin{equation}
\label{Ham_Q4b}
    H_{\text{Q}4}^{\text{a}2}(x(z),y(z);z) = x^3 y + d_{20}(z) x^2 + d_{11}(z) xy + y^2+ d_{10}(z) x + d_{01}(z) y \,.
\end{equation}
This is equivalent to the Hamiltonian $H^{\text{qsi-P}_{\text{II}}}_{3}(x_3,y_3;z)$ for the equation of quasi-Painlev\'e II type presented in the authors' article~\cite[sec.\ 4.1]{MDAKec}. Explicitly, the Hamiltonian  $H^{\text{qsi-P}_{\text{II}}}_{3}(x_3,y_3;z)$ in~\cite{MDAKec} is mapped into~\eqref{Ham_Q4b} after a transformation of the variables $x_3 \mapsto y$, $y_3 \mapsto x$ and a shift first in the variable $y$ and then in the variable~$x$. 

The conditions such that the system associated with the Hamiltonian~\eqref{Ham_Q4b} is of quasi-Painlev\'e type are: 
\begin{equation}
\fd{d_{20}}=0 \,, \qquad \sd{d_{11}}= 0\,, 
\end{equation}
i.e.\ $d_{20}(z)=d_{20}$, $d_{11}(z) = az +b$, with $d_{20}$, $a$, $b \in \mathbb{C}$ constants. The $2$-form is given by 
\begin{equation}
    \omega = \rd y \wedge \rd x = u_4 \, \rd u_4 \wedge \rd v_4  = 3V_{13} \, \rd U_{13} \wedge \rd V_{13}  \,, 
\end{equation}
and in the procedure of building the long cascade, we have to consider the 3-fold covering~\eqref{eq:generic_covering} for the coordinate chart $(U_7,V_7) \mapsto (\widehat{U}_7,\widehat{V}_7)$ as
\begin{equation}
    {V}_7 = \frac{\widehat{V}_7}{\widehat{U}_7}\,, \qquad {U}_7 = \widehat{U}_7^3\,. 
\end{equation}
The behaviour of the solutions at the movable singularities through $E_4$ is the same as in~\eqref{eq:lead_ord_quartic_4_general_E4}, whereas along the exceptional curve $E_{13}$ emerging after the final blow-up in the cascade starting at $p_5$ in figure~\ref{fig:quartic_4_subcases_cascades} (right), we obtain the following: 
\begin{equation}
    E_{13}\colon ~ x(z) = \frac{i}{\sqrt{2}}(z-z_*)^{-1/2} +\mathcal{P}_h\big( (z-z_*)^{1/2} \big)    \,, \qquad y(z) = \frac{i}{2\sqrt{2}}(z-z_*)^{-3/2}+\mathcal{P}_h\big( (z-z_*)^{1/2} \big)   \,. 
\end{equation}
The intersection diagram associated with this case is depicted on the right of figure~\ref{fig:quartic_4_subcases_inter_diag}.

\subsection{Case Q5: 1 quadruple point}\label{sec:1_quadrupole}
\begin{figure}[t]
    \centering
    \includegraphics[width=.8\textwidth]{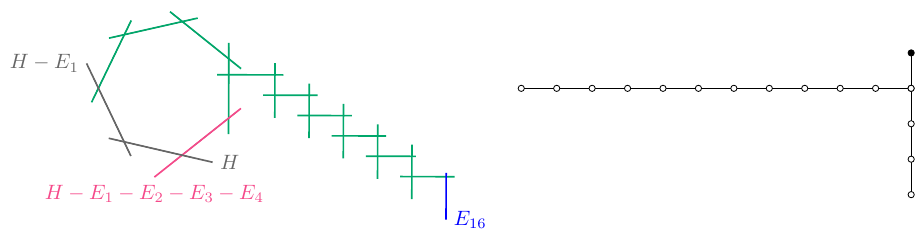}
    \caption{Rational surface constructed for $H_{\text{Q}5}$ (left) and its corresponding intersection diagram (right). }
    \label{fig:quartic_5}
\end{figure}

In this case the Hamiltonian can be brought into the standard form
\begin{equation} \label{eq:Ham_Q5}
\begin{split} 
H_{\text{Q}5}(x(z),y(z);z) &= x^4  + e_{21}(z)\, x^2 y + e_{12}(z)\, x y^2 + y^3 + e_{20}(z)\, x^2 + e_{11}(z)\, x y + e_{02}(z)\, y^2 \\
&~~+ e_{10}(z)\, x + e_{01}(z)\, y,
\end{split} 
\end{equation}
where we have eliminated the $e_{30}(z)\, x^3$ term by an appropriate shift in $x$ and re-scaled $y$ to set $e_{03}=1$. Thus, for $e_{03}(z) \neq 0$, we are essentially in the situation of the Hamiltonian presented in \cite{Kecker2016}, for which the defining manifold was computed in \cite{KeckerFilipuk}. However, here we consider an additional intermediate change of variables to keep the Hamiltonian polynomial, and for this reason we provide the cascade of points in appendix~\ref{app:blow-up_Q5} specifying the additional transformation. The rational surface associated with this system and the related intersection diagram are depicted in figure~\ref{fig:quartic_5}. 
This is obtained after a single cascade of $16$ blow-ups, the condition for the quasi-Painlev\'e case being
\begin{equation}
\sd{}\!\big(e_{12}^2 - 3 \, e_{21}\big)=0.
\end{equation}
The $2$-form after the final blow-up is given by 
\begin{equation}
    \omega = \rd y \wedge \rd x = 3\,u_{16}^4\,\rd u_{16} \wedge \rd v_{16} \,, 
\end{equation}
and the behaviour of the solutions through the only exceptional curve after the last blow-up of the cascade is
\begin{equation}
   E_{16} \colon ~ x(z) = \frac{1}{5^{3/5}}(z-z_*)^{-3/5}+\mathcal{P}_h\big( (z-z_*)^{1/5} \big)\,, \qquad y(z)= - \frac{1}{5^{4/5}} (z-z_*)^{-4/5} + \mathcal{P}_h\big( (z-z_*)^{1/5} \big) \,. 
\end{equation}

\subsubsection{\texorpdfstring{Degeneration $\text{Q5} \to \text{Q5.a1}$}{case Q5.a1}}\label{sec:Q5a}

\begin{figure}[t]
    \centering
    \includegraphics[width=.56\textwidth]{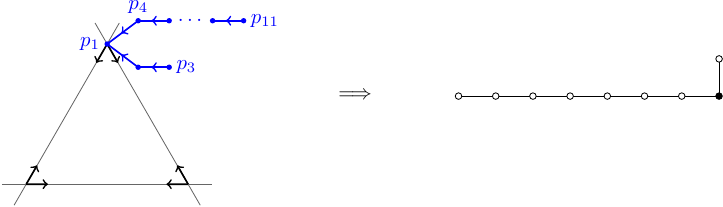}
    \caption{Cascades of blow-ups for $H_{\text{Q}5}^{\text{a}1}$ (left) and its corresponding intersection diagram (right). }
    \label{fig:quartic_5a}
\end{figure}

If $e_{03} = 0$, analogously to the way we proceed in section~\ref{sec:case_Q4_subcase_a}, we can re-scale the variables $x$ and $y$ so that the coefficient $e_{12}(z) = -1$. The Hamiltonian for this sub-case then becomes
\begin{equation} \label{eq:Ham_Q5.a1}
\begin{split} 
H^{\text{a1}}_\text{Q5}(x(z),y(z);z) &= x^4  + e_{21}(z)\, x^2 y - x y^2 + e_{20}(z)\, x^2 + e_{11}(z)\, x y + e_{02}(z)\, y^2 + e_{10}(z)\, x + e_{01}(z)\, y,
\end{split} 
\end{equation}
and the associated system is of quasi-Painlev\'e type if the following conditions are satisfied
\begin{equation}
\begin{split} 
    \fd{}\!\left(8\, \fd{e_{02}}+e_{21}\, \fd{e_{21}}\right)&=0\,, \\
    \fd{}\!\left(e_{01}+ e_{02} \,e_{11}+ e_{02}^2 e_{21}+\fd{e_{02}}\right)&=0\,.
    \end{split} 
\end{equation}
To obtain the 2-form in the final charts in a suitable way, we consider an additional intermediate change of variables with the form of a 2-fold covering at the step 3 for the branch $\{p_1, p_4, \dots, p_{11}\}$ in figure~\ref{fig:quartic_5a} as 
\begin{equation}
    {u}_3 = \frac{\widehat{u}_3}{\widehat{v}_3}\,, \qquad {v}_3= \widehat{v}_3^2\,.
\end{equation}
The $2$-form at the final step in the two cascades is 
\begin{equation}
    \omega = \rd y \wedge \rd x = \rd U_3 \wedge \rd V_3 = -2u_{11}^2\,\rd u_{11} \wedge \rd v_{11} \,,
\end{equation}
The behaviour of the solutions going through the exceptional curve $E_3$, here emerging after the final blow-up in the small branch presented in figure~\ref{fig:quartic_5a}, is 
\begin{equation}
    E_3 \colon ~ x(z)= e_{02}(z_*)+\mathcal{P}_h \big( (z-z_*)^1 \big)\,, \qquad y(z) = (z-z_*)^{-1} + \mathcal{P}_h\big( (z-z_*)^1 \big) \,,
\end{equation}
while along the exceptional curve $E_{11}$, after the last blow-up in the longer branch of figure~\ref{fig:quartic_5a}, we have
\begin{equation}
    E_{11}\colon ~ x(z)= 9^{-1/3}(z-z_*)^{-2/3} + \mathcal{P}_h\big( (z-z_*)^{1/3} \big) \,, \qquad y(z)= \frac{1}{3}(z-z_*)^{-1} + \mathcal{P}_h\big( (z-z_*)^{1/3} \big) \,.
\end{equation}
\subsubsection{\texorpdfstring{Degeneration coalescence $\text{Q5.a1} \to \to \text{Q5.b2}$ ($\text{P}_{\text{II}}$) and coalescence $\text{Q5.b2} \to \text{Q5.b3}$ ($\text{P}_{\text{I}}$)}{Q5.b2 e Q5.b3}}\label{sec:Q5b}
\begin{figure}[t]
    \centering
    \includegraphics[width=.56\textwidth]{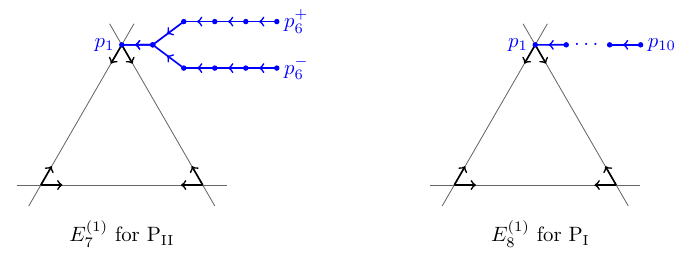}
    \caption{Cascades of blow-ups for $H_{\text{Q}5}^{\text{b2}}$ for $\text{P}_{\text{II}}$ (left) and $H_{\text{Q}5}^{\text{b2}}$ for $\text{P}_{\text{I}}$ (right). }
    \label{fig:quartic_5b_P2_P1}
\end{figure}

With the further degeneration induced by considering $e_{03}=0=e_{12}$ in~\eqref{eq:Ham_Q5}, we find a new Hamiltonian system for Painlev\'e $\text{II}$, which we already announced in~\cite{MDAKec}. The Hamiltonian in this case becomes 
\begin{equation}\label{eq:Ham_5b_general}
    H_{\text{Q}5}^{\text{b2}}(x(z),y(z);z) = x^4  + e_{21}(z)\, x^2 y  + e_{20}(z)\, x^2 + e_{11}(z)\, x y + e_{02}(z)\, y^2 + e_{10}(z)\, x + e_{01}(z)\, y, 
\end{equation}
and in order to trace back this expression to that of the Hamiltonian function for the system underlying $\text{P}_{\text{II}}$, we consider the transformation of variables $(x(z),y(z))$ as 
\begin{equation} \label{eq:case_5b_change_variables}
\begin{cases}
    x \mapsto a(z)\, x + b(z) \,, \\ y \mapsto c(z)\, y + d(z)\, x + e(z)\, x^2 + f(z)\,.
\end{cases} 
\end{equation}
By solving a system in six unknowns, we can determine the expressions for the $z$-dependent coefficients $a(z)$, $b(z)$, $\dots$, $f(z)$ in~\eqref{eq:case_5b_change_variables} such that the Hamiltonian takes the form: 
\begin{equation}\label{eq:case_5b_Ham_P2}
    H_{\text{Q}5}^{\text{b2}}(x(z),y(z);z) = x^4  + \widetilde{e}_{21}\, x^2 y  + (\widetilde{e}_{02})^2\, y^2 + \widetilde{e}_{10}(z)\, x + \widetilde{e}_{01}(z)\, y\,, 
\end{equation}
with the coefficients $\widetilde{e}_{02}$, $\widetilde{e}_{21}$ constants and with the constraint
\begin{equation} \label{eq:constraint_coeff_P2}
    \widetilde{e}_{21}^2 - 4\,\widetilde{e}_{02}^2 \neq 0 ~~\implies ~~ \widetilde{e}_{21}\neq \pm 2\, e_{02}\,,
\end{equation}
guaranteeing the split into two branches for the cascade of blow-ups ending in the points $p_6^+$ and $p_6^-$, respectively, shown in figure~\ref{fig:quartic_5b_P2_P1} (left).

In order to have the system in the final charts after the blow-ups of the points $p_6^{\pm}$ (left of figure~\ref{fig:quartic_5b_P2_P1}), we introduce an additional change of variables $(U_2,V_2) \mapsto (\widetilde{U}_2,\widetilde{V}_2)$ in the system before the splitting, the twist 
\begin{equation}
    \widetilde{U}_2 = \frac{1}{U_2} \,, \qquad \widetilde{V}_2 = V_2\,, 
\end{equation}
such that the 2-form in the last charts reads as 
\begin{equation}
    \omega = \rd y \wedge \rd x = - \rd U_6^+ \wedge \rd V^+_6 = - \rd U_6^- \wedge \rd V^-_6 \,. 
\end{equation}
The conditions for the Hamiltonian system derived from~\eqref{eq:case_5b_Ham_P2} to satisfy the Painlev\'e property are
\begin{equation}
    \fd{\widetilde{e}_{10}} = 0 \,, \qquad  \sd{\widetilde{e}_{01}} = 0 \,,
\end{equation}
hence the with $\widetilde{e}_{10}(z)=\widetilde{e}_{10}$ and $\widetilde{e}_{01}(z)=gz+h$, with $\widetilde{e}_{10}$, $g$, $h \in \mathbb{C}$ constants. The sketch of the cascades of blow-ups is at the left of figure~\ref{fig:quartic_5b_P2_P1}, and its well known intersection diagram is the Dynkin diagram $E_7^{(1)}$. Comparing this result with the discussion in section~\ref{sec:C2}, we get two different surface diagrams associated with two different structures of cascades of blow-ups (centre of figure~\ref{fig:cubic_cascades} and left of figure~\ref{fig:quartic_5b_P2_P1}) leading to the same minimal intersection diagram $E_7^{(1)}$. 
In particular, with the choice of coefficients 
\begin{equation}
    \widetilde{e}_{21}=-2 \,, \quad \widetilde{e}_{02} = -\frac{\sqrt{3}}{2}\,, \quad \widetilde{e}_{10}= \frac{1+4\alpha}{6} \,, \quad \widetilde{e}_{01}= -\frac{z}{4}\,,
\end{equation}
and a change of variables $\big(x(z) \mapsto y_4(z),y(z) \mapsto x_4(z) \big)$, we get the Hamiltonian $H_4^{\text{P}_{\text{II}}}$ \cite[eq.\ (3.4)]{MDAKec} in the variables $(x_4,y_4)$.

The coalescence of the two branches ending with the blow-ups of the points $p_{6}^{\pm}$ (left of figure~\ref{fig:quartic_5b_P2_P1}) is obtained by 
considering again the change of variables~\eqref{eq:case_5b_change_variables} in~\eqref{eq:Ham_5b_general} and solving the system to determine the coefficients $a(z)$, $b(z)$, $\dots$, $f(z)$ in~\eqref{eq:case_5b_change_variables} such that the Hamiltonian takes the form:
\begin{equation}\label{eq:Ham_5b_P1}
     H_{\text{Q}5}^{\text{b3}}(x(z),y(z);z) = x^4  + 2\,\widehat{e}_{02} \, x^2 y  + (\widehat{e}_{02})^2\, y^2 + \widehat{e}_{11}\, xy + \widehat{e}_{10}(z)\, x+ \widehat{e}_{01}\,y \,, 
\end{equation}
where $\widehat{e}_{02}$, $\widehat{e}_{11}$, $\widehat{e}_{01}$ are constants. 
Note that in this case we are studying the case complementary to~\eqref{eq:constraint_coeff_P2}, where we have 
\begin{equation}
    \widehat{e}_{21} = 2\,\widehat{e}_{02}\,. 
\end{equation}
The cascade of blow-ups for this case  is depicted at the right of figure~\ref{fig:quartic_5b_P2_P1}. Notice that in order to obtain a suitable expression for the symplectic form corresponding to the system derived after the final blow-up in the cascade (the point $p_{10}$ at the right of figure~\ref{fig:quartic_5b_P2_P1}), we need to implement two intermediate additional changes of variables: the twist at step 2
\begin{equation}
    {U}_2 = \frac{1}{\widetilde{U}_2} \,, \qquad {V}_2 = \widetilde{V}_2\,, 
\end{equation}
and the 2-fold covering at step 4,
\begin{equation}
    {U}_4 = \widehat{U}_4^2 \,, \qquad {V}_4 = \frac{\widehat{V}_4}{\widehat{U}_4} \,. 
\end{equation}
The 2-form in the final chart is then 
\begin{equation}
    \omega = \rd y \wedge \rd x = -2\, \rd U_{10} \wedge \rd V_{10} \,.  
\end{equation}
The condition for the Hamiltonian system derived from the Hamiltonian~\eqref{eq:Ham_5b_P1} to be of Painlev\'e type is 
\begin{equation}
    \sd{\widehat{e}_{10}} = 0\,, 
\end{equation}
such that $\widehat{e}_{10}(z)=kz + \ell$, with $k$, $\ell \in \mathbb{C}$ suitable constants, $x$ satisfies the well-known equation $\sd{x} = 6x^2 + z$.
The intersection diagram associated with the surface type is the Dynkin diagram for $E_8^{(1)}$, as already discussed in section~\ref{sec:C3}. So, with~\eqref{eq:Ham_5b_P1} we have found a quartic Hamiltonian giving rise to $\text{P}_{\text{I}}$.
\begin{remark}
We obtain a further degeneration in the case where in~\eqref{eq:Ham_5b_general} we let the coefficients $e_{12}=0$, $e_{02}=0$. The very degenerate Hamiltonian is given by 
\begin{equation}
    H^{\text{deg}}_{\text{Q}5}(x(z),y(z);z) = x^4 +{e}_{21}(z)\, x^2y + {e}_{11}(z)\, xy + {e}_{10}(z)\, x+ {e}_{01}(z)\,y \,, 
\end{equation}
giving rise to a decoupled system in $x$ and $y$, where the equation in $x$ takes the form of a Riccati equation. 

\end{remark}

\section*{Conclusion}
\label{sec:summary}
\addcontentsline{toc}{section}{\nameref{sec:summary}}

Starting from a general quartic polynomial, we have classified quasi-Painlev\'e Hamiltonian systems in terms of the intersection diagrams obtained for their defining manifold by an $n$-point blow-up of $\mathbb{CP}^2$, where $n \leq 16$, with equality in the most general cases. Through coalescence of base points and subsequent degeneration of the Hamiltonian, we obtain various sub-cases with a number $n < 16$ of blow-ups. Among these sub-cases we find various quasi-Painlev\'e systems whose solutions exhibit movable algebraic poles of the leading order $(z-z_\ast)^{-1/k}$ at a singularity $z_\ast \in \mathbb{C}$, where $k \in \{1,2,3,4,5\}$, and mixed types of poles and/or algebraic poles can occur in the same system. In the sub-cases where for all types of movable singularity we have $k=1$ we recover known Painlev\'e equations. Among the Hamiltonians of degree $3$ and $4$ we have recovered systems equivalent to the Painlev\'e equation $\text{I}$ - $\text{V}$. The Painlev\'e $\text{VI}$ equation will be found among the systems with Hamiltonian of degree $5$ or higher.     

We note that in the generic case for the homogeneous quartic part of the Hamiltonian with $4$ initial base points (section~\ref{sec:4_simple_points}) the intersection diagram itself does not suffice to uniquely determine the Hamiltonian structure, as by a M\"obius transformation we can only fix three of the base points and a free parameter remains in the Hamiltonian. When going to higher-degree polynomial Hamiltonian systems, we expect it to be necessary to introduce further parameters to uniquely describe the space of initial conditions.

It will be an interesting task to classify quasi-Painlev\'e Hamiltonian systems of higher degree and we expect to find many new, interesting systems. While in the degree $4$ case we found all intersection diagrams of (genuine) quasi-Painlev\'e systems (i.e.\ with at least one cascade of blow-ups to giving rise to algebraic poles) to have one and only one $-3$ curve (with all others being $-2$ curves), this will no longer be the case for higher degrees. While in the Painlev\'e cases, with all curves of self-intersection $-2$, the diagrams are extended Dynkin diagrams representing affine Weyl groups, the meaning of the diagrams with $-3$ curves remains to be discovered.

Below, we summarise the surface diagrams obtained for each sub-case in this scheme:
\begin{equation*}
    \begin{tikzpicture}[scale=0.46]
\path (6.4,6) node (Q5) {\hyperref[sec:1_quadrupole]{Q5}} -- ++(9,.1) node (Q3bi) {\hyperref[sec:Q3b]{Q3.b1}} -- ++(4,0) node (Q3bii) {\hyperref[sec:Q3b]{Q3.b2}}-- ++(4,0) node (Q3biii) {\hyperref[sec:Q3b]{Q3.b3}}; 
\path (Q5) -- ++(-8,0) node (Q5a) {\hyperref[sec:Q5a]{Q5.a1}};
\path (Q5a) -- ++(3,-5) node (Q5bi) {\hyperref[sec:Q5b]{Q5.b2}} -- ++(6,0) node (Q5bii) {\hyperref[sec:Q5b]{Q5.b3}};
\path (Q5) -- ++(3,5) node (Q3) {\hyperref[sec:2_double_points]{Q3}} -- ++(6,0) node (Q3ai) {\hyperref[sec:Q3a]{Q3.a1}} -- ++(4,0) node (Q3aii) {\hyperref[sec:Q3a]{Q3.a2}};
\path (Q5) -- ++(-3,5) node (Q4) {\hyperref[sec:1_triple_1_simple_point]{Q4}} -- ++(-4,0) node (Q4ai) {\hyperref[sec:case_Q4_subcase_a]{Q4.a1}} -- ++(-6,0) node (Q4aii) {\hyperref[sec:case_Q4_subcase_a]{Q4.a2}};
\path (Q5) -- ++(0,10) node (Q2) {\hyperref[sec:1_double_2_simple]{Q2}} -- ++(4,0) node (Q2ai) {\hyperref[sec:Q2a]{Q2.a1}} -- ++(4,0) node (Q2aii) {\hyperref[sec:Q2a]{Q2.a2}};
\path (Q2) -- ++(-4,1) node (Q1) {\hyperref[sec:4_simple_points]{Q1}};
\node[below=-.1 of Q1] (a) {\includegraphics[scale=.4]{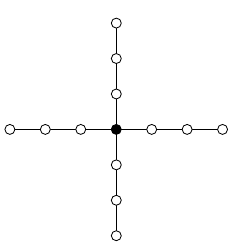}};
\node[below=-.1 of Q5] (a) {\includegraphics[scale=.4]{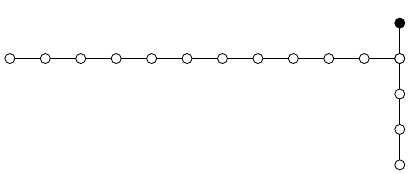}};
\node[below=-.1 of Q2] (a) {\includegraphics[scale=.4]{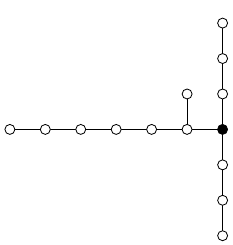}};
\node[below=-.1 of Q2ai] (a) {\includegraphics[scale=.4]{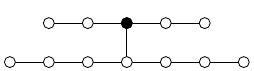}};
\node[below=-.1 of Q2aii] (a) {\includegraphics[scale=.4]{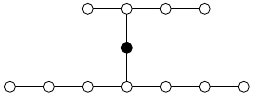}};
\node[below=-.1 of Q4] (a) {\includegraphics[scale=.4]{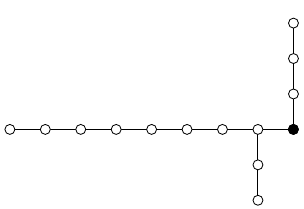}};
\node[below=-.1 of Q4ai] (a) {\includegraphics[scale=.4]{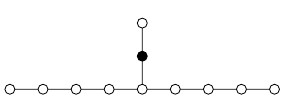}};
\node[below=-.1 of Q4aii] (a) {\includegraphics[scale=.4]{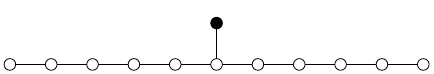}};
\node[below=-.1 of Q3] (a) {\includegraphics[scale=.4]{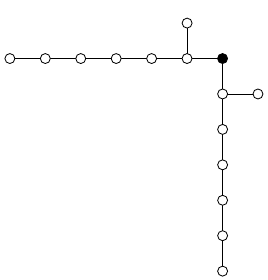}};
\node[below=-.1 of Q3ai] (a) {\includegraphics[scale=.4]{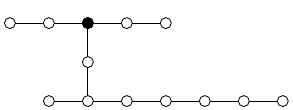}};
\node[below=-.1 of Q3aii] (a) {\includegraphics[scale=.4]{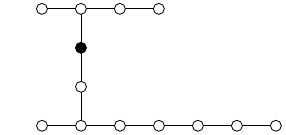}};
\node[below=-.1 of Q3bi] (a) {\includegraphics[scale=.35]{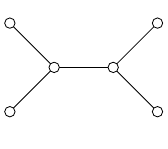}};
\node[below=-.1 of Q3bii] (a) {\includegraphics[scale=.35]{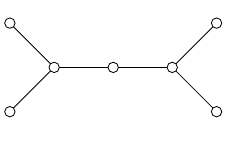}};
\node[below=-.1 of Q3biii] (a) {\includegraphics[scale=.35]{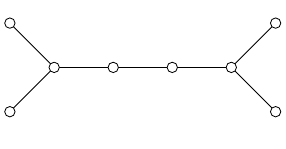}};
\node[below=-.1 of Q5a] (a) {\includegraphics[scale=.4]{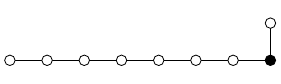}};
\node[below=-.1 of Q5bi] (a) {\includegraphics[scale=.35]{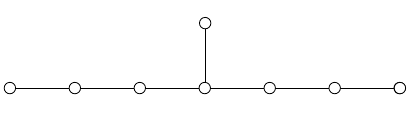}};
\node[below=-.1 of Q5bii] (a) {\includegraphics[scale=.35]{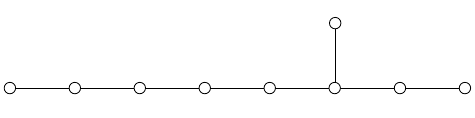}};
\path (Q1) -- ++(0.8,0) coordinate (q1ar); 
\path (Q2) -- ++(-.25,-3.9) coordinate (q2arl); 
\path (Q2) -- ++(.25,-3.9) coordinate (q2arr);
\path (Q3) -- ++(-0.3,-3.9) coordinate (q3ar); 
\path (Q4) -- ++(.3,-3.9) coordinate (q4ar); 
\path (Q5) -- ++(-.8,0) coordinate (q5ar); 
\path (Q5a) -- ++(0,-2) coordinate (q5aB); 
\path (Q5a) -- ++(0,-5) coordinate (q5aBB); 
\path (Q3ai) -- ++(0,-2.5) coordinate (q3aiar); 
\path (Q4ai) -- ++(0.15,.8) coordinate (q4aiP); 
\path (Q4aii) -- ++(0.15,.8) coordinate (q4aiiP); 
\path (Q5bi) -- ++(0,.8) coordinate (q5biP); 
\path (Q5bii) -- ++(0,.8) coordinate (q5biiP); 
\path (Q3bi) -- ++(.6,.8) coordinate (q3biP); 
\path (Q3bii) -- ++(0,.8) coordinate (q3biiP); 
\path (Q3biii) -- ++(0,.8) coordinate (q3biiiP); 
\draw[->,thick,cyan!80!gray] (q1ar) -- (Q2);
\draw[->,thick,cyan!80!gray] (q2arr) -- (Q3);
\draw[->,thick,cyan!80!gray] (q2arl) -- (Q4);
\draw[->,thick,cyan!80!gray] (q3ar) -- (Q5);
\draw[->,thick,cyan!80!gray] (q4ar) -- (Q5);
\draw[->,thick,cyan!80!gray] (Q2ai) -- (Q2aii);
\draw[->,thick,cyan!80!gray] (Q3ai) -- (Q3aii);
\draw[->,thick,cyan!80!gray] (Q4ai) -- (Q4aii);
\draw[->,thick,cyan!80!gray] (Q3bi) -- (Q3bii);
\draw[->,thick,cyan!80!gray] (Q3bii) -- (Q3biii);
\draw[->,thick,cyan!80!gray] (Q5bi) -- (Q5bii);
\draw[->,thick,orange!80!gray] (Q2) -- (Q2ai);
\draw[->,thick,orange!60!gray] (Q3) -- (Q3ai);
\draw[->,thick,orange!60!gray] (Q4) -- (Q4ai);
\draw[->,thick,orange!60!gray] (q5aB) -- (q5aBB);
\draw[->,thick,cyan!80!gray] (q5aBB) -- (Q5bi);
\draw[->,thick,orange!60!gray] (q3aiar) -- (Q3bi);
\draw[->,thick,orange!60!gray] (q5ar) -- (Q5a);
\node[blue] (a) at (q5biP) {\footnotesize $\text{P}_{\text{II}}$};
\node[below=.6 of Q5bi] (b) {\footnotesize $E_7^{(1)}$};
\node[blue] (a) at (q5biiP) {\footnotesize $\text{P}_{\text{I}}$};
\node[below=.6 of Q5bii] (b) {\footnotesize $E_8^{(1)}$};
\node[blue] (a) at (q4aiP) {\footnotesize $\text{quasi-P}_{\text{IV}}$};
\node[blue] (a) at (q4aiiP) {\footnotesize $\text{quasi-P}_{\text{II}}$};
\node[blue] (a) at (q3biP) {\footnotesize $\text{P}_{\text{V}}$};
\node[below=.6 of Q3bi] (b) {\footnotesize $D_5^{(1)}$};
\node[blue] (a) at (q3biiP) {\footnotesize $\text{P}_{\text{III}}$};
\node[below=.6 of Q3bii] (b) {\footnotesize $D_6^{(1)}$};
\node[blue] (a) at (q3biiiP) {\footnotesize $\text{P}_{\text{III}}$};
\node[below=.6 of Q3biii] (b) {\footnotesize $D_7^{(1)}$};
\path (Q3bii) -- ++(0,11) coordinate (leg) -- ++(2.8,0.05) coordinate (coal);
\path (leg) -- ++(0,-.75) coordinate (leg1) -- ++(2.94,0.05) coordinate (deg);
\draw[->,thick,cyan!80!gray] (leg) -- ++(1,0); 
\draw[->,thick,orange!60!gray] (leg1) -- ++(1,0); 
\node[black] (a) at (coal) {\footnotesize coalescence};
\node[black] (a) at (deg) {\footnotesize degeneration};
\end{tikzpicture}
\end{equation*}

\section*{Acknowledgements} 
The authors acknowledge funding by UK Research and Innovation under the EPSRC New Investigator Award EP/W012251/1, for the project entitled Geometric Aspects of Complex Differential Equations, at the University of Portsmouth from 2022 - 24. We also wish to thank Anton Dzhamay, Alex Stokes and Hidetaka Sakai for enlightening disssions when visiting Portsmouth. MD would like to thank Marco Bertola for interesting discussions. We also express our sincere appreciation to the anonymous reviewer for their many helpful comments.

\newpage
\begin{appendices}

\section{Summary of the classification scheme}\label{app:tables}
\setlength{\tabcolsep}{10pt}
\noindent
Cubic case genus 1:
{\small
\begin{equation*}
\arraycolsep=7pt\def\arraystretch{2.2}
    \begin{array}{c c c c c}
          \text{Case} & \text{Hamiltonian} & \text{Power series} & \text{Newton polygon} & \text{Surface type} \\
          \hline \\[-4.5ex]
         \hyperref[sec:C1]{\text{C1}} & \begin{aligned}
                 x(x-y)y + \alpha_{11}(z) \,xy \\ 
                 + \alpha_{10}(z) \,x + \alpha_{01}(z) \,y 
             \end{aligned} & \left( 1,1,1\right) & \includegraphics[width=.1\textwidth,valign=c]{newton_polygon_C1.pdf} & \includegraphics[width=.10\textwidth,valign=c]{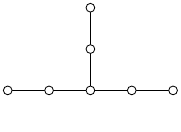}  \\[4.5ex]
         \hline \\[-4.5ex]
         \hyperref[sec:C2]{\text{C2}} & \begin{aligned}
             xy^2 + \beta_{20}(z) \,x^2 \\ + \beta_{10}(z) \,x + \beta_{01}(z) \,y 
         \end{aligned} & \left( 1,1\right) & \includegraphics[width=.1\textwidth,valign=c]{newton_polygon_C2.pdf} & \includegraphics[width=.16\textwidth,valign=c]{E7_nolab.pdf} \\[4.5ex]
         \hline \\[-4.5ex]
          \hyperref[sec:C3]{\text{C3}} & \begin{aligned}
              y^3 + \gamma_{20}(z)\, x^2 \\
              + \gamma_{11}(z) xy + \gamma_{01}(z) \,y 
          \end{aligned} & \left( 1\right) & \includegraphics[width=.1\textwidth,valign=c]{newton_polygon_C3.pdf} & \includegraphics[width=.2\textwidth,valign=c]{E8_nolab.pdf} \\[4.5ex]
         \hline 
    \end{array}
\end{equation*}}

\vspace{3ex}

\noindent
Quartic case genus 1:
{\small
\begin{equation*}
    \begin{array}{c c c c c}
          \text{Case} & \text{Hamiltonian} & \text{Power series} & \text{Newton polygon} & \text{Surface type} \\[2ex] 
          \hline \\[-2ex]
         \hyperref[sec:Q3b]{\text{Q3.b1}} & \begin{aligned}
                 x^2y^2 -\widetilde{c}(z)^2\left( x^2 + \,y^2 \right)\\ 
                 + c_{11}(z) \,xy + c_{10}(z)\, x + c_{01}(z)\, y
             \end{aligned} & \left( 1,1,1,1\right) & \includegraphics[width=.1\textwidth,valign=c]{newton_polygon_Q3b1.pdf} & \includegraphics[width=.08\textwidth,valign=c]{D5_nolab.pdf}  \\[6ex] 
          \hline \\
         \hyperref[sec:Q3b]{\text{Q3.b2}} & \begin{aligned}
             x^2 y^2 - \exp(2z) \left( x^2 + y^2 \right) \\
             + c\, xy + \exp(z) (a \,x + b\, y)
         \end{aligned} & \left( 1,1,1 \right) & \includegraphics[width=.1\textwidth,valign=c]{newton_polygon_Q3b2.pdf} & \includegraphics[width=.11\textwidth,valign=c]{D6_nolab.pdf} \vspace{2ex} \\
         \hline \\
          \hyperref[sec:Q3b]{\text{Q3.b3}} & \begin{aligned}
              x^2 y^2 - x^2 + c_{11}(z) \,xy \\
              + c_{10}(z)\, x + c_{01}(z)\, y
          \end{aligned} & \left( 1,1\right) & \includegraphics[width=.1\textwidth,valign=c]{newton_polygon_Q3b3.pdf}& \includegraphics[width=.13\textwidth,valign=c]{D7_nolab.pdf} \vspace{2ex} \\
         \hline \\
         \hyperref[sec:Q5b]{\text{Q5.b2}} & \begin{aligned}
             x^4  + e_{21}(z)\, x^2 y  + e_{20}(z)\, x^2 \\
             + e_{11}(z)\, x y + e_{02}(z)\, y^2 \\ 
             + e_{10}(z)\, x + e_{01}(z)\, y
         \end{aligned} & \left( 1,1 \right) & \includegraphics[width=.1\textwidth,valign=c]{newton_polygon_Q5b1.pdf} & \includegraphics[width=.15\textwidth,valign=c]{E7_nolab.pdf} \vspace{2ex} \\[1ex]
         \hline \\
         \hyperref[sec:Q5b]{\text{Q5.b3}} & \begin{aligned}
             x^4  + 2\,\widehat{e}_{02} \, x^2 y  + (\widehat{e}_{02})^2\, y^2 \\
             + \widehat{e}_{11}\, xy + \widehat{e}_{10}(z)\, x+ \widehat{e}_{01}\,y
         \end{aligned} & \left( 1 \right) & \includegraphics[width=.1\textwidth,valign=c]{newton_polygon_Q5b2.pdf} & \includegraphics[width=.2\textwidth,valign=c]{E8_nolab.pdf} \vspace{2ex} \\
         \hline 
    \end{array}
\end{equation*}
}

\newpage
Quartic case genus 2:

{\small
\begin{equation*}
    \begin{array}{c c c c c}
          \text{Case} & \text{Hamiltonian} & \text{Power series} & \text{Newton polygon} & \text{Surface type} \\[2ex]
          \hline \\[-2ex]
         \hyperref[sec:Q2a]{\text{Q2.a1}} & \begin{aligned}
                 x^2 (x - y) \,y + b_{21}(z)\, x^2 \,y \\
                 +  b_{20}(z)\, x^2 + b_{11}(z)\, xy \\
                 + \widetilde{b}_{02}^2(z)\, y^2 + b_{10}(z)\, x + b_{01}(z)\, y
             \end{aligned} & \left( \frac{1}{2},\frac{1}{2},1,1\right) & \includegraphics[width=.1\textwidth,valign=c]{newton_polygon_Q2a1.pdf} & \includegraphics[width=.12\textwidth,valign=c]{Q2a1.pdf}  \\[5ex]
         \hline \\
         \hyperref[sec:Q2a]{\text{Q2.a2}} & \begin{aligned}
             x^2 (x - y) \,y + b_{21}(z)\, x^2 \,y \\
             +  b_{20}(z)\, x^2 + b_{11}(z)\, xy \\
             + b_{10}(z)\, x + b_{01}(z)\, y
         \end{aligned} & \left( \frac{1}{2},\frac{1}{2},1\right) & \includegraphics[width=.1\textwidth,valign=c]{newton_polygon_Q2a2.pdf} & \includegraphics[width=.12\textwidth,valign=c]{Q2a2.pdf} \vspace{2ex} \\
         \hline \\[-2ex]
          \hyperref[sec:Q3a]{\text{Q3.a1}} & \begin{aligned}
              x^2y^2 -\widetilde{c}(z)^2\left( y^3 + \,x^2 \right) \\
              + c_{11}(z) \,xy + c_{02}(z)\, y^2 \\
              + c_{10}(z)\, x + c_{01}(z)\, y 
          \end{aligned} & \left( \frac{1}{3},1,1\right) & \includegraphics[width=.1\textwidth,valign=c]{newton_polygon_Q3a1.pdf}& \includegraphics[width=.13\textwidth,valign=c]{Q3a1.pdf} \vspace{2ex} \\
         \hline \\
         \hyperref[sec:Q3a]{\text{Q3.a2}} & \begin{aligned}
             x^2y^2 -c_{03}(z) \left(y^3 + x\right)  \\
             + c_{11}(z) \,xy + c_{02}(z)\, y^2 + c_{01}(z)\, y
         \end{aligned} & \left( \frac{1}{3},1\right) & \includegraphics[width=.1\textwidth,valign=c]{newton_polygon_Q3a2.pdf} & \includegraphics[width=.13\textwidth,valign=c]{Q3a2.pdf} \vspace{2ex} \\[1ex]
         \hline \\
         \hyperref[sec:case_Q4_subcase_a]{\text{Q4.a1}} & \begin{aligned}
             x^3 y + x y^2 + d_{20}(z) x^2 \\
             + d_{11}(z) xy + d_{02}(z) y^2 \\
             + d_{10}(z) x + d_{01}(z) y 
         \end{aligned} & \left( \frac{1}{2},\frac{1}{2},1\right) & \includegraphics[width=.1\textwidth,valign=c]{newton_polygon_Q4a1.pdf} & \includegraphics[width=.15\textwidth,valign=c]{Q4a1.pdf} \vspace{2ex} \\
         \hline \\
         \hyperref[sec:case_Q4_subcase_a]{\text{Q4.a2}} & \begin{aligned}
             x^3 y + d_{20}(z) x^2 + y^2  \\
             + d_{11}(z) xy + d_{10}(z) x + d_{01}(z) y 
         \end{aligned} & \left( \frac{1}{2},\frac{1}{2}\right) & \includegraphics[width=.1\textwidth,valign=c]{newton_polygon_Q4a2.pdf} & \includegraphics[width=.2\textwidth,valign=c]{Q4a2.pdf} \vspace{2ex} \\
         \hline \\
         \hyperref[sec:Q5a]{\text{Q5.a1}} & \begin{aligned}
             x^4  + e_{21}(z)\, x^2 y - x y^2 \\
             + e_{20}(z)\, x^2 + e_{11}(z)\, x y \\
             + e_{02}(z)\, y^2 + e_{10}(z)\, x + e_{01}(z)\, y
         \end{aligned} & \left( \frac{1}{3},1\right) & \includegraphics[width=.1\textwidth,valign=c]{newton_polygon_Q5a1.pdf} & \includegraphics[width=.15\textwidth,valign=c]{Q5a1.pdf} \vspace{2ex} \\
         \hline 
    \end{array}
\end{equation*}
}

\newpage

Quartic case genus 3:

{\small
\begin{equation*}
    \begin{array}{c c c c c}
          \text{Case} & \text{Hamiltonian} & \text{Power series} & \text{Newton polygon} & \text{Surface type} \\[2ex]
          \hline 
         \hyperref[sec:4_simple_points]{\text{Q1}} & \begin{aligned}
                 xy(x-y)(x-t\,y) + a_{21}(z)\, x^2y  \\
                 + a_{12}(z)\, xy^2 + a_{20}(z)\, x^2 + a_{11}(z)\, xy\\
                  + a_{02}(z)\, y^2 + a_{10}(z)\, x + a_{01}(z)\, y
             \end{aligned} & \left( \frac{1}{2},\frac{1}{2},\frac{1}{2},\frac{1}{2}\right) & \includegraphics[width=.1\textwidth,valign=c]{newton_polygon_Q1.pdf} & \includegraphics[width=.10\textwidth,rotate=45,valign=c]{Q1.pdf}  \\
         \hline \\
         \hyperref[sec:1_double_2_simple]{\text{Q2}} & \begin{aligned}
             x^2 (x - y) \,y + b_{21}(z)\, x^2 \,y \\
             + b_{03}(z)\, y^3 + b_{20}(z)\, x^2 + b_{11}(z)\, xy \\ 
             + b_{02}(z)\, y^2   + b_{10}(z)\, x + b_{01}(z)\, y
         \end{aligned} & \left( \frac{1}{2},\frac{1}{2},\frac{1}{3}\right) & \includegraphics[width=.1\textwidth,valign=c]{newton_polygon_Q2.pdf} & \includegraphics[width=.11\textwidth,valign=c]{Q2.pdf} \vspace{2ex} \\
         \hline \\
          \hyperref[sec:2_double_points]{\text{Q3}} & \begin{aligned}
              x^2y^2 + c_{30}(z)\,x^3 \\
              + c_{03}(z)\,y^3 + c_{20}(z)\,x^2 + c_{11}(z) \,xy \\ 
              + c_{02}(z)\, y^2 + c_{10}(z)\, x + c_{01}(z)\, y
          \end{aligned} & \left( \frac{1}{3},\frac{1}{3}\right) & \includegraphics[width=.1\textwidth,valign=c]{newton_polygon_Q3.pdf}& \includegraphics[width=.13\textwidth,valign=c]{Q3.pdf} \vspace{2ex} \\
         \hline \\
         \hyperref[sec:1_triple_1_simple_point]{\text{Q4}} & \begin{aligned}
             x^3 y + d_{12}(z) x y^2 + d_{03}(z) y^3 \\ 
             + d_{20}(z) x^2 + d_{11}(z) xy + d_{02}(z) y^2 \\
             + d_{10}(z) x + d_{01}(z) y
         \end{aligned} & \left( \frac{1}{2},\frac{1}{4}\right) & \includegraphics[width=.1\textwidth,valign=c]{newton_polygon_Q4.pdf} & \includegraphics[width=.15\textwidth,valign=c]{Q4.pdf} \vspace{2ex} \\[1ex]
         \hline \\
         \hyperref[sec:1_quadrupole]{\text{Q5}} & \begin{aligned}
             x^4  + e_{21}(z)\, x^2 y + y^3 \\
             + e_{12}(z)\, x y^2  + e_{20}(z)\, x^2 + e_{11}(z)\, x y\\
              + e_{02}(z)\, y^2 + e_{10}(z)\, x + e_{01}(z)\, y
         \end{aligned} & \left( \frac{1}{5}\right) & \includegraphics[width=.1\textwidth,valign=c]{newton_polygon_Q5.pdf} & \includegraphics[width=.2\textwidth,valign=c]{Q5.pdf} \vspace{2ex} \\
         \hline 
    \end{array}
\end{equation*}
}

\vspace{6ex}

\section{Blow-ups for the general case Q5} \label{app:blow-up_Q5}
We consider the cascade of $16$ blow-ups required for the Hamiltonian system associated with the Hamiltonian~\eqref{eq:Ham_Q5} to be regularised. Although the space of initial conditions was already determined in~\cite{KeckerFilipuk}, we list here the sequence of points at which to blow-up the surface, since here we consider an additional intermediate change of variables. 

In the compactified version of the complex plane there is one base point only, namely 
\begin{equation}
    p_1 \colon (U_0, V_0) = (0,\,0)\,.
\end{equation}
At the step $4$ we have implemented the 3-fold covering $(u_4,v_4) \mapsto (\widehat{u}_4,\widehat{v}_4)$
\begin{equation}
    u_4 = \frac{\widehat{u}_4}{\widehat{v}_4} \,, \qquad v_4 = \widehat{v}^3\,. 
\end{equation}
Here and in the following, we use the notation $'$ for differentiation with respect to the independent variable, $z$.
From the point $p_1$ the following cascade of blow-ups emerges 
\begin{equation}
\begin{split}
   & p_1  \,\,\leftarrow\,\, p_2 \colon\left( u_1,v_1 \right) = \left( 0 
 \,, 0 \right) \,\,\leftarrow\,\, p_3 \colon\left( u_2,v_2 \right) = \left( 0 
 \,, 0\right) \,\,\leftarrow\,\,  p_4 \colon\left( u_3,v_3 \right) = \left( 0 
 \,, 0\right) \,\,\leftarrow\,\,  \\[1ex]
 & \,\,\leftarrow\,\, p_5 \colon\left( \widehat{u}_4,\widehat{v}_4 \right) = \left( 0 
 \,, -1\right)  \,\,\leftarrow\,\, p_6 \colon\left( u_5,v_5 \right) = \left( 0 
 \,,  -\frac{e_{12}(z)}{3}  \right) \,\,\leftarrow\,\, \\[1ex]
 & \,\,\leftarrow\,\, p_7 \colon\left( u_6,v_6 \right) = \left( 0 
 \,,  -\frac{e_{12}(z)^2}{9}+-\frac{e_{21}(z)}{3}  \right) \,\,\leftarrow\,\,  p_8 \colon\left( u_7,v_7 \right) = \left( 0 
 \,,  -2 \,\frac{e_{12}(z)^3}{81} +\frac{e_{12}(z)\,e_{21}(z)}{9}  \right) \,\,\leftarrow\,\, \\[.5ex]
 & \,\,\leftarrow\,\, p_9 \colon\left( u_8,v_8 \right) = \left( 0 
 \,,  -\frac{e_{02}(z)}{3}  \right) \,\,\leftarrow\,\,  p_{10} \colon\left( u_9,v_9 \right) = \big( 0 
 \,, f_9(z)   \big) \,\,\leftarrow\,\,  p_{11} \colon\left( u_{10},v_{10} \right) = \big( 0 
 \,,  f_{10}(z)  \big) \,\,\leftarrow\,\, \\[.5ex]
 & \,\,\leftarrow\,\, p_{12} \colon\left( u_{11},v_{11} \right) = \left( 0 
 \,,  \frac{1}{5}\left(\frac{e_{21}'(z)}{3}-\frac{(e_{12}(z)^2)'}{9} \right)  \right) \,\,\leftarrow\,\,  p_{13} \colon\left( u_{12},v_{12} \right) = \big( 0 
 \,,  f_{12}(z)  \big) \,\,\leftarrow\,\, \\[.5ex]
 & \,\,\leftarrow\,\, p_{14} \colon\left( u_{13},v_{13} \right) = \big( 0 
 \,,  f_{13}(z)   \big) \,\,\leftarrow\,\,  p_{15} \colon\left( u_{14},v_{14} \right) = \big( 0 
 \,,  f_{14}(z)   \big) \,\,\leftarrow\,\,  p_{16} \colon\left( u_{15},v_{15} \right) = \big( 0 
 \,,   f_{15}(z)  \big) 
 \end{split} 
\end{equation}
with the functions $f_i(z)$ are 
{\small 
\begin{align} 
    f_9(z)  &= \frac{2}{3}\left( \frac{e_{12}(z)}{3} \right)^{\!5} - \frac{5}{3}\,\frac{e_{12}(z)^3 e_{21}(z)}{81} + \frac{e_{12}(z)}{3} \left( \frac{e_{21}(z)^2}{3} - 2\,\frac{e_{02}(z)}{3} \right) + \frac{e_{11}(z)}{3}\,,  \\[2ex]
    \begin{split} 
    f_{10}(z) & = \frac{7}{9} \left( \frac{e_{12}(z)}{3} \right)^{\!6} +  \frac{7}{3}    \left( \frac{e_{12}(z)}{3} \right)^{\!4} \frac{e_{21}(z)}{3} + 2\left(\frac{e_{12}(z)}{3} \right)^{\!2} \left( \left( \frac{e_{21}(z)}{3} \right)^{\!2} - \frac{e_{02}(z)}{3} \right) - \frac{1}{3}\left( \frac{e_{21}(z)}{3} \right)^{\!3}  \\[.5ex]
    &~~ + \frac{e_{02}(z)\,e_{21}(z)}{9} - \frac{e_{11}(z)\,e_{12}(z)}{9} - \frac{e_{12}'(z)}{18} - \frac{e_{20}(z)}{3}\,,  
    \end{split} 
    \end{align} }
    {\small 
    \begin{equation} 
    \begin{split} 
    f_{12}(z) & = -\frac{11}{9}\left( \frac{e_{12}(z)}{3} \right)^{\!8} + \frac{44}{9} \left( \frac{e_{12}(z)}{3}  \right)^{\!6} \frac{e_{21}(z)}{3} + \frac{1}{3}\! \left( \frac{e_{12}(z)}{3}  \right)^{\!4} \!\left( 10\,\frac{e_{02}(z)}{3} - 20\left(\frac{e_{21}(z)}{9} \right)^{\!2} \right) \\[.5ex]
    &~~+ \frac{5}{3}\left(\frac{e_{12}(z)}{3}\right)^{\!3}\frac{e_{11}(z)}{3} + \left( \frac{e_{12}(z)}{3} \right)^{\!2} \left(  \frac{10}{3} \left( \frac{e_{12}(z)}{3} \right)^{\!3} - 5\, \frac{e_{02}(z)\,e_{21}(z)}{9} + \frac{e_{20}(z)}{3} - \frac{e_{12}'(z)}{9} \right) \\[.5ex]
    &~~+ \frac{e_{12}(z)}{3} \left( 2\, \frac{e_{11}(z)\,e_{21}(z)}{9}+\frac{1}{4}\,\frac{e_{21}'(z)}{3} \right) - \frac{1}{3} \left( \frac{e_{21}(z)}{3} \right)^{\!4}  + \left(\frac{e_{21}(z)}{3}\right)^{\!2} \frac{e_{02}(z)}{3} \\[.5ex]
    &~~ + \frac{e_{21}}{3} \left( \frac{1}{4}\,\frac{e_{21}'(z)}{3} - \frac{e_{20}(z)}{3} \right) - \left( \frac{e_{02}(z)}{3} \right)^{\!2} + \frac{e_{01}(z)}{3} \,,
    \end{split} 
    \end{equation} }
    \vspace{2ex}
    {\small 
    \begin{equation} 
    \begin{split} 
    &\hspace*{-7ex}f_{13}(z)  =-\frac{130}{81} \left( \frac{e_{12}(z)}{3} \right)^{\!\!9}\! +\frac{65}{9}\! \left(\frac{ e_{12}(z)}{3}\right)^{\!\!7} \!\frac{ e_{21}(z)}{3} +\frac{7}{3}\left(\frac{e_{12}(z)}{3}\right)^{\!\!5} \!\left(2\,\frac{e_{02}(z)}{3}-5\left(\frac{ e_{21}(z)}{3} \right)^{\!\!2}\right)\! \\[.5ex]
    &\hspace*{-3ex}~~-\frac{7}{3} \frac{e_{11}(z)}{3} \left(\frac{e_{12}(z)}{3}\right)^{\!\!4} +\left(\frac{e_{12}(z)}{3}\right)^{\!\!3} \! \left(\frac{70}{9}\! \left( \frac{e_{21}(z)}{3}\right)^{\!\!3}\!-\frac{28}{3} \!\left( \frac{e_{02}(z)\, e_{21}(z)}{9}+\frac{e_{12}'(z)}{45} \right)+\frac{4}{3} \frac{e_{20}(z)}{3}\right) \\[.5ex]
    &\hspace*{-3ex}~~+\left(\frac{e_{12}(z)}{3}\right)^{\!\!2} \!\left(4 \,\frac{e_{11}(z) \,e_{21}(z)}{9}-\frac{e_{21}'(z)}{45}\right) -\frac{ e_{11}(z)}{3} \!\left(\frac{ e_{21}(z)}{3}\right)^{\!\!2}+\frac{ e_{21}(z) \,e_{21}'(z)}{45}  \\[.5ex]
    &\hspace*{-3ex}~~+\frac{e_{12}(z)}{3} \left( \frac{e_{21}(z)}{3} \!\left(4\, \frac{e_{21}(z)\, e_{02}(z)}{9} - \frac{5}{3}\, \frac{e_{21}(z)^3}{9} - \frac{11}{45} \frac{e_{12}'(z)}{9}-\frac{2}{3}\, \frac{e_{20}(z)}{3}\right)\! +\frac{e_{01}(z)}{3}-2 \!\left(\frac{ e_{02}(z)}{3}\right)^{\!\!2}\right)\\[.5ex]
    &\hspace*{-3ex}~~-\frac{e_{02}'(z)}{9}+\frac{e_{02}(z) \,e_{11}(z)}{9}-\frac{e_{10}(z)}{3} \,, 
    \end{split} 
    \end{equation} }
    \vspace{2ex}
    {\small
    \begin{equation} 
    \begin{split} 
    &\hspace*{-7ex}f_{14}(z)  =\frac{1}{10}  \frac{e_{12}(z)^4 \,e_{12}'(z)}{9} -\frac{1}{10} \frac{e_{12}(z)^3 \,e_{21}'(z)}{6} +\frac{e_{12}(z)}{3} \left(\frac{11}{10} \, \frac{e_{21}(z) \, e_{21}'(z)}{3}-\frac{e_{02}'(z)}{9}\right)\\[.5ex]
    &\hspace*{-3ex}~~+\left(\frac{e_{21}(z)^2}{18}-\frac{1}{10} e_{12}(z)^2\, e_{21}(z) \right) \frac{e_{12}'(z)}{2}  +\frac{ e_{11}'(z)}{6}-\frac{e_{02}(z) e_{12}'(z)}{9}  \,,
    \end{split} 
    \end{equation} } 
    \vspace{2ex}
    {\small
    \begin{equation}
    \begin{split} 
    &\hspace*{-7ex} f_{15}(z)  = \frac{238}{81} \!\left( \frac{e_{12}(z)}{3}\right)^{\!\!11}\!\!-\frac{1309}{27} \!\left( \frac{e_{12}(z)}{3}\right)^{\!\!9} \frac{e_{21}(z)}{3} +\left(\frac{e_{12}(z)}{3}\right)^{\!\!7} \!\! \left(\frac{308}{9} \frac{ e_{21}(z)^2}{9}-\frac{88}{9} \frac{e_{02}(z)}{3}\right)  \\[.5ex]
    &\hspace*{-3ex}~~+\left(\frac{e_{12}(z)}{3}\right)^{\!\!6}\!\frac{44}{9} \frac{e_{11}(z) }{3}  -\left(\frac{e_{12}(z)}{3}\right)^{\!\!5} \!\left(\frac{308}{9} \! \left(\frac{e_{21}(z)}{3}\right)^{\!\!3}\!-\frac{88}{3} \frac{e_{02}(z) \,e_{21}(z)}{9}-\frac{44}{9} \frac{e_{12}'(z)}{3}+\frac{8}{3} \frac{e_{20}(z)}{3}\right)\\[.5ex]
    &\hspace*{-3ex}~~-\!\left(\frac{e_{12}(z)}{3}\right)^{\!\!4}\! \left(\frac{40}{3} \frac{e_{11}(z) \,e_{21}(z)}{9}+\frac{8}{3} \frac{e_{21}'(z)}{3}\right) +\frac{e_{21}(z)}{3}  \left(\frac{2}{3} \frac{e_{02}'(z)}{3}+\frac{2}{3} \frac{e_{02}(z) \, e_{11}(z)}{9}-\frac{e_{10}(z)}{3}\right)  \\[.5ex]
    &\hspace*{-3ex}~~+\!\left(\frac{e_{12}(z)}{3}\right)^{\!\!3}\! \left(\frac{20}{9}\frac{e_{21}(z)}{3}\!\left(\frac{7\,e_{21}(z)^3}{27} -\frac{4\,e_{02}(z)\, e_{21}(z)}{9} +\frac{e_{20}(z)}{3}-\frac{43}{30} \frac{e_{12}'(z)}{3} \right)\! -\frac{5}{3} \!\left(\frac{e_{01}(z)}{3}+\frac{4\, e_{02}(z)^2}{9} \right)\!\right)\\[.5ex]
    &\hspace*{-3ex}~~+\!\left(\frac{e_{12}(z)}{3}\right)^{\!\!2}\! \left(10\, \frac{ e_{11}(z) \, e_{21}(z)^2}{27}-\frac{5}{9} \!\left( \frac{e_{02}'(z)}{3}-\frac{ e_{02}(z)\, e_{11}(z)}{9} \right)\!+\frac{e_{10}(z)}{3}+ \left(e_{21}(z)^2\right)'\right)  \\[.5ex] 
    &\hspace*{-3ex}~~+\frac{e_{12}(z)}{3} \left[\frac{e_{21}(z)^2}{9}\!\left(-\frac{7}{9}\frac{ e_{21}(z)^4}{81}+\frac{20}{9} \frac{e_{02}(z) \,e_{21}(z)^2}{27} + \frac{23}{18} \frac{e_{12}'(z)}{9}-\frac{4}{9} \frac{e_{20}(z)}{3}\right) +\frac{e_{11}'(z)}{9}+\frac{e_{11}(z)^2}{27}
    \right.   \\[.5ex]
    &\hspace{7ex}~~ \left. +e_{21}(z) \left(\frac{2 e_{01}(z)}{27}-\frac{5 e_{02}(z)^2}{81}\right) -\frac{11}{3} \frac{e_{02}(z)\, e_{12}'(z)}{3}+\frac{2}{3} \frac{e_{02}(z)\, e_{20}(z)}{3}\right]   \\[.5ex]
    &\hspace*{-3ex}~~+\frac{e_{11}(z)}{3}\!\left(\frac{5}{6} \frac{ e_{12}'(z)}{3}+\frac{1}{9}  e_{20}(z)-\frac{4}{243} e_{21}(z)^3\right)\!+\!\left(\frac{e_{02}(z)}{3}-\frac{e_{21}(z)^2}{9} \right)\! \frac{e_{21}'(z)}{3}-\frac{e_{12}''(z)}{18}-\frac{e_{20}'(z)}{3}\,.
    \end{split} 
\end{equation}}

\end{appendices}

\bibliographystyle{style}
\bibliography{thebiblio}

\end{document}